\documentclass[a4paper,fleqn,usenatbib,useAMS]{mnras}
\usepackage{graphicx}	
\usepackage{amsmath}	
\usepackage{amssymb}    
\usepackage{amssymb}	
\usepackage{multicol}	
\usepackage{bm}		
\usepackage{pdflscape}	
\usepackage{hyperref}
\usepackage{ulem} 
\usepackage{lineno} 
 
\newcommand{\redmapper}{redMaPPer}

\newcommand{\hmsun}{h^{-1} {\rm M}_{\odot}}

\newcommand{\vel}{\,{\rm km\,s^{-1}}}
\newcommand{\de}{\text{d}}

\newcommand{\Mpc}{\rm{Mpc}}

\newcommand{\LCDM}{$\Lambda$CDM }

\newcommand{\ob}{^{\rm ob}}
\newcommand{\true}{^{\rm true}}
\newcommand{\Mmin}{M_{\rm min}}
\usepackage[usenames]{color}
\definecolor{purple}{RGB}{150,0,200}

\newcommand{\planck}{{\it Planck}}

\newcommand{\avg}[1]{\langle #1 \rangle}
\newcommand{\Var}{\mbox{Var}}
\newcommand{\hMpc}{h^{-1}\ \Mpc}
\newcommand{\hMsun}{h^{-1}\ {\rm M}_{\odot}}

\newcommand{\Planck}{{\it Planck}}

\usepackage[T1]{fontenc}
\usepackage{ae,aecompl}

\usepackage{txfonts}

\title[SDSS redMaPPer Cluster Cosmology]{Dark Energy Survey Year 1 Results: Methods for Cluster Cosmology and Application to the SDSS}

\author[DES Collaboration]{
\parbox{\textwidth}{
\Large
M.~Costanzi,$^{1}$
E.~Rozo,$^{2}$
M.~Simet,$^{3}$
Y.~Zhang,$^{4}$
A.~E.~Evrard,$^{5,6}$
A.~Mantz,$^{7}$
E.~S.~Rykoff,$^{7,8}$
T.~Jeltema,$^{9}$
D.~Gruen,$^{7,8}$
S.~Allen,$^{10}$
T.~McClintock,$^{2}$
A.~K.~Romer,$^{11}$
A.~von der Linden,$^{12}$
A.~Farahi,$^{13}$
J.~DeRose,$^{10,7}$
T.~N.~Varga,$^{14,1}$
J.~Weller,$^{15,14,1}$
P.~Giles,$^{11}$
D.~L.~Hollowood,$^{9}$
S.~Bhargava,$^{16}$
A.~Bermeo-Hernandez,$^{16}$
X.~Chen,$^{6}$
T.~M.~C.~Abbott,$^{17}$
F.~B.~Abdalla,$^{18,19}$
S.~Avila,$^{20}$
K.~Bechtol,$^{21}$
D.~Brooks,$^{18}$
E.~Buckley-Geer,$^{4}$
D.~L.~Burke,$^{7,8}$
A.~Carnero~Rosell,$^{22,23}$
M.~Carrasco~Kind,$^{24,25}$
J.~Carretero,$^{26}$
M.~Crocce,$^{27,28}$
C.~E.~Cunha,$^{7}$
L.~N.~da Costa,$^{22,23}$
C.~Davis,$^{7}$
J.~De~Vicente,$^{29}$
H.~T.~Diehl,$^{4}$
J.~P.~Dietrich,$^{15,30}$
P.~Doel,$^{18}$
T.~F.~Eifler,$^{31,32}$
J.~Estrada,$^{4}$
B.~Flaugher,$^{4}$
P.~Fosalba,$^{27,28}$
J.~Frieman,$^{4,33}$
J.~Garc\'ia-Bellido,$^{34}$
E.~Gaztanaga,$^{27,28}$
D.~W.~Gerdes,$^{5,6}$
T.~Giannantonio,$^{35,36,1}$
R.~A.~Gruendl,$^{24,25}$
J.~Gschwend,$^{22,23}$
G.~Gutierrez,$^{4}$
W.~G.~Hartley,$^{18,37}$
K.~Honscheid,$^{38,39}$
B.~Hoyle,$^{14,1}$
D.~J.~James,$^{40}$
E.~Krause,$^{31}$
K.~Kuehn,$^{41}$
N.~Kuropatkin,$^{4}$
M.~Lima,$^{42,22}$
H.~Lin,$^{4}$
M.~A.~G.~Maia,$^{22,23}$
M.~March,$^{43}$
J.~L.~Marshall,$^{44}$
P.~Martini,$^{38,45}$
F.~Menanteau,$^{24,25}$
C.~J.~Miller,$^{5,6}$
R.~Miquel,$^{46,26}$
J.~J.~Mohr,$^{15,30,14}$
R.~L.~C.~Ogando,$^{22,23}$
A.~A.~Plazas,$^{32}$
A.~Roodman,$^{7,8}$
E.~Sanchez,$^{29}$
V.~Scarpine,$^{4}$
R.~Schindler,$^{8}$
M.~Schubnell,$^{6}$
S.~Serrano,$^{27,28}$
I.~Sevilla-Noarbe,$^{29}$
E.~Sheldon,$^{47}$
M.~Smith,$^{48}$
M.~Soares-Santos,$^{49}$
F.~Sobreira,$^{50,22}$
E.~Suchyta,$^{51}$
M.~E.~C.~Swanson,$^{25}$
G.~Tarle,$^{6}$
D.~Thomas,$^{20}$
and R.~H.~Wechsler$^{10,7,8}$
\begin{center} (DES Collaboration) \end{center}
}
}

\usepackage{eso-pic}
\AddToShipoutPictureBG*{%
  \AtPageUpperLeft{%
    \hspace{0.75\paperwidth}%
    \raisebox{-3.5\baselineskip}{%
      \makebox[0pt][l]{\textnormal{DES-2017-0319}}
}}}%

\AddToShipoutPictureBG*{%
  \AtPageUpperLeft{%
    \hspace{0.75\paperwidth}%
    \raisebox{-4.5\baselineskip}{%
      \makebox[0pt][l]{\textnormal{FERMILAB-PUB-18-554}}
}}}%

\hypersetup{draft}

\begin{document}
\label{firstpage}
\pagerange{\pageref{firstpage}--\pageref{lastpage}}
\maketitle
\begin{abstract}
We perform the first blind analysis of cluster abundance data. Specifically, we derive cosmological constraints from the abundance and weak-lensing signal of \redmapper\ clusters of richness $\lambda\geq 20$ in the redshift range $z\in[0.1,0.3]$ as measured in the Sloan Digital Sky Survey (SDSS).  We simultaneously fit for cosmological parameters and the richness--mass relation of the clusters.  For a flat $\Lambda$CDM cosmological model with massive neutrinos, we find $S_8 \equiv \sigma_{8}(\Omega_m/0.3)^{0.5}=0.79^{+0.05}_{-0.04}$.  This value is both consistent and competitive with that derived from cluster catalogues selected in different wavelengths.  Our result is also consistent with the combined probes analyses by the Dark Energy Survey (DES) and the Kilo-Degree Survey (KiDS), and with the Cosmic Microwave Background (CMB) anisotropies as measured by \planck. We demonstrate that the cosmological posteriors are robust against variation of the richness--mass relation model and to systematics associated with the calibration of the selection function.  In combination with Baryon Acoustic Oscillation (BAO) data and Big-Bang Nucleosynthesis (BBN) data \citep{cookeetal16}, we constrain the Hubble rate to be $h=0.66\pm 0.02$, independent of the CMB.  Future work aimed at improving our understanding of the scatter of the richness--mass relation has the potential to significantly improve the precision of our cosmological posteriors.  The methods described in this work were developed for use in the forthcoming analysis of cluster abundances in the DES.  Our SDSS analysis constitutes the first part of a staged-unblinding analysis of the full DES data set.
\end{abstract}

\setcounter{footnote}{1}


\section{Introduction}
\label{sec:intro}

Galaxy clusters form from the high density peaks of the initial matter distribution.  As such, they bear the imprints of the statistical properties of the matter density field and its growth \citep[for reviews, see e.g.][]{Allen2011,Kravtsov2012}. The abundance of galaxy clusters has been used since the late 90's to constrain the mean matter density of the Universe, $\Omega_m$, and the amplitude of the density fluctuations in terms of $\sigma_8$, the present-day rms of the linear density field in spheres of $8 \hMpc$ radii \citep[e.g.][for early works]{Eke1998,Henry2000,Borgani2001,Pierpaoli2001,Reiprich2002,Henry2004}.
Constraints on $\sigma_8$ are especially powerful in combination with measurements of the amplitude of the matter power spectrum at high redshift --- e.g. from CMB data --- enabling us to study the growth of density perturbations over cosmic time.  These studies allow us to place constraints on parameters such as the total neutrino mass, the dark energy equation of state, and parameters governing modified gravity models.

Current studies using cluster catalogs selected in the X-ray, optical, and millimeter wavelengths, provide consistent constraints on $\sigma_8$ and $\Omega_m $ \citep[e.g.][]{Vikh2009,Rozo2010,Mantz2010,Mantz2015,PlanckSZ2016,SPTSZ2016}.  These data have been also used in combination with $H_0$ and Baryon Acoustic Oscillation (BAO) priors to place competitive constraints on the dark energy equation of state parameter, modification of gravity and neutrino masses \citep[e.g.][]{Burenin2012,Mantz2010,Mantz2015,Cataneo2015,PlanckSZ2016,SPTSZ2016}.
Ongoing --- e.g. the Dark Energy Survey (DES)\footnote{https://www.darkenergysurvey.org}, the Hyper
Suprime-Cam\footnote{http://hsc.mtk.nao.ac.jp/ssp/} --- and forthcoming --- the Large Synoptic Survey
Telescope\footnote{https://www.lsst.org/}, Euclid\footnote{http://sci.esa.int/euclid/}, eRosita\footnote{http://www.mpe.mpg.de/eROSITA} --- wide-area surveys aim to use clusters samples with an order of magnitude more systems than previous analyses in order to improve upon current constraints.

The most critical difficulty that cluster abundance studies must confront is the fact that cluster masses are not easily measured, forcing us to rely on observational proxies that correlate with mass. Specifically, while it is possible to predict the abundance of dark matter halos as a function of mass in an arbitrary cosmology with percent level accuracy \citep[e.g.][]{sheth1999,Tinker2008,Crocce2010,Bocquet2016,McClintock2018}, halo masses themselves are not directly observable.  At present, cosmological constraints from cluster abundance analyses at all wavelengths are limited by the uncertainty in the calibration of the relation between the cluster mass and the observable property used as a mass proxy, be it richness (i.e. count of member galaxies), X-ray luminosity, or the thermal Sunyaev-Zeldovich signal.  

Currently, weak gravitational lensing measurements provide the gold-standard technique for estimating cluster masses \citep[see e.g.][for a discussion]{vdlinden2014}. The weak lensing signal, i.e. the tangential alignment of background galaxies around the foreground cluster due to gravitational lensing, is a well-understood effect, sensitive to both dark and baryonic matter. Moreover, in contrast to other techniques  (e.g. velocity dispersion and hydrostatic mass measurements), weak lensing mass measurements do not rely on assumptions about the dynamical state of the cluster. Despite these advantages, many sources of systematic error do affect this type of measurement,  e.g. shear and photometric redshift biases, halo triaxiality, and projection effects.  These systematics represent a significant amount of the total error budget of many recent studies \citep[e.g.][]{wtgI,hoekstraetal15,Simet2016,Melchior2017,medezinksietal18,miyatakeetal18}.  Not surprisingly, as the statistical uncertainty continues to decrease, these systematics have come to dominate the total error budget \citep[e.g.][]{desy1wl}.

In this work we combine cluster abundances and stacked weak lensing mass measurements from the Sloan Digital Sky Survey data release 8 \citep[SDSS DR8,][]{Aihara2011} to simultaneously constrain cosmology and the richness--mass relation of galaxy clusters.  Our cluster sample is selected using the red sequence Matched-filter Probabilistic Percolation algorithm \citep[redMaPPer;][]{Rykoff2014}, and the stacked weak lensing mass estimates are presented in \citet{Simet2016}.
The analysis is similar in spirit to that of \citet{Rozo2010} but with significant updates, particularly with regards to the modeling of the cluster selection function.  Our observables are the number of clusters and the mean cluster mass in bins of richness --- our mass proxy --- and redshift.
We explicitly account for the small cosmological dependence of the recovered weak lensing masses in our analysis. To avoid confirmation bias, the bulk of the analysis has been performed blind: the values of the cosmological parameters sampled by the Monte Carlo Markov Chains (MCMC) were randomly displaced by an amount that was unknown to us, and were shifted back only after a broad set of validation tests were passed.  No changes were done to the analysis pipeline post-unblinding.  This is the first cosmology analysis to be performed blind in the cosmological parameters.  The methods presented in this paper were developed for the forthcoming cosmological analysis of the DES Y1 \redmapper\ cluster catalog (DES Collaboration in prep.).

This paper is organized as follows. In Section \ref{sec:datasets} we introduce the data used for this study. In Section \ref{sec:methods} we present the model used to perform the cosmological analysis, including expectation values for the two observables, modeling of the systematics and likelihood model. We validate our model by means of synthetic data in Section \ref{sec:res:mock_data}. We detail our blinding procedure in Section \ref{sec:blind}.The results of our analysis are presented in Section \ref{sec:res} and \ref{sec:ext}. Finally we conclude in Section \ref{sec:summary}. 

\section{Data}
\label{sec:datasets}

\subsection{Cluster and Weak Lensing Catalogs}
\label{sec:datasets:cl_catalog}

Both the cluster and weak lensing shear catalogs used in this analysis are based on the Sloan Digital Sky Survey data release 8 \citep[SDSS DR8,][]{Aihara2011}. A summary of the data employed in this analysis is presented in Table \ref{tab:data}. Throughout the paper, all masses are given in units of ${\rm M}_\odot / h$, where $h=H_0/100\vel \Mpc^{-1}$, and refer to an overdensity of 200 with respect to the mean. We use "$\log$" and "$\ln$" to refer to the logarithm with base $10$ and $e$ respectively.


\begin{table*}
    \centering
    \footnotesize
    \caption{Summary of the data and systematic corrections adopted in the analysis. We use the same redshift range, $0.1\leq z \leq 0.3$, for all the richness bins.
The second column lists the observed number counts and their uncertainties, the latter estimated as the square root of the diagonal terms of the best-fit model covariance matrix (see Section~\ref{sec:methods:like}). The numbers between parenthesis correspond to the number counts corrected for the miscentering bias factors listed in the third column (see Section \ref{sec:datasets:abundances}). The values of the mean cluster mass reported here, $\log(\bar{M}^{\rm WL}_{200,m})$, assume $\Omega_{\rm m}=0.30$; the two uncertainties shown correspond to the statistical and systematic error, respectively. We assume systematic errors to be fully correlated between the cluster bins. The fifth column lists the slopes which define the cosmological dependence of the weak lensing mass estimates (Eq. \ref{eqn:MvsOm}).
}
    \label{tab:data}
   \begin{tabular}{lcccc}
    \hline 
	$\Delta \lambda\ob ;\, z \in [0.1,0.3]$			&	Number counts	& $\gamma^{\rm Misc}$&	$\log(\bar{M}^{\rm WL}_{200,m}) [{\rm M}_\odot /h]$ & $\de \log M^{\rm WL}/\de \Omega_m$	\\
    \hline \vspace{-3mm}\\
	$[20,27.9)$			& $3604 \,(3711) \pm 100$& $1.030\pm0.011$ &$ 14.111 \pm 0.024 \pm 0.026$ & $-0.65$			\vspace{0.5mm} \\
    $[27.9,37.6)$		& $1740 \,(1788) \pm 61$& $1.028\pm0.011$ &$ 14.263 \pm 0.030 \pm 0.024$ & $-0.66$			\vspace{0.5mm} \\
    $[37.6,50.3)$		& $942 \,(978) \pm 41$ & $1.039\pm0.014$ &$ 14.380 \pm 0.033 \pm 0.026$ & $-0.68$			\vspace{0.5mm} \\
    $[50.3,69.3)$		& $461 \,(476) \pm 27$ & $1.034\pm0.015$ &$ 14.609 \pm 0.036 \pm 0.028$ & $-0.77$			\vspace{0.5mm} \\
    $[69.3,140)$		& $217 \,(223) \pm 18$ & $1.028\pm0.016$ &$ 14.928 \pm 0.029 \pm 0.036$ & $-0.65$			\vspace{0.5mm} \\
   \hline \vspace{-3mm}\\
    \end{tabular}
\end{table*}


We use photometrically selected galaxy clusters identified in the SDSS DR8 with 
the \redmapper\ cluster finding algorithm \citep{Rykoff2014}. In brief, \redmapper\
utilizes all five bands ({\it ugriz}) of the SDSS imaging to self-calibrate a model
for red-sequence galaxies.  This model is then used to
identify galaxy clusters as red galaxy overdensities, while simultaneously estimating 
the probability that each red galaxy is a cluster member.
The cluster richness $\lambda$ is the sum of the membership probabilities over all the red galaxies within an empirically calibrated scale radius, $R_\lambda$. Due to the probabilistic approach adopted by \redmapper\, an individual galaxy can contribute to the richness of more than one cluster. We use the cluster richness $\lambda$ as an observational mass-proxy in our analysis,
calibrating the relation between galaxy richness and cluster mass using weak lensing data.
Of the $\sim 14,000 \deg^2$ covered by SDSS DR8, we restrict ourselves to the $\sim 10,000 \deg^2$
of high quality contiguous imaging defined by the Baryon Oscillation Spectroscopic
Survey (BOSS) experiment \citep{Dawson2013}.  Typical 
photometric redshift uncertainties for SDSS \redmapper\ clusters are $\sigma_z/(1 + z) \lesssim  0.01$. 
We further restrict ourselves to the redshift range $z\in[0.1,0.3]$.  The lower redshift limit ensures that galaxies
are sufficiently dim that the SDSS photometric pipeline produces reliable galaxy magnitudes,
while the upper redshift limit is set by requiring that the cluster catalog be volume limited.
We also apply a richness threshold $\lambda\geq 20$.  In numerical simulations this richness threshold ensures that 99\% of the \redmapper\ galaxy clusters can be unambiguously mapped to individual dark matter halos \citep{farahietal16}. In this work, we use v5.10 of the SDSS \redmapper\ cluster catalog \citep{Rozo2015b}.

The weak lensing mass estimates employed in this analysis are a slight update from those presented in \citet{Simet2016}, and rely on the shear catalog presented in \citet{Reyes2012}.  This catalog covers $\approx 9,000\ \deg^2$ of the SDSS footprint
and contains 39 million galaxies, corresponding to a source density of 
$1.2\ \mbox{gal/arcmin$^2$}$.  Shear estimates were derived from the SDSS imaging
using the re-Gaussianization algorithm of \citet{hirataseljak2003} and the 
appropriately calibrated responsivity to convert the measured shape distortions into 
shear estimates.  The multiplicative shear bias appropriate for this catalog was characterized 
in \citet{mandelbaumetal12} and \citet{mandelbaumetal13}.  
In a recent work, \citet{mandelbaumetal2017} revised the multiplicative shear estimate
of this source catalog due to a previously undetected bias arising from
the impact of neighboring galaxies and unrecognized blends.  Our error budget
accounts for this additional source of error.  The photometric redshifts
for the sources in the shear catalog were obtained using the
Zurich Extragalactic Bayesian Redshift Analyzer \citep[ZEBRA,][]{feldmannetal2006},
and the associated systematic uncertainties were calibrated in \cite{nakajimaetal2012}.


\begin{figure*}
\begin{center}
    \includegraphics[width= \textwidth]{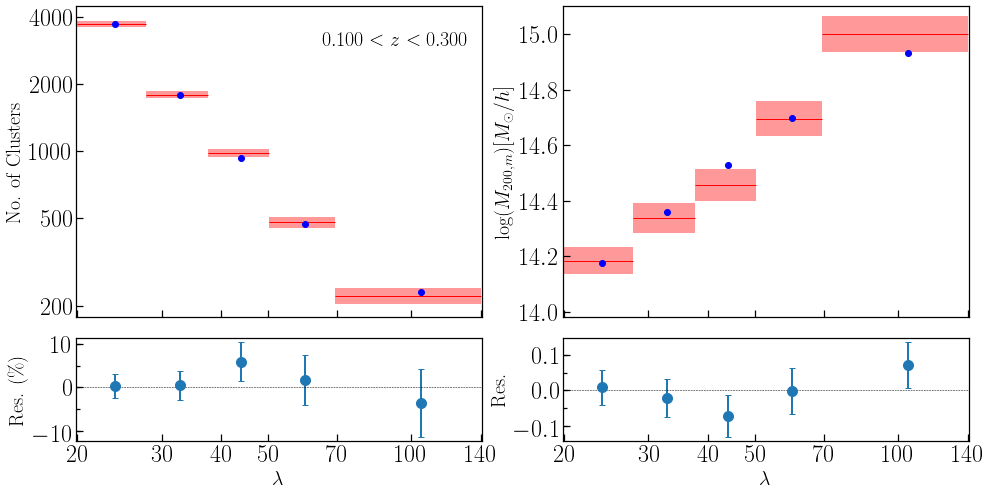}
\end{center}
\caption{Observed ({\it shaded area}) and best-fit model ({\it dots}) of cluster number counts ({\it left} panel) and mean cluster masses ({\it right} panel) in the five richness bins considered. The model predictions have been computed as described in Section~\ref{sec:methods} using the best-fit values derived from our analysis. The $y$ extent of the {\it shaded} areas is given by the square root of the diagonal terms of the corresponding covariance matrix. The lower panels show the percent residual (\textit{left}) and the residual (\textit{right}) of our best-fit model to the data.}
\label{fig:NC}
\end{figure*}


\subsection{Cluster Number Counts Data}
\label{sec:datasets:abundances}

Following the weak lensing analysis of \citet{Simet2016}, we collect our galaxy clusters in five richness bins and a single redshift bin (see Table \ref{tab:data}).  The richness limits of our bins are $\lambda\ob = [20,27.9,37.6,50.3,69.3,140]$.  The two key observational systematics in our analysis are photometric redshift errors and cluster miscentering.  Photometric redshift uncertainties are forward-modeled as described in section~\ref{sec:methods:model}.

Turning to cluster centering, in this work we assume that the correct center of a galaxy cluster 
is always coincident with a bright cluster galaxy (though not necessarily the brightest).  
This assumption is motivated by the fact that modern halo finders \citep[e.g.][]{Behroozi2013} define the
center of dark matter halos as the position of the dominant dark matter substructure within the halo.
It is expected that such a substructure will host a bright cluster galaxy, and therefore a proper
comparison of theory to data should rely on identifying the correct central galaxy for each cluster. 
In \redmapper\!, this is accomplished through an iteratively self-trained algorithm that combines luminosity
and local galaxy density information to determine which cluster galaxy is the best central candidate in the 
cluster. This algorithm is demonstrably superior to centering clusters on the brightest cluster-member
galaxy \citep{hikageetal17}.  Nevertheless,
miscentering systematics still arise from the fact that \redmapper\ does not always successfully identify the correct
central galaxy in each cluster.  As we shall see shortly, the uncertainty in the systematic corrections due 
to cluster miscentering are very nearly negligible, so rather than incorporating these corrections into
our forward-model, we have opted for the simpler route of applying a correction to the observed data-vector,
and adding the corresponding systematic uncertainty in the abundances to the covariance matrix characterizing the
uncertainty in our measurements.

Our miscentering correction is based on the analyses in Zhang et al. (in prep.) and von der Linden et al. (in prep.).  To characterize the probability of a \redmapper\ cluster being miscentered, we use subsamples of \redmapper\ clusters
with either {\it Chandra} or {\it Swift} X-ray imaging.  Specifically, we compared the \redmapper\ central galaxies to the X-ray derived cluster centers. Doing so, we demonstrated that $71.5\% \pm 6\%$ of \redmapper\ galaxy clusters are correctly centered.  This value is somewhat lower than but consistent with the estimate of \citet{hikageetal17} on the basis of the weak lensing profile of \redmapper\ clusters. In addition, we also characterized the radial offset distribution of the miscentered clusters using a 2-dimensional Gaussian to describe the distribution of positional offsets for the miscentered clusters.

To determine the impact that cluster centering has on the recovered cluster richness, we measured the richness of all 
\redmapper\ clusters at the location of the second most likely central galaxy, and characterized how 
cluster richness changes as a function of the centering offset.   As expected,
we find that miscentering systematically underestimates cluster richnesses, albeit with
some scatter.  Both the bias and scatter increase as the radial offset increases (Zhang et al., in prep.).  

Having characterized (i) the fraction of miscentered clusters; (ii) the distribution of radial offsets
of miscentered clusters; and (iii) how the richness of a cluster changes when it is miscentered by a given
radial offset, we can readily estimate the impact of miscentering on the cluster abundance function.  Specifically, we assigned richness values to halos in a numerical simulation using the model of \citet{costanzi18a}. The assigned halo richnesses are then scattered using our centering model.  We compute the ratio $\gamma^{\rm Misc}=N^{\rm cent}/N^{\rm miscent}$ between $N^{\rm cen}$, the number of clusters in a bin in the absence of miscentering, and $N^{\rm miscent}$, the number of clusters in a bin including the impact of miscentering.  We multiply the cluster abundance data vector by $\gamma^{\rm Misc}$ to correct for the impact of miscentering on the observed cluster abundances. From our Monte Carlo realizations, we are 
able to characterize both the mean correction and the corresponding covariance matrix.
We apply the $\gamma^{\rm Misc}\sim 1.030$ (see Table \ref{tab:data}) systematic correction to
the number of galaxy clusters we observe, and add the corresponding covariance matrix 
to the covariance matrix of the cluster counts.  The uncertainty associated with cluster
miscentering in the abundance function is $\approx 1.3\%$.  That is, cluster miscentering is
sub-dominant to Poisson noise in all richness bins.  Its impact on the cosmological posterior is
easily subdominant to the uncertainty in our calibration of the richness--mass relation.

The shaded regions in the top-left panel of Figure~\ref{fig:NC} shows the miscentering-corrected 
cluster abundances.  The width of the regions along the $y$-axis is set by the diagonal entries of the corresponding
covariance matrix, whereas their width along the $x$-axis reflects the width of the richness bin.  Our best-fit model is shown as the blue points. The bottom-left panel shows the corresponding percent residuals.


\subsection{Weak Lensing Cluster Masses}
\label{sec:datasets:wl_data}

\subsubsection{Measurement}
\label{sec:wlmeasurement}

We calculate the mean mass of clusters in a richness bin using the 
stacked weak lensing mass profiles of the clusters as described in \citet{Simet2016}.  
These profiles are modeled using a Navarro, Frenk, and White profile \citep[NFW,][]{NFW}, accounting for the effects of cluster miscentering, halo triaxiality, and projection effects.
The concentration of the best-fit NFW profiles is modeled using the mass-concentration relation presented in \citet{Diemer2015} with a free parameter for the amplitude.

In this work, we follow the methodology of \citet{Simet2016} to estimate the mean
cluster mass in each bin with three critical exceptions:
\begin{enumerate}
\item We correct for the $3\%\pm 3\%$ multiplicative shear bias due to undetected blends that was first identified
in \citet{mandelbaumetal2017}.  The original multiplicative bias in \citet{Simet2016} had a systematic
uncertainty of $\pm 5\%$ (top-hat).  To that we add a second bias term with a top-hat prior of width $3\% \pm 3\%$.  The sense of the correction is to {\it increase} the recovered weak lensing masses.  The standard deviation of the resulting distribution is $0.034$, i.e. roughly equivalent to a 3.4\% Gaussian scatter.
\item We update the centering priors employed in the weak lensing analysis
based on the work of Zhang et al. and von der Linden et al. (in preparation).  In \citet{Simet2016}, the fraction of correctly
centered galaxy clusters was estimated as $f=80\% \pm 7\%$ on the basis of published X-ray data.
The analyses in Zhang et al. and von der Linden et al. (in preparation) benefit from significantly improved statistics and
custom X-ray analyses specifically designed to calibrate this source of systematic uncertainty.
The revised fraction of correctly centered clusters is $f=0.715 \pm 0.06$.
The radial offset of miscentered clusters is described by a two dimensional Gaussian of width
$\tau=(0.29\pm 0.04)R_\lambda$, where $R_\lambda$ is the cluster radius assigned by \redmapper.
\item Rather than simultaneously modeling all galaxy clusters to derive a scaling relation that describes all richness bins, we estimate the mean mass of each individual richness bin.  To account for the finite width of the bin, we fit a scaling relation to each individual bin, and then use the posterior of the scaling relation parameters to determine the mean cluster mass in that bin.  It is these mean cluster masses that we use as the weak lensing observables in our cosmological analysis.
\end{enumerate}

The recovered mean mass in any single bin is only weakly dependent on the scatter assumed in the mass--richness relation when performing the fits.  The degeneracy of the weak lensing posterior takes the form  $\log \avg{M|\lambda,\sigma}= \log M + 0.06\sigma_{\ln M|\lambda}^2$.  That is, 
varying $\sigma_{\ln M|\lambda}$ over the range $[0.0,0.5]$ modifies the recovered masses
by an amount ranging from 0 to 0.015.  We adopt a fiducial correction of $0.007\pm 0.007$.
Following the analysis described in \citet[][see section 5.5]{Simet2016},
we further estimate the systematic uncertainty in our weak lensing masses due to modeling the
lensing profile with a pure NFW halo, without accounting for a two-halo term, or due to deviations
from the NFW profile \citep[e.g.][]{diemerkravtsov14}.  Briefly, we artificially populate
dark matter halos in an $N$-body simulation by drawing from the richness--mass relation obtained by
inverting our best fit mass--richness relation \citep{evrardetal14}.  We then compute the
corresponding cluster--mass correlation function, and project along the line of sight 
to obtain a predicted weak lensing profile, which we fit in the same way as we fit the
data. 
The recovered biases for each of our richness bins varies from 2\% on the low mass end
to 3\% on the high mass end.  We apply these corrections to our data, and 
assign a systematic uncertainty on this correction equal to half the magnitude of the correction.

Our results are summarized in Table~\ref{tab:data}, where we collect the best weak lensing
estimates for the mean mass of the galaxy clusters in each richness bin.  The logarithm of the mean mass, $\log \bar{M}$, for each of our five richness bins is shown in the top-right panel 
of Figure~\ref{fig:NC}, along with the best-fit model from our cosmological analysis.
The bottom-right panel shows the corresponding residuals. We again
emphasize that all aspects of this weak lensing analysis are as per \citet{Simet2016}
except as detailed above; our results are a minor update to that work.

\subsubsection{Weak Lensing Systematic Error Budget}

Cluster cosmology has long been limited by systematic uncertainties in mass calibration.  Here, we summarize the observational systematics we have
explicitly accounted for in our analysis and their contributions to the error budget of the mass estimates.  For further details on how these systematics
were estimated, we refer the reader to \citet{Simet2016}.  Table~\ref{tab:err_budget} summarizes the sources of error detailed below.

\begin{itemize}
\item Multiplicative shear and photo-$z$ bias ($\pm 5\%$ top-hat).  This bias reflects the 
total multiplicative systematic uncertainty arising from two distinct sources of 
error, each contributing roughly the same amount of uncertainty: 
1) systematic uncertainties in shear calibration as evaluated from realistic
image simulations \citep{reyesetal12,mandelbaumetal12,mandelbaumetal13}, and 2) photometric
redshift systematics \citep{nakajimaetal2012}.  This $\pm 5\%$ uncertainty is on the amplitude of the lensing
profile.  The equivalent mass error is obtained by multiplying by $\approx 1.3$, resulting
in a $\pm 6.5\%$ top-hat error in the mass.\footnote{The factor of 1.3 can be estimated by boosting a \citet{NFW} profile by a constant factor, estimating the new $R_{200m}$ aperture, and then verifying the ratio of the boosted mass to the original mass.  The end result is that the boost to the mass is 1.3 times the boost to the profile.} In practice, the mass measurements are marginalized
over the top-hat amplitude prior above.
\item Blending of lensing sources ($3\%\pm 3\%$ top hat). This refers
to the systematic bias introduced by the blending of sources discovered by \citet{mandelbaumetal2017},
and characterized using deep imaging from the Hyper-Suprime Camera.  Again, the 3\% uncertainty is
in the correction of the amplitude of the profile, leading to a $\approx 1.3\times 3\% = 3.9\%$ error
in the mass.
\item Cluster triaxiality and projection effects (3\%, Gaussian). Cluster triaxiality and projection effects both introduce correlated cluster scatter between cluster richness and weak lensing masses.  
As detailed in \citet[][section 5.3 and 5.4]{Simet2016}, we can readily compute the necessary corrections using the formalism laid out in \citet{evrardetal14}.  The estimate of the impact of triaxiality derived in this way is in excellent agreement with the simulation-based calibration of \citet{dietrichetal14}.  The quoted 3\% error is the uncertainty in the mass.
\item Cluster centering ($\leq 1\%$).  Cluster centering is explicitly incorporated into the weak lensing mass estimates via forward modeling.  The miscentering parameters have informative priors obtained from studies of the sub-sample of \redmapper\ galaxy clusters with high-resolution X-ray data (Zhang et al. and von der Linden et al., in preparation).  The weak lensing posterior fully accounts for the systematic uncertainty associated with cluster miscentering. The quoted 1\% error is the uncertainty in the mass.
\item Modeling systematics ($\sim 2\%$, Gaussian, richness dependent).
Modeling systematics refers to biases in the inferred cluster mass due to our
observational procedure. We explicitly calibrated this systematic using numerical simulations as detailed in the previous section.  The quoted 2\% error is the uncertainty in the mass.
\item Impact of the scatter on the recovered weak lensing masses (1.6\%, Gaussian).  This error
refers to the sensitivity of the weak lensing masses to the assumed scatter in the mass--richness relation detailed in section~\ref{sec:wlmeasurement}.  The error $\pm 0.007$ in $\log_{10} M$ corresponds to a 1.6\% error on the mass.
\item Systematic uncertainty in the boost factor corrections (likely negligible).  Misidentification of cluster galaxies as lensing sources dilutes the lensing signal.  However, these galaxies are obviously clustered around the cluster.  Consequently, measurements of the source--cluster correlation function allow one to directly measure the contamination rate of the source galaxy population as a function of radius, and thus estimate the boost factor needed to correct the lensing signal.  We explicitly checked that the statistical uncertainty associated with this measurement is negligible compared to the systematic errors in the multiplicative shear bias and photoz corrections.   However, small-scale features in the cluster or source selection function that are not adequately modeled by the associated random catalogs could lead to systematic uncertainties in the boost factor correction.  Here, we assume these errors are negligible.
\end{itemize}

The total systematic error budget for our observational estimate of the mean mass of the galaxy clusters
in a given richness bin is $\approx 6\%$ (Gaussian).  Two additional, less well-understood 
potential systematics remain.  These are:
\begin{itemize}
\item Intrinsic alignments: Radial alignments of member galaxies that are mistakenly selected as lensing sources
will bias the corresponding cluster mass. \citet{Simet2016} derives a 6\% upper limit for the
impact of intrinsic alignments, but does not correct for this effect nor enlarge the systematic error budget 
since this intrinsic alignment signal has yet to be detected in the low-luminosity cluster member galaxies that
can be potentially misidentified as lensing sources.  Note intrinsic alignments are uni-directional: explicitly modeling intrinsic alignments would increase the recovered cluster masses.  
\item Baryonic impact on the recovered cluster masses.  In \citet{Simet2016} it is argued that baryonic
effects were unlikely to have a large impact on the recovered cluster masses for two reasons.  First,
baryonic effects are strongest in the cluster core, which we exclude from our fits.  Second, we do not
impose priors on the amplitude and slope of the concentration--mass relation, allowing for the concentration to
absorb the effects that baryons have on the cluster lensing profile, so long as an NFW profile remains 
a reasonable description of the halo profiles \citep{schalleretal15a}.  This hypothesis has
received additional support from the work of \citet{hensonetal17}, who
found that while baryonic physics can clearly impact the lensing profile of galaxy clusters,
the bias incurred from fitting said profiles is independent of baryonic physics, provided the
concentration--mass relation of the clusters is allowed to vary.  Based on Figure 11 in that work,
it is difficult to imagine baryonic effects introducing biases larger than $\approx 3\%$.
\end{itemize}

We do not increase our systematic error budget on the expectation that both of these effects
will eventually be proven to be sub-dominant to the systematic errors quoted above.
We are not aware of any unaccounted-for systematic effects impacting the recovered weak lensing masses at
a level that exceeds that or is comparable to the systematics detailed above.  A complete summary of the error budget associated with the weak lensing mass calibration of the SDSS \redmapper\ clusters is detailed in Table~\ref{tab:err_budget}.

\begin{table*}
    \centering
    \footnotesize
    \caption{
    Error budget for the weak lensing mass calibration data used in our cosmological analysis.
    The ``Gaussian equivalent'' error for top-hat systematics refers to the square root of the 
    variance of the top-hat prior of the appropriate effect.  For instance, a Gaussian prior for shear and
    photo-$z$ biases of $\pm 2.8\%$ has the same variance as a top-hat prior of $\pm 5\%$.
    The statistical error quoted in the table refers to the uncertainty in the amplitude of the mass--richness relation. The statistical uncertainty on any individual mass estimate is larger.}
    \label{tab:err_budget}
   \begin{tabular}{ll}
    \hline 
	Source			&	Associated Error	\\
    \hline \vspace{-3mm}\\
	Shear and Photo-$z$ Bias & 6.5\% top-hat (3.8\% Gaussian equivalent) 		\\
    Source blending		& 3.9\% top-hat (2.3\% Gaussian equivalent)	\\
    Cluster triaxiality and projections	& 3\% Gaussian \\
    Cluster centering		& $\leq 1\%$  \\
    Modeling systematics	& 2.0\% Gaussian (richness dependent) \\
    Scatter corrections & 1.6\% Gaussian \\
   \hline \vspace{-3mm}\\
   Total Systematic Error &  6.0\% Gaussian \\
   Statistical Error & 4.8\% Gaussian \\
    \hline \vspace{-3mm}\\
	{\bf Total} &   7.7\%
    \end{tabular}
\end{table*}

\subsubsection{Cosmology Dependence of the Recovered Masses}

The weak lensing masses derived above depend
on the angular diameter distance to the sources and the mean matter density at the redshift of the clusters.  
These in turn depend on cosmological parameters.  Assuming a flat $\Lambda$CDM model, and using mass estimates in units of ${\rm M}_\odot / h$, the only cosmological
parameter that can appreciably impact the resulting weak lensing mass measurements is the 
matter density $\Omega_m$.  Consequently, we repeat our analysis along a grid of values
for $\Omega_m$ ranging from $\Omega_m=0.24$ to $\Omega_m=0.36$ while 
setting $\Omega_\Lambda = 1-\Omega_m$.
As in \citet{Simet2016}, we find the recovered
log-masses scale as a linear function in the matter density $\Omega_m$, i.e.
\begin{equation}
\label{eqn:MvsOm}
\log \bar{M}^{\rm WL}(\Omega_m) = \left . \log \bar{M}^{\rm WL} \right \vert_{\Omega_m = 0.3} +
	\left( \frac{d\log M^{\rm WL}}{d \Omega_m} \right)\left(\Omega_m - 0.3 \right )
\end{equation}

The slopes derived from fitting Equation \ref{eqn:MvsOm} to the data are listed in Table \ref{tab:data},
and are used in our cosmological analysis to re-scale $\log \bar{M}^{\rm WL}$ at each step of the MCMC by the appropriate $\Omega_m$ value.\footnote{Our posterior extends to matter densities below $\Omega_m=0.24$.
We verified \it a posteriori \rm that the linear matter density scaling extends to $\Omega_m=0.15$.}


\section{Theory and Methods}
\label{sec:methods}

In the previous section, we described how we constructed the observational vectors
(cluster abundances and weak lensing masses) employed in our analysis.  We now turn
to describe how we model the expectation values of these observables and derive their covariance in order to simultaneously constrain both cosmology and the richness--mass relation of the \redmapper\ clusters.
In what follows, all quantities labeled with ``ob'' denote quantities inferred from observation, 
while quantities
labeled with ``true'' indicate intrinsic halo properties. 
$P(Y|X)$ denotes the conditional probability of $Y$ given $X$. 


\subsection{Expectation values}
\label{sec:methods:model}

\subsubsection{Base Model}

Let $\Delta \lambda\ob_i$ denote richness bin $i$, and $\Delta z\ob_j$ denote the redshift
bin $j$. The expectation value of the number 
of galaxy clusters $N(\Delta \lambda\ob_i, \Delta z\ob_j)$ is given by
\begin{multline}
\label{eqn:NC}
 \langle N(\Delta \lambda\ob_i, \Delta z\ob_j) \rangle =  \int_0^{\infty}  \de z\true\ 
  {\Omega_{\rm mask}(z\true)} \frac{\de V}{\de z\true \de \Omega}(z\true)    \\ \times \langle n(\Delta \lambda\ob_i,z\true) \rangle     \int_{\Delta z\ob_j} \de z\ob\ P(z\ob|z\true,\Delta\lambda\ob_i) ,
\end{multline}
where $\de V /(\de z\true \de \Omega)$ is the comoving volume element per unit redshift and solid angle, and $\Omega_{\rm mask}(z\true)$ is the effective survey area at redshift $z$.
The survey area depends on redshift because galaxy clusters are not point-like: whether a cluster 
is formally within the survey area or not depends not just on the location of the galaxy cluster in the
sky, but also on how the survey boundaries (including star holes and any other masked regions) intersect
the projected area of the cluster in the sky.  To estimate the survey area, we randomly place clusters 
in the sky, and compute the fraction of the galaxy cluster that does not fall within the survey
footprint.  The footprint of the cluster survey is defined by the collection of all points for which at
least 80 per cent of the cluster falls within the photometric survey boundaries.  
This 80\% criterion is the fiducial choice for all \redmapper\ runs, and is chosen as a compromise between 
requiring clusters not be heavily masked, and losing a minimal amount of area due to masking.
In principle, this masking criteria implies that the survey area depends
on cluster richness (via the scale radius $R_\lambda$), but we find this dependence to be negligible ($\leq 1\%$ over the redshift
range $z\in[0.1,0.3]$ employed in this study).

The second integral of Eq.~\ref{eqn:NC} accounts for the uncertainty in the photometric redshift estimate. We model $P(z\ob|z\true,\Delta\lambda\ob_i)$ --- the probability of assigning to a cluster at redshift $z\true$ a photometric redshift $z\ob$ --- with a Gaussian distribution having mean $z\true$ and a redshift and richness-dependent variance.
The variance is set by the reported photometric redshift uncertainty in the \redmapper\ cluster catalog.  Specifically, we fit a third-order polynomial to the \redmapper\ photometric redshift errors as a function of redshift for galaxy clusters in each of our five richness bins (we find this is sufficient to fully describe our data).
Photometric redshift uncertainties range from $\approx 0.005$ at $z\approx 0.15$ to
$\approx 0.014$ at $z\approx 0.3$, with richer clusters having somewhat smaller photometric redshift uncertainties than low richness clusters.

As shown in Figure~9 of \citet{Rykoff2014}, the \redmapper\ photometric redshifts are excellent: they are  nearly unbiased, and the reported photometric redshift uncertainties are both small and they accurately describe the width of the photometric redshift offsets relative to the spectroscopic cluster redshifts (where available).  Using the specific cluster sample employed
in this work $(\lambda \geq 20, z \in [0.1,0.3])$, we evaluate the systematic bias of the
\redmapper\ photometric redshift by fitting a Gaussian distribution to the redshift offset
$z_\lambda - z_{\rm spec}$, where 
$z_\lambda$ is our photometric cluster redshift estimate, and $z_{\rm spec}$ is the spectroscopic
redshift of the central galaxy assigned to the cluster (when available). We find a mean bias of 
$0.002$, i.e. $z_\lambda = z_{\rm spec} + 0.002$.  
Likewise, from a Gaussian fit to the distribution of scaled errors $(z_\lambda - z_{\rm spec})/\sigma_z$
we find that the reported photometric redshift uncertainties in \redmapper\ should be boosted by
a factor of 1.014.  In both cases, the statistical uncertainties in the estimates are negligible.
We have verified that the above systematic errors are completely negligible 
by running two versions of our analysis, one without applying these corrections, and
one after applying these corrections.  The impact on the posteriors is negligible.
For specificity, from here on out we apply the above corrections, that is,
we set $\avg{z\ob|z\true}=z\true+0.002$ and increase the photometric redshift
errors by a factor of $1.014$.

Finally, $\langle n(\Delta \lambda\ob_i,z\true) \rangle$ in Eq.~\ref{eqn:NC}
is the expected number density of halos in the richness bin $\Delta \lambda\ob_i$.
This quantity is given by
\begin{equation}
\label{eqn:dndz}
 \langle n(\Delta \lambda\ob_i,z\true) \rangle = \int_{0}^{\infty} \de M\ n(M,z\true) 
 \int_{\Delta \lambda\ob_{\rm i}} \de \lambda\ob\ P(\lambda\ob | M,z\true) \, ,
\end{equation}
where $P(\lambda\ob | M,z\true)$ denotes the probability that a halo of mass $M$ at redshift $z\true$ is observed with a richness $\lambda\ob$ (see \autoref{sec:irmr}) and $n(M,z\true)$ is the halo mass function which is assumed to follow the form of \citet{Tinker2008}:
\begin{equation}
n(M,z) = \frac{3}{4\pi R^3(M)} \frac{\de \ln \sigma(M)^{-1}}{\de M} f^{\rm Tinker}(\sigma(M),z)
\end{equation}

Several studies have explored how the mass function from $N$-body simulations should be extended in order to incorporate the effects of massive neutrinos \citep[e.g.][]{Brand2010,Villa2014,Castorina2014,Liu2017}. A common finding of these studies is that massive neutrinos play a negligible role in the collapse of dark matter halos, while they suppress the growth of matter density fluctuations on scales smaller than the neutrino free-streaming length. Here, we account for these effects following the prescription of \cite{Costanzi2013}: (i) we neglect the density neutrino component in the relation between mass and scale --- i.e. $M \propto (\rho_{\rm cdm} + \rho_{\rm b})R^3$ --- and (ii) use only the cold dark matter and baryon power spectrum components to compute the variance of the density field, $\sigma^2(M)$.  

We can use similar arguments to the ones used to derive Eq. \ref{eqn:NC} to compute the expectation value for the mean mass of galaxy clusters within a specific richness and redshift bin. This is given by:
\begin{equation}
\langle \bar{M}(\Delta \lambda\ob_i, \Delta z\ob_j) \rangle = \left[ \frac{\langle M^{\rm tot}(\Delta \lambda\ob_i, \Delta z\ob_j) \rangle}{\langle N(\Delta \lambda\ob_i, \Delta z\ob_j)\rangle} \right] \, ,
\label{eqn:avgM}
\end{equation}
i.e. the ratio of the expected total mass inside the bin over the total number of clusters inside said bin (Eq.~\ref{eqn:NC}).  Note that observationally, we stack $\Delta\Sigma$, not $M$, and since
the two are not linearly related to each other --- $\Delta \Sigma$ is proportional to the integrated mass density profile --- it is not necessarily the case that the recovered
weak lensing mass is identical to the mean mass of the clusters.  However, as described in section~\ref{sec:wlmeasurement}, we calibrate the relation between the recovered weak lensing
mass and the mean mass in a bin so as to ensure our measurement is an unbiased proxy of the mean
mass of the clusters in a bin.

The total mass of all clusters in a bin is calculated via
\begin{multline}
\langle M^{\rm tot}(\Delta \lambda\ob_i, \Delta z\ob_j) \rangle=\int_0^{\infty}  \de z\true\  {\Omega_{\rm mask}} \frac{\de V}{\de z\true \de \Omega}  (z\true)  \\ \langle Mn(\Delta \lambda\ob_i,z\true) \rangle  \int_{\Delta z\ob_j} \de z\ob\ P(z\ob|z\true)\, ,
\end{multline}
where the total mean mass per unit volume in the $i$-th richness bin is given by: 
\begin{equation}
\label{eqn:dnMdz}
 \langle Mn(\Delta \lambda\ob_i,z\true) \rangle = \int_{0}^{\infty} \de M\ M n(M,z\true) 
 \int_{\Delta \lambda\ob_{\rm i}} \de \lambda\ob\ P(\lambda\ob | M,z\true) \, .
\end{equation}

In practice, the integrals over the observed redshift in the numerator and denominator of equation~\ref{eqn:avgM}
are each weighted by the appropriate weak lensing weight per clusters $w_{\rm WL}(z)$,
where $w$ is the mean weight applied to sources as a function of redshift --- $w \propto \langle \Sigma_{\rm crit}^{-1} \rangle^2 / (1+z)^2$ --- times the number
of sources per cluster.
We have found including this weight changes the predicted
masses by less than 1\%.  Nevertheless, our fiducial result includes this additional redshift
weighting.


\subsubsection{The Observed Richness--Mass Relation}
\label{sec:irmr}

Turning to the probability distribution $P(\lambda\ob|M,z\true)$, 
in addition to the stochastic nature of the relation between cluster richness and halo mass,
the observed richness of a galaxy cluster is subject to projection effects. 
Indeed, there are now multiple lines in support of the existence of
projection effects in the SDSS \redmapper\ cluster 
catalog \citep{farahietal16,zuetal17,buschwhite17,sohnetal17}.
Following \cite{costanzi18a}, we model $P(\lambda\ob | M,z\true)$ as the convolution of 
two probability distributions:
\begin{equation}
\label{eqn:PlobM}
 P(\lambda\ob | M,z\true) = \int_0^{\infty} \de \lambda\true\ P(\lambda\ob | \lambda\true,z\true) P(\lambda\true | M,z) \, .
\end{equation}
The first term inside the integral accounts for projection effects and observational noise in
the richness estimates. 
The second term inside the integral accounts for the stochastic relation between
halo mass and the intrinsic halo richness $\lambda\true$.   
 
Below, we start describing the model adopted for the intrinsic richness--mass relation $P(\lambda\true | M,z)$. The probability distribution $P(\lambda\ob | \lambda\true,z\true)$ was the focus of a detailed numerical study in a companion paper \citep[][]{costanzi18a}.  A brief overview of that work is presented in the subsequent subsection.

\subsubsection*{The Intrinsic Richness--Mass Relation}

Different parameterizations for $P(\lambda\true | M,z)$ have been proposed in the literature, typically assuming a log-normal distribution with the expectation value for the richness modeled as a power law \citep[see e.g.][]{Rozo2010,Mana2013,Murata2017}. In this analysis we opt for a model which more closely resembles halo occupation distribution functions typically used to study galaxy clustering \citep{berlindweinberg02,bullock02}. A recent review of halo occupation modeling and other approaches to modeling the galaxy--halo connection can be found in \cite{wechslertinker18} (see also e.g. the discussion in section 4 of \citealt{jiangvandenbosh04} and Appendix D of \citealt{reddick13}).  The total richness of a halo of mass $M$ is given by $\lambda\true= \lambda^{\rm cen} + \lambda^{\rm sat}(M)$, where $\lambda^{\rm cen}$ is the number of central galaxies (either zero or unity), and $\lambda^{\rm sat}$ is the number of satellite galaxies in the cluster \citep{Kravtsov2004,zhengetal05}.  We model the expectation value of $\lambda^{\rm cen}$ as a simple step function, $\avg{\lambda^{\rm cen}|M} = 1$ for $M\geq M_{\rm min}$, and $\avg{\lambda^{\rm cen}|M} = 0$ otherwise.  While in practice these step functions have a finite width, we expect all clusters in our sample to have masses $M \gg M_{\rm min}$, so that the step-function approximation should be easily sufficient.  

Turning to the satellite galaxy population, it has long been known that the scatter in the number of satellites is super-Poissonian at large occupations numbers \citep{boylankolchinetal10}, but is close to Poissonian otherwise.  More recently, the number of satellites has been shown to be sub-Poissonian at very low occupancy \citep{maoetal15,jiangvandenbosh04}.  Since we are interested in galaxy clusters, we ignore this small deviation from Poisson statistics at low $N$ in our analysis. We add variance to a Poisson distribution by modeling $P(\lambda^{\rm sat}|M)$ as the convolution of a Poisson distribution with a Gaussian distribution.  Operationally, this is equivalent to assuming the number of satellites galaxies in a halo of mass $M$ is a Poisson realization of some expectation value $\mu$, where $\mu$ exhibits halo-to-halo scatter, e.g. due to formation history \citep{maoetal15}.  We model the halo-to-halo scatter as a Gaussian with variance $\sqrt{\Var(\mu)} = \sigma_{\rm intr}\avg{\lambda^{\rm sat}|M}$.    This additional halo-to-halo scatter enables us to recover the super-Poisson scatter in the halo occupation at large occupancy numbers.

In detail, the expectation value of the satellite contribution to $\lambda\true$ is given by \citep{Kravtsov2004,zehavietal11}
\begin{equation}
\label{eqn:RMR}
\langle \lambda^{\rm sat} | M\rangle = \left( \frac{M-M_{\rm min}}{M_1 - M_{\rm min}} \right)^\alpha \, ,
\end{equation}
Here, $M_{\rm min}$ is the minimum mass for a halo to form a central galaxy, while $M_1$ is the characteristic mass at which halos acquire one satellite galaxy. Our parameterization enforces $\avg{\lambda^{\rm sat}|M}=0$
when $M \leq M_{\rm min}$.  Finally, the expectation value of the Gaussian component is set to zero, while the variance of the Gaussian term is set to $\sigma_{\rm intr}\avg{\lambda^{\rm sat}|M}$.  

The convolution of a Poisson distribution with a Gaussian distribution does not have an analytic closed form.  However, we have found that a skew-normal distribution is an excellent approximation to the resulting distribution. Its model parameters --- skewness and variance --- depend on $\langle \lambda^{\rm sat} | M\rangle$ and $\sigma_{\rm intr}$, and are obtained by fitting the  skew-normal model to the appropriate Gaussian-Poisson convolution (see Appendix \ref{app:skewnorm}). By creating a lookup table for these parameters, we can avoid having to numerically compute the convolution of the Poisson and Gaussian distributions, significantly increasing the computational efficiency of our model.  In Section~\ref{sec:res:model} we discuss how our choice of parameterization impacts the cosmological constraints derived from the SDSS \redmapper\ sample.


\subsubsection*{Modeling Observational Scatter in Richness Estimates}

The scatter in the distribution $P(\lambda\ob|\lambda\true)$ is due to observational scatter, i.e. noise on the estimated richness values due to photometric noise, uncertainties in background subtraction, and projection effects. The latter refers to the boosting of the richness of a galaxy cluster due to additional galaxies along the line of sight that are mistakenly associated with the galaxy cluster. In \cite{costanzi18a} we developed a formalism that quantitatively characterizes these effects, demonstrated the accuracy and precision of this formalism, and combined it with numerical simulations to calibrate the impact of projection effects and observational noise in the SDSS DR8 \redmapper\ catalog.  Here, we provide a succinct description of that work,
and refer the reader to \cite{costanzi18a} for details.

The scatter in $\lambda\ob$ due to background subtraction and photometric noise in the data is estimated as follows.  First, we inject synthetic clusters of known richness $\lambda\true$ and redshift $z\true$ at a random position in the survey footprint.  We then measure the recovered richness, and repeat the procedure 10,000 times to characterize the associated scatter.  The scatter is decomposed into a Gaussian component --- the observational noise --- and a decaying exponential component that characterizes the tail of high richness clusters due to projection effects.  The biases in the recovered richness, as well as the standard deviation of the Gaussian component, are recorded.  It is this Gaussian distribution that we use to model observational noise.

The tail due to projection effects is modeled explicitly as a sum over the true richnesses of galaxy clusters along the line of sight, weighted by a ``projection kernel'' that is calibrated directly from the data.  The validity of this model is explicitly demonstrated by comparing the impact of projection effects around random points in the data to the impact of projection effects we predict when applying our model to $N$-body simulations (see also \S \ref{sec:res:mock_data}).  Having validated the model, we measure the impact of projection effects at the location of massive halos in the $N$-body simulations, and use this calibrated measurement to characterize the tail due to projections arising from correlated large structures.

\begin{figure}
\begin{center}
    \includegraphics[width=0.48 \textwidth]{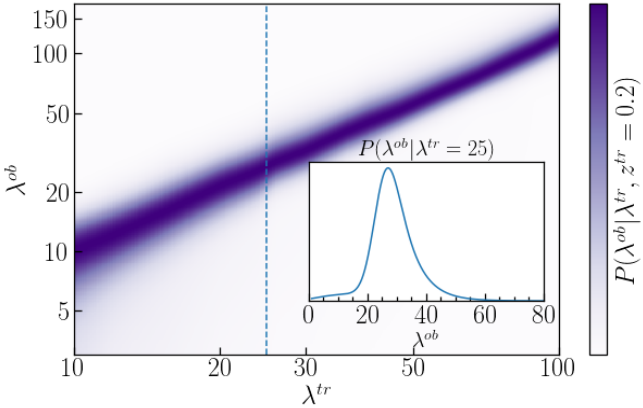}
\end{center}
\caption{$P(\lambda\ob|\lambda\true)$ as a function of the {\it true} richness at redshift $z=0.2$. The inset shows a section of $P(\lambda\ob|\lambda\true)$ for $\lambda\true=25$ ({\it dashed} line in the main plot). Note the non-Gaussian tail to high richness due to projection effects, and the low-richness tail due to percolation effects. 
}
\label{fig:Plob_ltr}
\end{figure}

The end result of this procedure is a calibrated model for $P(\lambda\ob|\lambda\true)$ composed of a Gaussian observational noise, an exponentially decaying noise contribution due to projection effects, and a tail that extends to $\lambda\ob=0$ which accounts for incompleteness due to percolation effects:
low-richness clusters along the same line of sight of a richer objects can lose some of their galaxies to the richer object, leading to a severe underestimation for their richness. Figure \ref{fig:Plob_ltr} shows the distribution $P(\lambda\ob|\lambda\true)$ as a function of $\lambda\true$ at redshift $z=0.2$. The inset shows a cross-section of the distribution for $\lambda\true=25$. The three main components of the distribution are obvious by eye: a Gaussian kernel due to observational noise, a large tail to high richness due to projection effects, and a low-richness tail due to percolation effects.

We note that the calibration of $P(\lambda\ob|\lambda\true)$ depends on the input cosmology and richness--mass relation parameters adopted to generate the synthetic cluster catalog. However, we verified in \cite{costanzi18a} that this assumption has no impact on the posterior of the cosmological parameters.  In section \ref{sec:res} we explicitly test the robustness of our cosmological constraints to different calibrations of $P(\lambda\ob|\lambda\true)$.


\subsection{Mass-Function Systematics}
\label{sec:methods:hmf_nuis}

The modeling of the cluster counts and mean cluster masses depends on the halo mass function. 
Here, we use the \citet{Tinker2008} halo mass function. \cite{Tinker2008} report their mass function
formula to be accurate at the $\approx 5\%$ level, but we do not have a robust estimate of the 
associated systematic uncertainty as a function of mass.  
Moreover, a number of studies comparing different halo finders and fitting functions 
suggest a systematic uncertainty of the order of $10\%$ \citep[e.g.][]{Knebe2013,Hoffmann2015,Despali2016}.
To characterize the systematic uncertainty in the halo mass function in dark matter only simulations 
we introduce two nuisance parameters $q$ and $s$ relating the \citet{Tinker2008} mass function to the true mass
function via
\begin{equation}
\label{eqn:hmfnuis}
n(M,z)=n(M,z)^{\rm Tinker} \left(s \log(M/M^*)+q \right) \, ,
\end{equation}
where the pivot mass is set to $\log(M^*)=13.8 \,\hmsun$.  Note that if $q=1$ and $s=0$, then the halo mass function is given by the \citet{Tinker2008} formula.   We set the priors on $q$ and $s$ using the ensemble of simulations developed as part of the Aemulus project \citep{derose18}.  This is a set of 40 $N$-body simulations spanning a range of cosmologies  in the redshift range $0.0<z<1.0$.  Each simulation box has a length $L=1050\ \hMpc$ and contains $1400^3$ particles.  The particle mass is cosmology dependent, and averages $\approx 3.5\times 10^{10}\ \hmsun$.  The cosmologies sampled by the simulation spans the $4\sigma$ confidence interval from WMAP \citep{wmap9} and {\it Planck} \citep{PlanckXIII2016} in combination with Baryon Acoustic Oscillation data \citep{anderson14} and the Union 2.1 Super-Nova (SN) data \citep{suzuki12}. Halo catalogs were generated using the {\tt ROCKSTAR} algorithm \citep{Behroozi2013}.
Further details of the simulation data as well as the convergence tests done to ensure the reliability of these simulations are presented in \citet{derose18}.

We fit the simulation halo abundance data for the nuisance parameters $s$ and $q$ at each snapshot of each of the 40 simulations by computing the ratio $n^{\rm Sims}(M,z)/n(M,z)^{\rm Tinker}$.
Next, we model the distribution of $(s,q)$ values as Gaussian, and fit for the 
mean values and covariance matrix describing the scatter of $(s,q)$ across all 320 snapshots.
We find $\bar{s}=0.037$ and $\bar{q}=1.008$, with a covariance matrix:
\begin{equation}
\label{eqn:Chmf}
{\rm C}(\bar{s},\bar{q}) = \begin{bmatrix} 0.00019 & 0.00024 \\ 0.00024 & 0.00038 \end{bmatrix} .
\end{equation}

The above matrix is the sum of the covariance matrix obtained from the fit of $s$ and $q$, i.e. the 
statistical calibration uncertainty in the mean values of the parameters, and the covariance matrix
describing the simulation-to-simulation scatter in $s$ and $q$.
That is, the above covariance matrix fully accounts for any possible deviations of the \citet{Tinker2008} halo mass function to the test set of 40 N-body simulations employed here.
Note too the variance of $q$ corresponds to a 6\% uncertainty in the amplitude of the halo mass function, 
consistent with the quoted precision in \citet{Tinker2008}.
The above covariance matrix and best fit values define the bivariate Gaussian priors used for the parameters
$s$ and $q$ in our cosmological analysis.  Future analyses should be benefit from the significantly higher precision that can be achieved using emulators \citep[e.g.][]{McClintock2018}.

The above analysis does {\it not} account for the impact of baryons on the halo mass function.  Several recent works have estimated the impact of baryonic feedback on total masses of halos and, thereby, the mass function \citep[e.g.][]{Cui2014,Velliscig2014,Bocquet2016,Springel2017}.  These works all find that baryonic impact decreases with increasing radial aperture.  For the specific mass definition we adopt, namely a 200 overdensity criterion relative to mean, the impact of AGN-based full physics is modest for massive ($M\gtrsim 10^{14}\ h^{-1}M_\odot$) halos.  In the IllustrisTNG simulation, full physics leads to a mean enhancement of $\sim 3\%$ while multiple methods analyzed by \citet{Cui2014} produce mainly decrements in $M_{200c}$ of roughly similar magnitude.  Due to the current uncertainty in modeling baryonic effects, we defer its inclusion into the error budget of the halo mass function to future work. For now, we simply note that a 3\% systematic uncertainty in the halo mass function is already sub-dominant to the precision of the Tinker mass function, as characterized by our model parameters $s$ and $q$.


\subsection{Covariance Matrix}
\label{sec:methods:like}

Having specified the expectation values for our observables (cf. Section~\ref{sec:methods:model}) we need to define the covariance matrix of our data vector in order to fully specify the likelihood function.
Here, we assume that the abundance and weak lensing data are uncorrelated.  This assumption is well justified: the weak lensing error budget is strongly dominated by shape noise, and the dominant systematic is the overall multiplicative shear and photo-$z$ bias of the source catalog.  None of these errors affect the abundance data, and are therefore clearly uncorrelated. 

Our Gaussian likelihood model takes the form
\begin{equation}
\label{eqn:like}
 \mathcal{L}({\bm d} |{\bm \theta}  ) \propto \frac{ \exp \left[ -\frac{1}{2} \left( {\bm d} -  {\bm m}({\bm \theta}) \right)^T {\bm C}^{-1} \left( {\bm d} -  {\bm m}({\bm \theta}) \right) \right]}{\sqrt{(2 \pi)^M {\rm det}(\bm C)}} \, .
\end{equation}
where ${\bm C}$ is the total covariance matrix detailed below, and ${\bm d}$ and ${\bm m}({\bm \theta})$ are respectively the data vectors (see Table \ref{tab:data}) and the expectation values for the number counts and $\log M$ (Equation \ref{eqn:NC} and \ref{eqn:avgM}, respectively).

In reality, the likelihood for the abundance data is a convolution of a Poisson error on the counts and a Gaussian error due to density fluctuations within the survey area ~\citep[e.g.][]{Hu2006,Takada2014}.   Such a convolution does not have an analytic closed form.  Here, we take care to ensure that all of our richness bins are well populated --- our least populated richness bin contains over 200 galaxy clusters --- so that the Poisson component can be adequately modeled with a Gaussian distribution. Consequently, our likelihood for the abundance data can be modeled as a Gaussian with a total covariance matrix having three distinct contributions:
\begin{enumerate}
    \item{A Poisson contribution due to the Poisson fluctuation in the number of halos at given mass in the survey volume.}
    \item{A sample variance contribution due to the fluctuations of the density field in the survey volume.}
    \item{A contribution due to uncertainty in the miscentering corrections detailed in section~\ref{sec:datasets:abundances}.}
\end{enumerate}

\begin{figure}
\begin{center}
    \includegraphics[width=0.45 \textwidth]{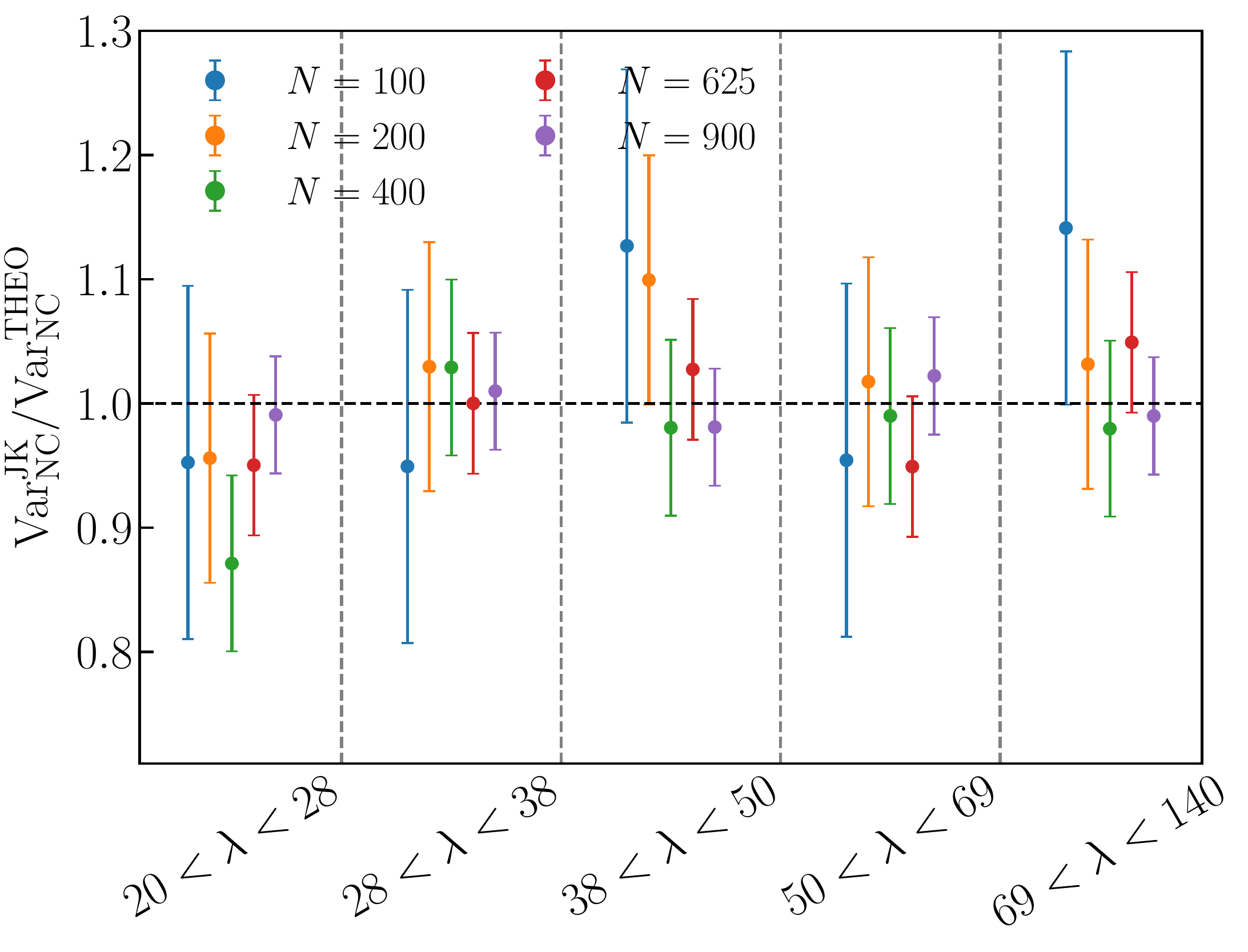}
\end{center}
\caption{Comparison between the analytically derived variance for the number counts and the one derived from the mock catalog. Different colors correspond to the different splitting schemes used to estimate the Jackknife covariance matrix. Error bars are estimated from the jackknife covariance matrices itself assuming the jackknife realizations to be Gaussian distributed.
}
\label{fig:cov}
\end{figure}

The Poisson and sample variance contributions to the covariance matrix are computed analytically at each step of the chain to properly account for their dependence on cosmology and model parameters. At high richness, the Poisson contribution dominates, with sample variance becoming increasingly important at low richness \citep{hukravtsov03}. The analytical expression used to derive these two terms is provided in Appendix~\ref{app:NC_cov}, and it is validated by means of mock catalogs. In particular, we use the mock cluster catalogs generated by \citet{costanzi18a} to calculate the variance in the abundance for each of the richness bins employed in our analysis. To do so, we split the full simulated survey area into $N$ patches, ranging from $N=100$ to $N=900$, and use these patches to estimate the variance within the survey using a Jackknife method.  Note that jackknife methods cannot adequately resolve modes that are larger than the jackknife region under consideration.  Consequently, for the purposes of comparing our analytic estimates to the jackknife covariance matrix, we truncate our sample variance estimates by setting matter power spectrum to zero below the wave-number $k_{\rm min}=\pi/R_{\rm max,N}$, where $R_{\rm max,N}$ is the radius of a spherical volume equal to the volume of the jackknife patch considered (see Eq. \ref{eqn:rms}). As we can see from Figure~\ref{fig:cov}, our analytic estimates are a good match to the observed variance in the simulations.  We have also explicitly verified using our synthetic cluster catalog that the sample variance contribution to the mean cluster mass is negligible relative to the statistical and systematic uncertainties of the weak lensing analysis.  We also find that the cross-correlation between number counts and the mean weak lensing cluster mass obtained from our jackknife estimates is consistent with zero.

Turning to the uncertainty in the weak lensing mass estimates, we use the posteriors from the stacked weak lensing analysis described in Section~\ref{sec:datasets:wl_data} (see Table \ref{tab:data}).  Critically, these posteriors are found to be nearly Gaussian in the log, that is, the posterior is well described by a log-normal distribution.  Consequently, we model the weak lensing likelihood as a Gaussian distribution in $\log M$. The errors include not just the statistical uncertainty of the measurement, but also systematic uncertainties due to shear and photo-$z$ biases (multiplicative shear bias), cluster projections, halo triaxiality, and miscentering effects. The overall shear and photo-$z$ multiplicative bias is shared across all richness bins.  Consequently, the variance associated with this error should be perfectly correlated across all bins.  

To determine the overall mass error in each bin associated with the multiplicative bias we proceed as in \citet[][section 5.7]{Melchior2017} and \citet[][section 5.6]{McClintock2018}.  First,
we repeat our weak lensing analysis while fixing the systematic parameters.  We then subtract the variance obtained from
these fixed-bias run to the total variance measured while floating the multiplicative bias
parameters.  The resulting variance is the variance attributable to the multiplicative bias.
We do this for all five richness bins, and verify that, as expected, the variance due to
multiplicative bias corrections to the $\Delta\Sigma$ profile, $\sigma_b^2$, is nearly richness-independent,
and roughly consistent with the expectation $\sigma_{\ln M} \approx 1.3\sigma_b$.  The 
systematic uncertainty on the log of the masses is $\sigma_{\log M}=0.026$.  This value
is larger than the $0.023$ systematic quoted in \citet{Simet2016} because of the new
contribution to the systematic error budget from blended sources, and the scatter correction.
The error is perfectly correlated across all richness bins,
so we set $C_{ij}=0.026^2$ along the non-diagonal elements of the covariance matrix.
This corresponds to the 6\% total systematic error in Table~\ref{tab:err_budget}.
The diagonal entries of the covariance matrix are taken directly from the posteriors
of our weak lensing analysis (see \citet{Simet2016} for details). 
Finally, as noted in Section~\ref{sec:datasets:wl_data}, using numerical simulations we
estimated that our fitting procedure introduces a bias that varies from 2\% at low richness
to 3\% at high richness.  We adopt a systematic uncertainty in this correction equal to half
the magnitude of the correction, and add this (sub-dominant) contribution to arrive
at the final covariance matrix employed for the weak lensing data.

In summary, our likelihood:
\begin{itemize}
\item Is Gaussian in the abundances, and includes Poisson, sample variance, and miscentering uncertainties.
\item Is Log-normal in the weak lensing masses, as per the posteriors of \citet{Simet2016}.  It also accounts
for the covariance due to shared multiplicative shear and photo-$z$ biases, blended sources, and cluster triaxiality and projection effects.  These systematics are assumed to be perfectly correlated across all richness bins.
\item Has no covariance between the abundance and weak lensing data.  We expect this to be an excellent 
approximation.
\end{itemize}
The posteriors from our analysis are fully marginalized over all sources of systematic uncertainty
described above.


\section{Validation Tests}
\label{sec:res:mock_data}

We validate our likelihood framework by placing cosmological constraints from 
a synthetic cluster catalog whose cosmology
and richness--mass relation is known a priori.
The mock data are generated starting from an $N$-body simulation run with $1400^3$ dark matter particles in a box of comoving size $L=1050 \Mpc\, h^{−1}$.  The code used is L-Gadget, a variant of Gadget \citep{Springel2005}. 
The simulation assumes a flat-\LCDM model with $\Omega_{\rm m} = 0.286$, $h = 0.7$, $\Omega_{\rm b} = 0.047$, $n_s = 0.96$, and $\sigma_8 = 0.82$.  Lightcone data, including a halo catalog down to $M_{200m}=10^{12.5} h^{-1} M_\odot$, is constructed from the simulation on the fly.  The halo finder is {\tt ROCKSTAR} \citep{Behroozi2013}. For further details, see DeRose et al, in preparation; this is one realization of the L1 box described in that work.

We build a synthetic cluster catalog as follows.  First, each halo is assigned a richness $\lambda\true$ drawn from the $P(\lambda\true|M,z)$ distribution detailed in Section \ref{sec:methods:model}. We set our fiducial model parameters to: $\alpha=0.704$, $\log M_{\rm min}=11.0$, $\log M_1=12.12$ and $\sigma_{\ln \lambda}=0.25$. These fiducial values have been chosen by inverting the mass--richness relation of \citet{Simet2016} to arrive at parameters for the richness--mass relation (Eq. \ref{eqn:RMR}).  The assigned richnesses are then modified to account for observational noise plus projection effects as described in \citet{costanzi18a}.  Specifically:
\begin{enumerate}
\item Starting at the richest halo, we compute the projected richness of each halo via 
\begin{equation}
\label{eqn:prj}
\lambda\ob_i = \lambda\true_i + \sum_{j\neq i}^N \lambda\true_j f^A_{ij} w(\Delta z_{ij},z_j)\,.
\end{equation}
Here, the index $j$ runs over all $j\neq i$ clusters, $f^A_{ij}$ is the fraction of the area of the $j$-th object which overlaps with the $i$-th cluster, and $w(\Delta z_{ij},z_j)$ is the data-calibrated weight which accounts for the redshift distance between the cluster $i$ and $j$.
\item A Gaussian noise is added to the projected richness to account for observational errors (including the
effects of background subtraction).
\item Having calculated the observed projected richness of the richest galaxy cluster, we update the richnesses of all remaining clusters via 
\begin{equation}
\lambda\true_j=\lambda\true_j(1- f^A_{ij} w(\Delta z_{ij},z_j)).
\end{equation}
That is, the richness of cluster $j$ is reduced by the amount it contributed to the richest cluster.
\item We now move on to the next richest cluster in the catalog, and iterate the above procedure to
compute the observed projected richness of every halo in the simulation.
\item Finally, we assign a photometric redshift to every cluster based on their spectroscopic redshifts
using the model in section~\ref{sec:methods:model}.
\end{enumerate}

Given our synthetic cluster catalog, we now compute our observable data vectors.  The cluster counts are measured using the same binning scheme as for the real data (see Table \ref{tab:data}).
As for the synthetic weak lensing masses, these are obtained from the mean halo mass in each richness bin.
We also add a log-normal noise to the mean weak lensing masses by drawing from the covariance matrix
described in Section~\ref{sec:methods:like}. To mimic the fact that the weak lensing masses are estimated assuming $\Omega_m=0.3$ while the simulation use the fiducial value $\Omega_m=0.286$, we invert Equation \ref{eqn:MvsOm} to arrive to our final mock data vector:
\begin{equation*}
\log \bar{M}^{\rm WL MOCK} = \left . \log \bar{M}^{\rm WL} \right \vert_{\Omega_m = 0.286} -
	\left( \frac{d\log M^{\rm WL}}{d \Omega_m} \right)\left(0.286 - 0.3 \right )
\end{equation*}

\begin{figure*}
\begin{center}
    \includegraphics[width=\textwidth]{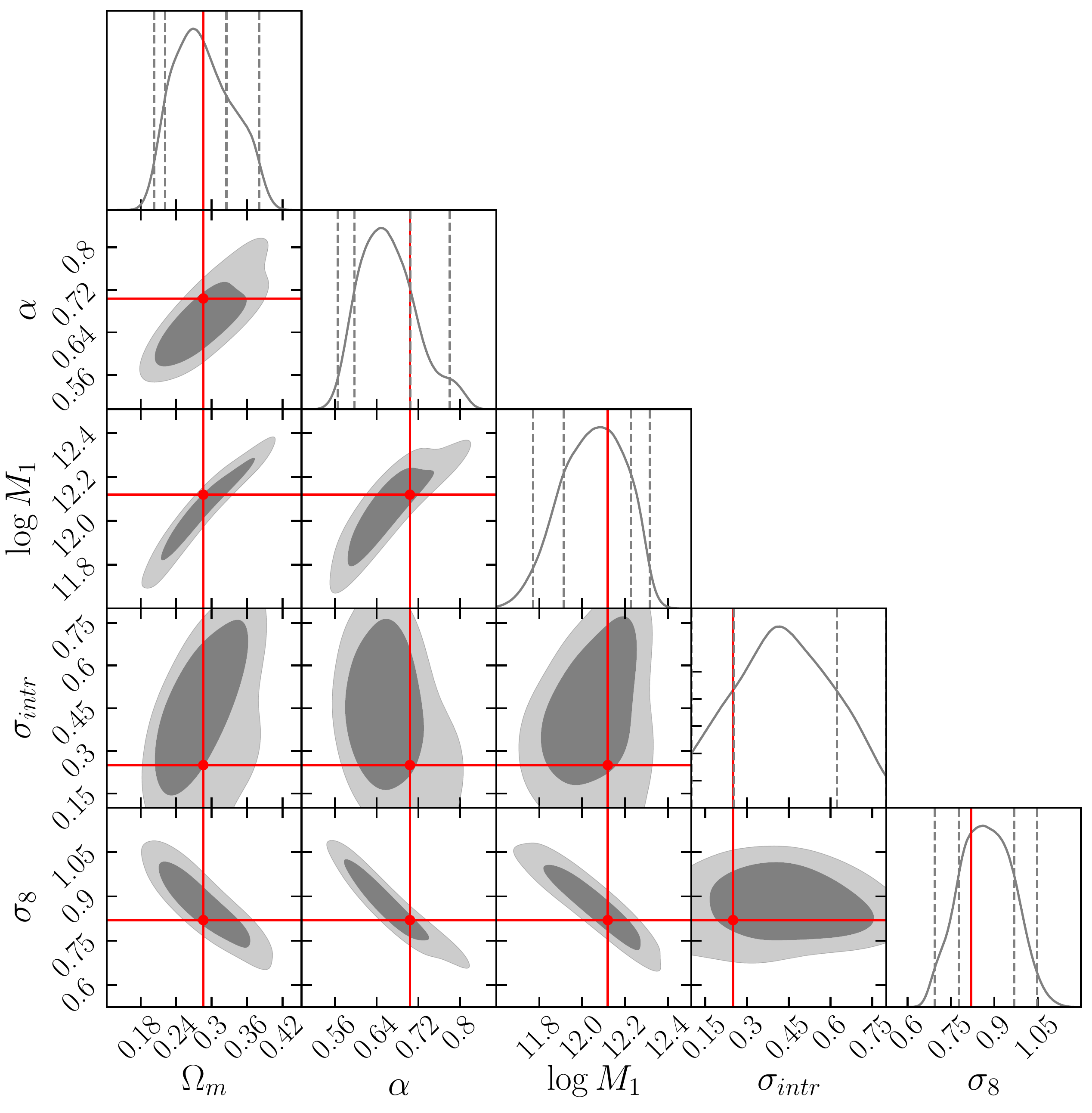}
\end{center}
\caption{ $68\%$ and $95\%$ confidence contours obtained running our pipeline on mock data. The input parameter values used to generate the simulation and the mock data catalog are shown in {\it red}. The {\it dashed} lines shown in the $1$-d marginalized distributions (diagonal of the triangle plot) correspond to the $0.025$, $0.16$, $0.84$ and $0.975$ quantiles of the distributions.  Not included in this plot is the parameter $\Mmin$, which is prior dominated.}
\label{fig:mock_res}
\end{figure*}

Figure~\ref{fig:mock_res} shows the constraints obtained by analyzing one realization of our synthetic data set with our pipeline.  This analysis varies the same parameters and assumes the same priors as the real data analysis (see Section~\ref{sec:res}). We can see that our analysis successfully recovers the true values of the parameters of our synthetic data set ({\it red} lines in the triangle plots).  Because this test relied on a single simulation, we could not use it to validate the width of our posteriors.

%


\section{Blinding and Unblinding}
\label{sec:blind}

In order to avoid confirmation biases our cosmological analysis was performed blinded.  By ``blinded'', we mean that the following protocols were followed:
\begin{enumerate}
\item The cosmological parameters in the MCMC were randomly displaced before being stored, and the random displacement was stored in binary (i.e. a not-human-readable format).
\item All modeling choices --- specifically which set of cosmological models and models for the scaling relations we would consider --- were made before unblinding.  In this work, we chose to focus exclusively on flat $\Lambda$CDM cosmologies with massive neutrinos.  Our choice to let neutrino mass vary follows the practice of the DES Year 1 combined probe analysis \citep{des17}.  
\item The set of scaling relation models considered in our analysis was fully specified before unblinding. In addition to our fiducial model for the scaling relation, we considered three additional sets of models for the scaling relation (see Figure~\ref{fig:mod_dep}). 
"Random-point-injection" refers to the $P(\lambda\ob|\lambda\true)$ distribution calibrated by inserting galaxy clusters at random locations in the sky, i.e. neglecting the contribution from correlated structures to the observed richness. Since we know this model underestimates projection effects, we take half the difference between our fiducial model and the "Random-point-injection" model as our estimate of the systematic uncertainty in our cosmological posteriors.
We further tested how allowing the intrinsic scatter of the richness--mass relation to scale with mass affected our cosmological posterior constraints.  Finally, to compare to existing works, we considered both a power-law scaling relation with constant scatter and a scaling relation modeled after the work of \citet{Murata2017}. 
\item All priors were set before unblinding.
\item The set of external data sets we would compare against was decided before unblinding.  Here, we distinguish between data sets with which we intend to combine our analysis and further test the flat $\Lambda$CDM model (see next item), and other large-scale-structure data sets that constrain the $S_8 \equiv \sigma_8(\Omega_{\rm m}/0.3)^{1/2}$ parameter.  For the latter, we simply compare the recovered $S_8$ values, rather than test for consistency across the full parameter space.  Specifically, we compare our $S_8$ constraints to those of {\it Planck} \citep{PlanckXIII2016}\footnote{After we unblind our analysis new \planck\ results have been released \citep{Planck2018}. To provide the most updated results we decided to consider the latest \planck\ constraints (hereafter \planck\ DR18) as CMB data set to compare against the SDSS clusters results.}, the DES Year 1 3x2 point analysis \citep{des17}, the KiDS-450+2dFLenS analysis \citep{kids2df2018}, and the KiDS-450+GAMA analysis \citep{kidsgama2018}.  In addition, we compare our constraint to several additional recent cluster abundance constraints, namely those from the {\it Planck}-SZ catalog \citep{PlanckSZ2016}, those from the South Pole Telescope \citep{SPTSZ2016} and Atacama Cosmology Telescope \citep{ACT2013}, and those from the Weighing the Giants team \citep{Mantz2015}.  Finally, we compare our constraints to those derived from the SDSS maxBCG cluster sample \citep{Rozo2010}.   For these purposes, we consider analyses $A$ and $B$ to be consistent if their central values of $S_8$ deviate by no more than $3\sigma_{\rm tot}$, where $\sigma_{\rm tot}^2 = \sigma_A^2+\sigma_B^2$.
\item The set of external data sets with which we intend to combine with the SDSS analysis (provided the two are deemed to be consistent, see metrics for consistency below) were fully specified. In addition to {\Planck} \citep{PlanckXIII2016}, we decided to compare our constraints to those derived from Baryon Acoustic Oscillations (BAO) as measured in the 6dF galaxy survey \citep{BAO6dF}, the SDSS Data Release 7 Main Galaxy sample \citep{BAOSDSSMain}, and the BOSS Data Release 12 \citep{BAOBoss}.  As for the {\it Planck} constraints, we relied on the TT+lowP data set.  With the expectation that our combined BAO and cluster abundance analysis (which included a prior on $\Omega_{\rm b} h^2$) would result in a tight constraint on the Hubble constant \citep[see e.g.][]{DESH02017}, we also planned on comparing our posterior in $h$ to that of \citet{riessetal16}.
\item The metrics for consistency with external data sets were selected before unblinding.  Specifically, consistency between two data sets $A$ and $B$ was established by testing whether the hypothesis $\bm{p}_A - \bm{p}_B = 0$ is acceptable  \citep[see method `3' in][]{charnocketal17}.  Here $\bm{p}_A$ and $\bm{p}_B$ are the model parameters of interest as constrained by data sets $A$ and $B$ respectively.  The two data sets were deemed to be consistent if the point $\bm{p}_A-\bm{p}_B=0$ falls within the 99\% confidence level of the multi-dimensional distribution of $\bm{p}_A-\bm{p}_B$.  If two data sets were found to be inconsistent with one another, we did not consider the combined analysis. We note that in order for this test to be the sharpest possible test, it is important to restrict one-self to parameter sub-spaces that are well constrained in both data sets.  To that end, in all cases we restrict the parameter space for comparison to the set of parameters whose posterior is at most half as uncertain as the prior of each individual data set.
\item No comparison of our cosmological constraints to any other data sets were performed prior to unblinding.
\end{enumerate}
The weak lensing analysis upon which our work relies was not performed blind \citep{Simet2016}, though our forthcoming analysis using data from the Dark Energy Survey data will have benefited from a blind weak lensing analysis.  Importantly, all relevant weak lensing priors --- specifically the multiplicative shear bias, photometric redshift correction, source dilution, etc --- were finalized well before the advent of our particular analysis: no tuning of any input catalog was done in response to the weak lensing analysis of \citet{Simet2016} or our own cosmological analysis.  

During the DES internal review process the question arose whether past experiences from people in our team --- particularly those involved with the development of the maxBCG cluster catalog \citep{maxbcg} --- had in any way informed the development of \redmapper, possibly leading to unintended biases in this way.  The \redmapper\ algorithm was {\it not} tuned based on previous experiences.  The algorithm iteratively self-trains the colors of red-sequence galaxies as a function of redshift without any cosmological assumptions.  The parameters chosen for the richness estimation --- particularly the aperture used --- were selected by minimizing the scatter in X-ray luminosity at fixed richness \citep{rykoffetal12}.  We do adopt a fiducial cosmology when defining our richness measurements (this is necessary to turn physical apertures into angular apertures), but since the richness--mass relation is not known a priori, it carries no cosmological information.

Our unblinding protocol was defined by the set of requirements detailed below.
\begin{enumerate}
\item{Our inference pipeline had to successfully recover the input cosmology in a synthetic data set.  For details, see Section \ref{sec:res:mock_data}.}
\item{All SDSS-only chains (including alternative models) were run demanding the fulfillment of the Gelman-Rubin criteria \citep{Gelman1992}  with $R-1 \leq 0.03$ being our convergence criteria.  }

\item{We had to demonstrate systematics uncertainties in our model for $P(\lambda\ob|\lambda\true,z)$ did not appreciably bias our cosmological posteriors.  Since the random-point-injection model for this distribution is an extreme model, we adopted half of the systematic shift in the values of the cosmological parameters between our fiducial model and this extreme model as our estimate of the associated systematic uncertainty.  Note this definition implies that the extreme random-point-injection model is consistent with our fiducial model at $2\sigma$ despite being clearly extreme.  We demanded that these systematic shifts be less than the corresponding statistical uncertainties.}

\item {We explicitly verified that the priors of all parameters which we expected would be well-constrained a priori are not informative, that is the posteriors of such parameters did not run into the priors within the 95\% confidence region. In case this condition was not fulfilled we planned to extend the relevant prior ranges until the requirement was met.}
Parameters that are prior dominated (i.e. their posterior runs into the prior) are $M_{min}$, $\sigma_{intr}$, $s$, $q$, $h$, $\Omega_b h^2$, $\Omega_\nu h^2$, and $n_s$.  All of these were expected to be prior dominated a priori, and all prior ranges were purposely conservative. Of these, the two that might be most surprising to the reader are $M_{min}$ and $\sigma_{intr}$, as these parameters help govern the richness--mass relation.  However, notice that $M_{min}$ is the mass at which halos begin to host a single central galaxy; since our cluster sample is defined with the richness threshold $\lambda\geq 20$, the mass regime of halos which host a single galaxy is not probed by our data set.  Likewise, our data vector is comprised only of the mean mass of galaxy clusters in a given richness bin, a quantity that is essentially independent of the scatter in the richness--mass relation \citep[see][]{desy1wl}. 

\item{The $\chi^2$ of the data for our best-fit model must be acceptable.
To this end we considered the best-fit $\chi^2$ distribution recovered from 100 mock data realizations generated from the best-fit model of the data.  We assumed these 100 trials were distributed according to a $\chi^2$ distribution, and fit for the effective number of degrees of freedom.  The number of degrees of freedom is not obvious a priori due to the presence of priors in the analysis.  We considered the $\chi^2$ of our data to be not acceptable if the probability to exceed the observed value was less than 1\%, after marginalization over the posterior for the effective number of degrees of freedom.  We emphasize that verifying an acceptable $\chi^2$ does not unblind the cosmological parameters. While our model did indeed have an acceptable $\chi^2$ (see section \ref{sec:res:ref} for details), our plan was to revisit our model and covariance matrix estimation procedures in the case of an unacceptable $\chi^2$ value.  This proved unnecessary.}

\item{Finally, this paper underwent internal review by the collaboration prior to unblinding.
All members of the DES cluster working group, as well as all internal reviewers, agreed that 
our analysis was ready to unblind before we proceeded.}
\end{enumerate}

After all of our unblinding requirements were satisfied, we proceeded to unblind our analysis. No changes to the analyses were made post unblinding.


\begin{table*}
    \centering
    \footnotesize
    \caption{Model parameters and parameter constraints from the joint analysis of \redmapper\ SDSS cluster abundance and weak lensing mass estimates. In the forth column we report our model priors: a range indicates a top-hat prior, while $\mathcal{N}(\mu,\sigma)$ stands for a Gaussian prior with mean $\mu$ and variance $\sigma^2$. In the fifth column are listed the maximum likelihood values of the $1$-d marginalized posterior along with the $1$-$\sigma$ errors. Parameters without a quoted value are those having marginalized posterior distribution corresponding to their prior. Note that systematic uncertainties in the lensing masses are contained in their covariance, and are therefore not explicitly modeled in the likelihood. For the priors on the nuisance parameters $s$ and $q$ we report here only the square root of the diagonal terms of the covariance matrix defined in Equation~\ref{eqn:Chmf}. The distribution characterizing the impact of observational noise, projection effects, and percolation includes several additional parameters that are held fixed.  The systematic uncertainty associated with uncertainties in these parameters is estimated by repeating our analysis using an extreme set of values for these parameters, as estimated using random clusters in the SDSS \citep{costanzi18a}.  Because these parameters are not marginalized over in our chains, they are not included in this table.  We stress that the impact of the associated systematic error is negligible relative to the recovered width of our cosmological posteriors. }
    \label{tab:res}
   \begin{tabular}{lcccc}
    \hline 
	Parameter			&	Description	& Prior &	Posterior	\\
    \hline \vspace{-3mm}\\
	$\Omega_m$		& Mean matter density 						& $[0.0,1.0]$ &$ 0.22^{+0.05}_{-0.04}$ 		\vspace{0.5mm} \\
    $\ln(10^{10} A_s)$		& Amplitude of the primordial curvature perturbations	& $[-3.0,7.0]$ &$ 3.97^{+0.67}_{-0.47}$ 		\vspace{0.5mm} \\
    $\sigma_8$		& Amplitude of the matter power spectrum	& $-$ &$ 0.91^{+0.11}_{-0.10}$ 		\vspace{0.5mm} \\
    $S_8=\sigma_8 (\Omega_m/0.3)^{0.5}$		& Cluster normalization condition	& $-$ &$ 0.79^{+0.05}_{-0.04}$ 		\vspace{0.5mm} \\
    $\log M_{min} [{\rm M}_\odot /h]$	& Minimum halo mass to form a central galaxy& $(10.0,14.0)$ &$ 11.2\pm 0.2$	\vspace{0.5mm} \\
    $\log M_1 [{\rm M}_\odot /h]$		& Characteristic halo mass to acquire one satellite galaxy &$[10M_{min},30M_{min}]$ &$ 12.42^{+0.16}_{-0.13}$\vspace{0.5mm} \\
    $\alpha$		& Power-law index of the richness--mass relation 				& $[0.4,1.2]$ &$ 0.65^{+0.05}_{-0.07}$ 		\vspace{0.5mm} \\
    $\sigma_{intr}$	& Intrinsic scatter of the richness--mass relation				& $[0.1,0.8]$ &$ <0.4$		\vspace{0.5mm} \\
    $s$				& Slope correction to the halo mass function& $\mathcal{N}(0.037,0.014)$ &$ -$	\vspace{0.5mm} \\
    $q$				& Amplitude correction to the halo mass function& $\mathcal{N}(1.008,0.019)$ &$ -$	\vspace{2.0mm} \\
   $h$ & Hubble rate & $\mathcal{N}(0.7,0.1)$   & $-$ \vspace{0.5mm}  \\
   $\Omega_b h^2$ & Baryon density & $\mathcal{N}(0.02208, 0.00052)$ & $-$ \vspace{0.5mm}  \\
   $\Omega_\nu h^2$ & Energy density in massive neutrinos & $[0.0006,0.01]$ & $-$  \vspace{0.5mm} \\
   $n_s$ & Spectral index & $[0.87, 1.07]$ & $-$ \vspace{0.5mm}  \\
   \hline \vspace{-3mm}\\   
    \end{tabular}
\end{table*}

\section{Results}
\label{sec:res}

\subsection{SDSS Cluster Abundances and Weak Lensing Data}
\label{sec:res:ref}

We model the abundance of galaxy clusters and their weak lensing masses assuming a flat $\Lambda$CDM
cosmological model, allowing for massive neutrinos.  The full set of cosmological parameters
we consider is: $\ln(10^{10}A_s)$, $\Omega_m$, $n_s$, $\Omega_b h^2$, $h$, and $\Omega_\nu h^2$.  Neutrinos
are included assuming three degenerate neutrino species.  We adopt
the same priors as in the DES Year 1 analysis of galaxy clustering and weak lensing \citep{des17},
with the exception of $h$, where we adopt the slightly more restrictive prior $h=0.7\pm 0.1$.
There are also two parameters ($s$ and $q$) associated with systematic uncertainties in the halo
mass function, and 4 parameters governing the richness--mass relation : $M_{min}$, $M_1$, $\alpha$,
and $\sigma_{intr}$.  The priors for all parameters are summarized in Table~\ref{tab:res}. 
We have explicitly verified that increasing the range of the priors adopted
for the richness--mass relation parameters does not adversely impact the recovered cosmological
constraints.

\begin{figure*}
\begin{center}
    \includegraphics[width=\textwidth]{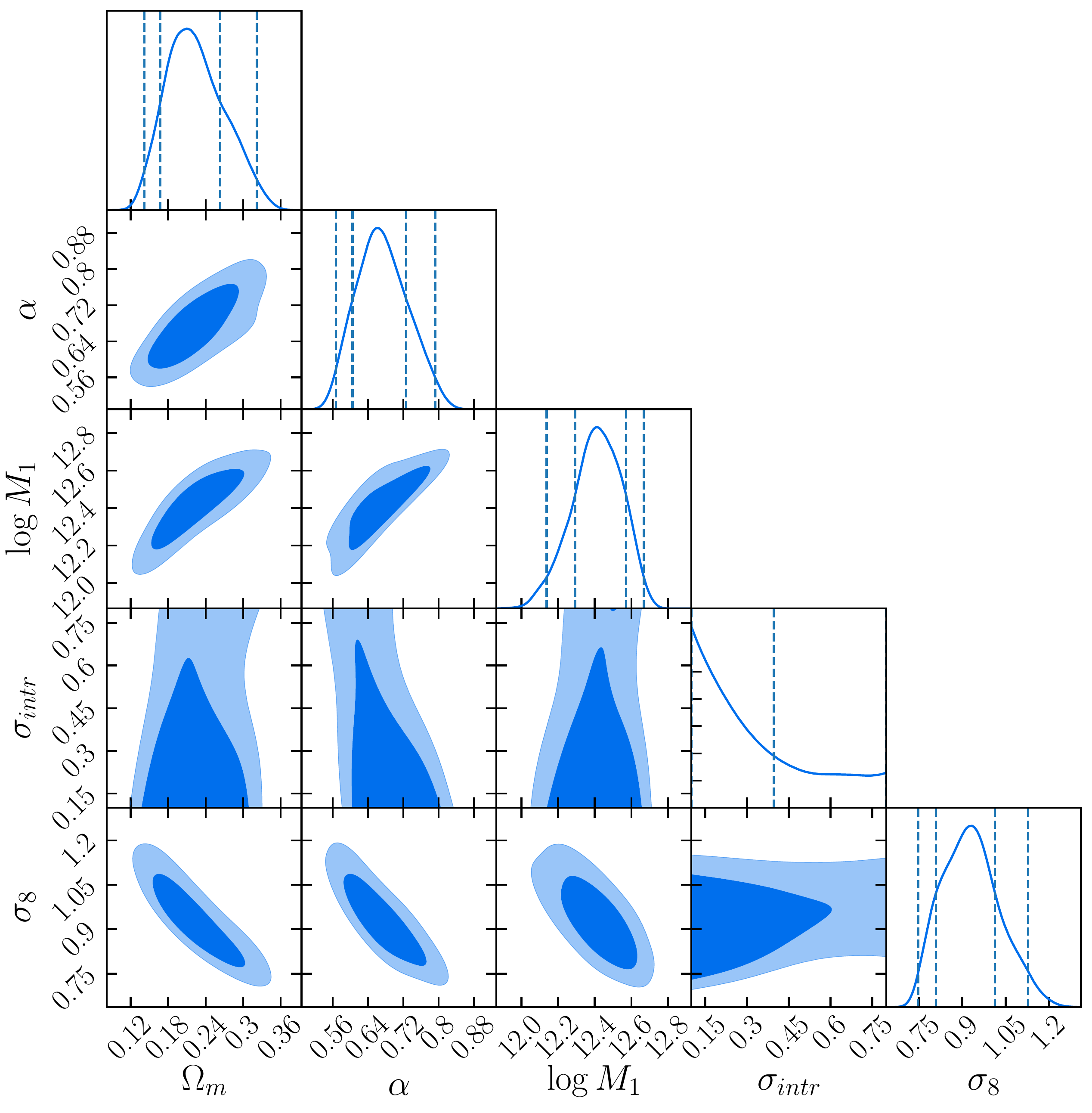}
\end{center}
\caption{Marginalized posterior distributions of the fitted parameter. The $2D$ contours correspond to the $68\%$ and $95\%$ confidence levels of the marginalized posterior distribution. The dashed lines on the diagonal plots correspond respectively to the $2.5$th,$16$th,$84$th and $97.5$th percentile of the $1$-d posterior distributions. The description of the model parameters along with their posterior are listed in Table \ref{tab:data}. Not included in this plot is the parameter $\Mmin$, which is prior dominated.}
\label{fig:data_res}
\end{figure*}

The result of our MCMC fitting procedure is shown in Figure \ref{fig:data_res}, while the marginalized posterior values are reported in Table \ref{tab:res}.  Parameters not shown in Figure \ref{fig:data_res} and without a quoted posterior in Table \ref{tab:res} are those whose posterior is equal to their prior.  Also shown in the table is the posterior for the so-called cluster normalization condition parameter $S_8 \equiv \sigma_8(\Omega_m/0.3)^{0.5}$.  In practice, the $\sigma_8$--$\Omega_m$ degeneracy in our cosmology analysis corresponds to
\begin{equation}
\sigma_8\left( \frac{\Omega_m}{0.3} \right)^{0.47} = 0.80 \pm 0.04.
\end{equation}
Nevertheless, unless otherwise specified in the text, from this point on we will focus on the cluster
normalization condition $S_8 \equiv \sigma_8(\Omega_m/0.3)^{0.5}$ as it has become a standard parameter 
in the literature.

\begin{figure}
\begin{center}
    \includegraphics[width=0.45 \textwidth]{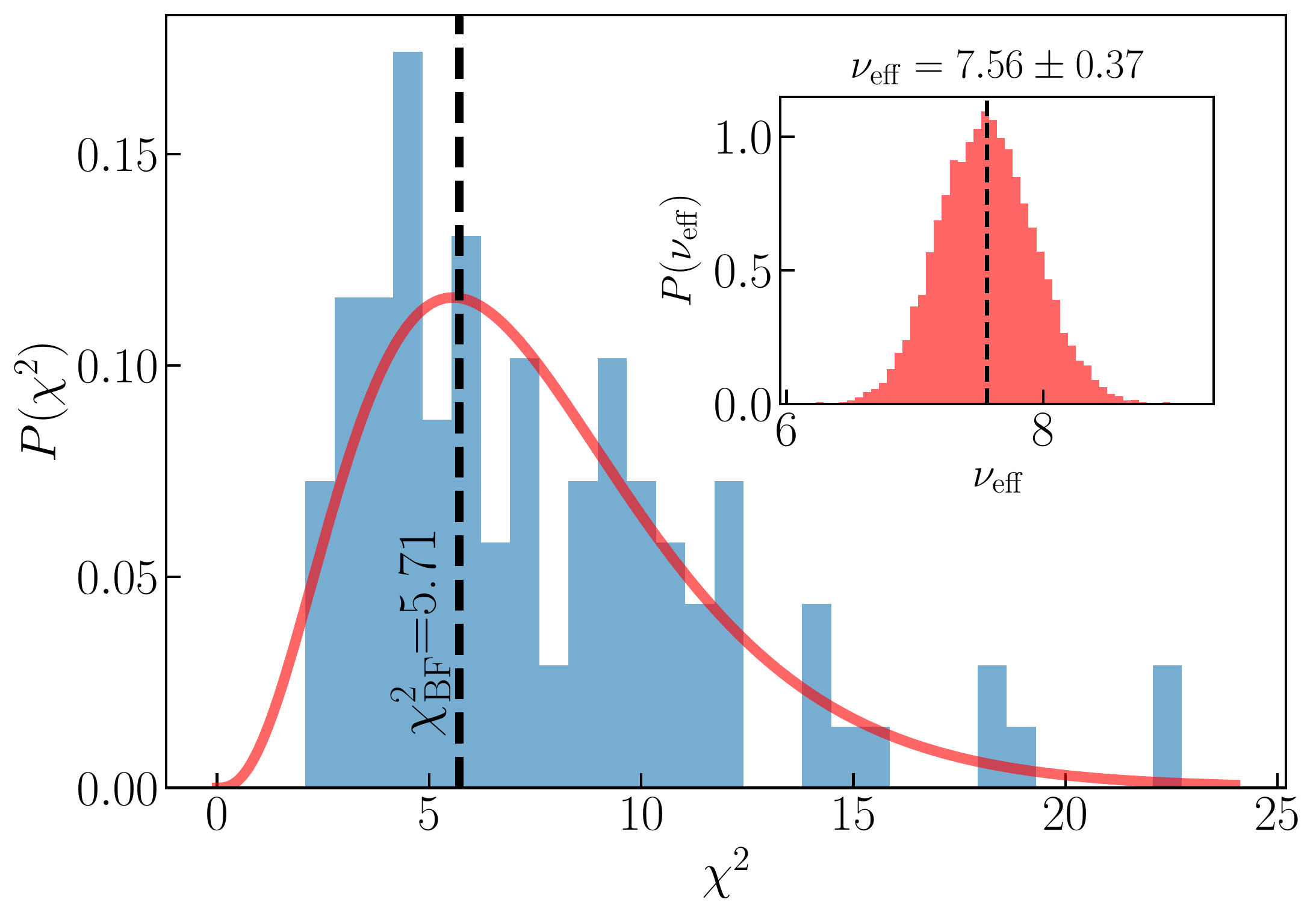}
\end{center}
\caption{Goodness of fit analysis. The \textit{blue} histogram shows the distribution of the best-fit $\chi^2$ values recovered from $100$ mock data realizations generated from the best-fit model of the data. The \textit{red} histogram in the inset plot shows the posterior distribution for the effective number of degrees of freedom obtained by fitting a $\chi^2$ distribution to the above $100$ $\chi^2$ values. The \textit{red solid} line represents the $\chi^2$ distribution for the best-fit model ($\nu_{\rm eff}=7.56$), while the vertical \textit{dashed} line corresponds to the $\chi^2$ value of the data.}
\label{fig:chisq}
\end{figure}

A comparison of our best-fit model with the data is shown in Figure \ref{fig:NC}.  The $\chi^2$ of our best-fit model is $\chi^2=5.71$. To assess the goodness of the fit we generated $100$ mock data vectors from our best-fit model of the data and, for each of them, we recovered the best-fit $\chi^2$ value. Assuming a $\chi^2$ distribution for the $100$ trials we fit for the effective number of degree of freedom, finding $\nu_{\rm eff}=7.56 \pm 0.37$ (see Figure \ref{fig:chisq}).  This corresponds to a probability to exceed of $p=0.64 \pm 0.04$. Thus, we conclude that our model provides a good fit to the data.

\begin{figure*}
\begin{center}
    \includegraphics[width=0.45 \textwidth]{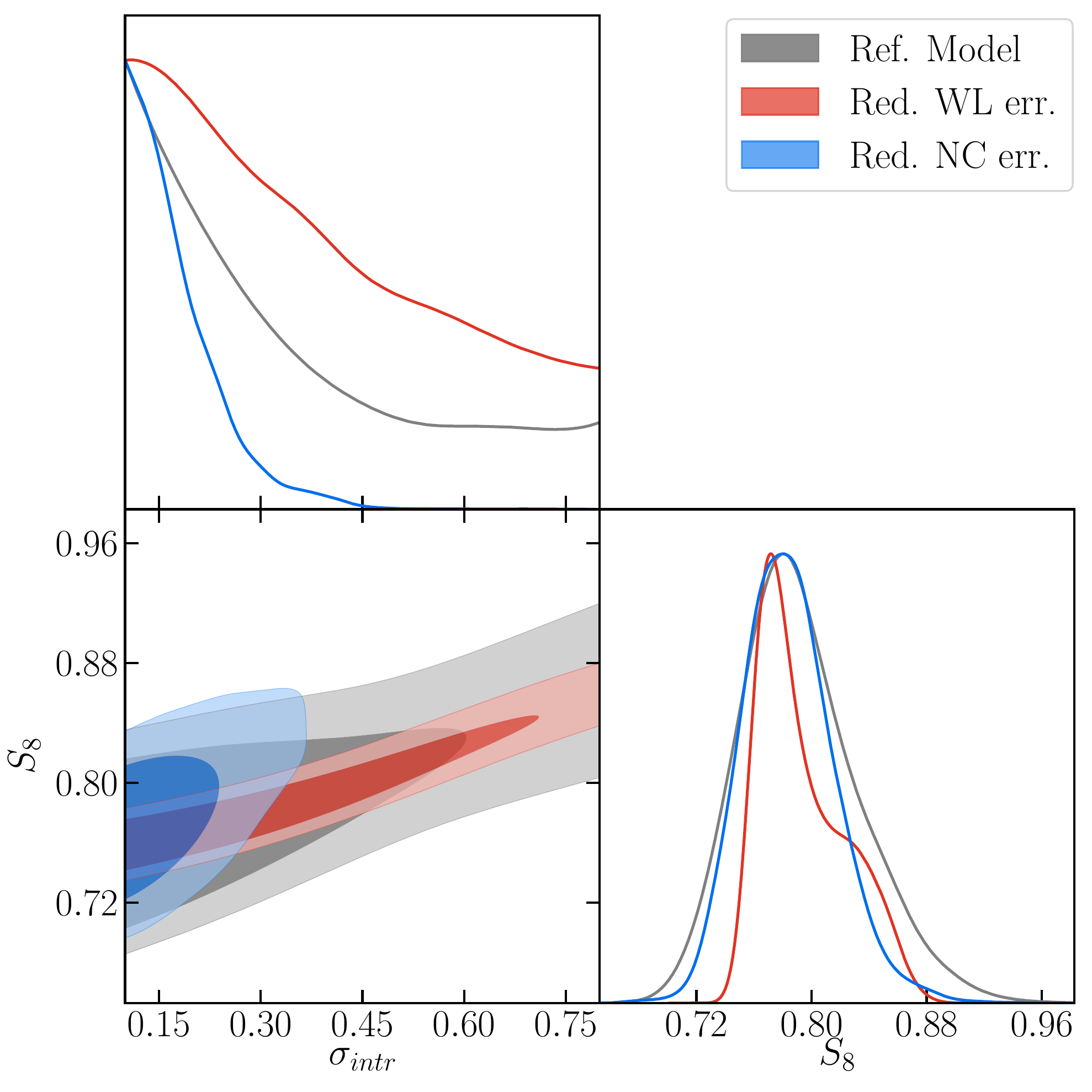}
    \includegraphics[width=0.45 \textwidth]{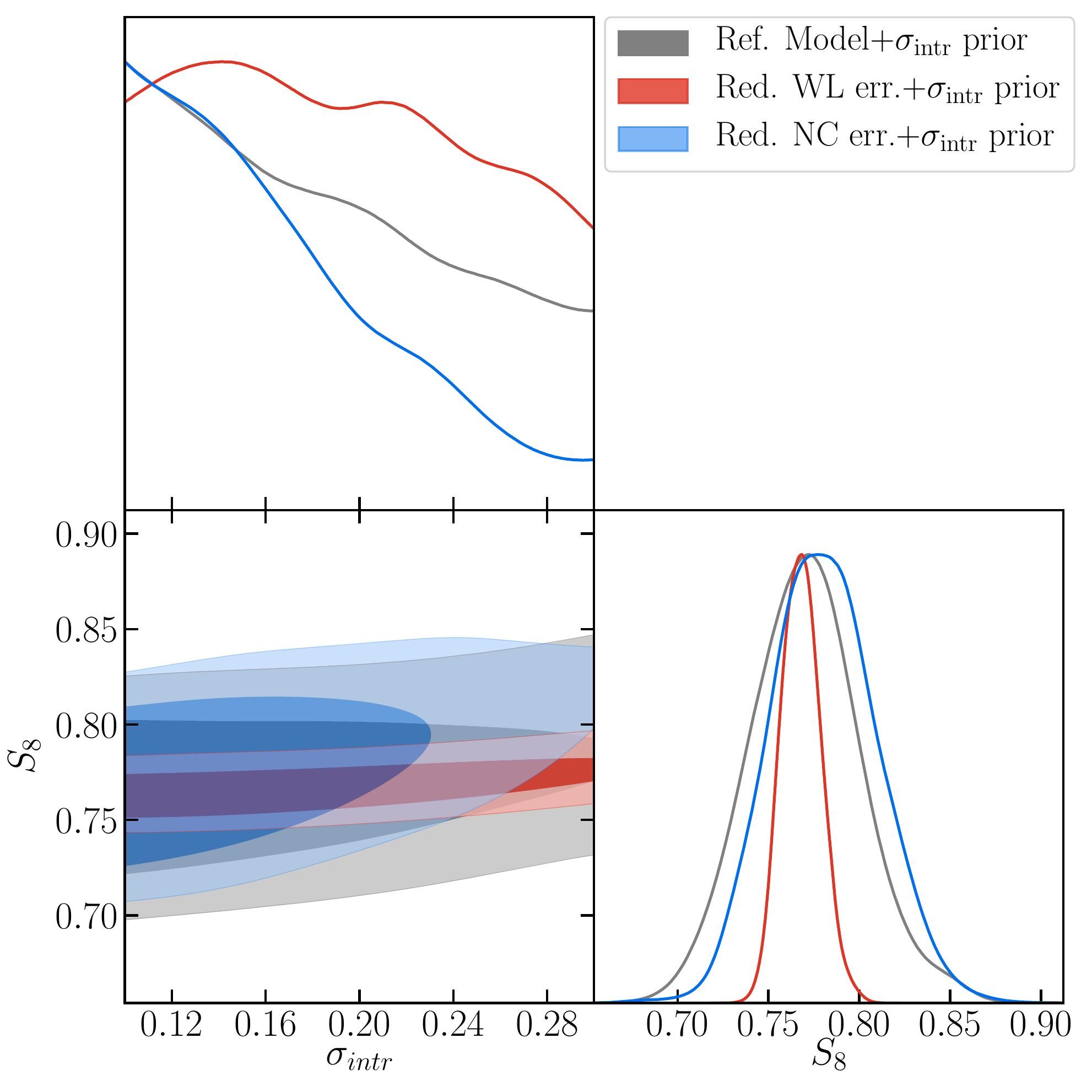}
\end{center}
\caption{Assessment of the error budget on $S_8$ associated with our two observables: cluster abundance data and weak lensing mass estimates. \textit{Left} panels: Comparison of the $(S_8,\sigma_{\rm intr})$ constraints derived in our reference model (\textit{gray} contours) and rescaling the weak lensing data (\textit{red} contours) or the number counts (\textit{blue} contours) covariance matrix by a factor $0.01$. \textit{Right} panels: same as the left panels but including a flat prior on the scatter parameter: $\sigma_{\rm intr} \in [0.1:0.3]$. See text for additional details and discussion.}
\label{fig:err_b}
\end{figure*}

We wish to determine whether the error budget in the cosmological parameter $S_8$ is dominated
by uncertainties in the abundance data
or the weak lensing data.  To determine this, we first compute the predicted abundance and weak lensing
data using our best-fit model.  We then run two additional chains using the predicted expectation
values as a synthetic data vector.  The key difference between the two chains is that for one 
we reduce the abundance covariance matrix by a factor of 100, while for the other chain we reduce the
covariance matrix of the weak lensing data by a factor of 100.  By comparing the cosmological posteriors
for these two chains we can determine if there is one observable which dominates our error budget.
In both cases the corresponding posterior on $S_8$ have an error bar $\sigma_{S_8} = 0.03$ (see \textit{left} panels of Figure \ref{fig:err_b}). Thus we conclude that our error budget on cosmological parameters is not dominated by either of the uncertainties associated with the two observables.  Rather, both observables contribute in comparable ways to the total error budget of $S_8$.

The balance between weak lensing errors and abundance uncertainties is surprising in light of the fact that all analyses to date have been dominated by uncertainties in the calibration of the mean of the observable--mass relation.  Nevertheless, this feature of our results can be easily understood.  The left panel of Figure~\ref{fig:err_b} demonstrates that there is a strong degeneracy between $S_8$ and $\sigma_{intr}$. Unlike previous analysis, which have incorporated well-motivated simulation-based priors on the scatter of the observable--mass relation, our analysis adopts a very broad prior on the scatter of the richness--mass relation.  This broad prior reflects the difficulty inherent to predicting properties of the richness--mass relation a priori. Since the scatter parameter impacts the detailed shape of the abundance function --- larger scatter leads to flatter abundance functions --- exquisitely precise measurements of the abundance function can break the degeneracy between scatter and $S_8$, leading to significant improvements in the $S_8$ constraints.  Conversely, even modest constraints of the scatter parameter $\sigma_{intr}$ can break the $S_8$--$\sigma_{intr}$ degeneracy, leading to tighter constraints.

We demonstrate the impact that a modestly-precise prior can have on our cosmological posteriors by redoing our analysis while imposing a flat prior $\sigma_{\rm intr} \in [0.1:0.3]$.  The corresponding posterior for $S_8$ is $S_8=0.77 \pm 0.03$.  If we now repeat our sensitivity analysis, and shrink the weak lensing mass errors, the width of the $S_8$ posterior decreases to 
$\sigma_{S_8} = 0.01$, while decreasing the abundance errors while holding the weak lensing errors fixed has a negligible impact on the posterior. These trends are illustrated in the right panel of Figure~\ref{fig:err_b}.  Evidently, external constraints on the scatter of the richness--mass relation of \redmapper\ clusters are extremely valuable from a cosmological perspective.

\subsection{Robustness to Assumptions About the Richness--Mass Relation}
\label{sec:res:model}

\begin{figure*}
\begin{center}
    \includegraphics[width=\textwidth]{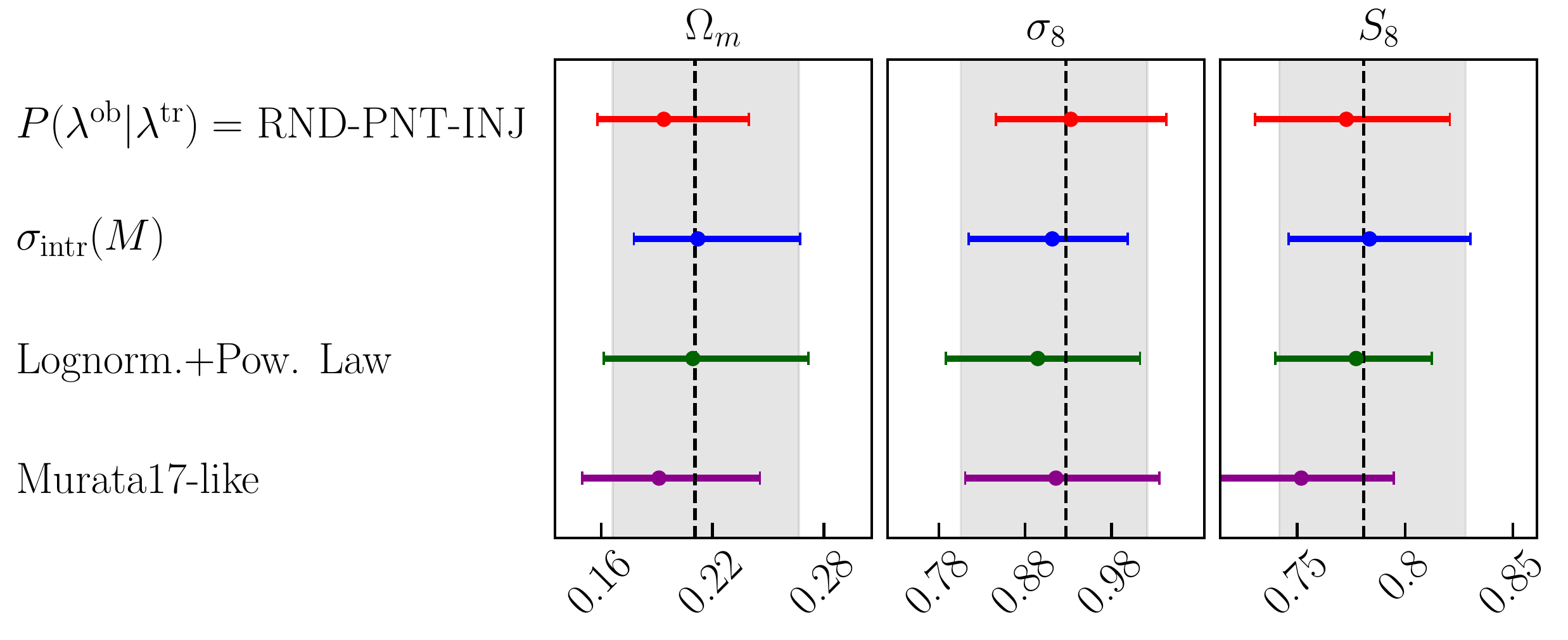}
\end{center}
\caption{Comparison of the $68\%$ confidence regions for $\Omega_m$, $\sigma_8$ and $S_8$ derived assuming different model for $P(\lambda\ob|M)$ (see Section \ref{sec:res:model}). The shaded area corresponds to the constraints derived using our reference model.
}
\label{fig:mod_dep}
\end{figure*}

Figure~\ref{fig:data_res} shows that there is strong covariance between cosmological parameters and parameters governing the richness--mass relation.  Consequently, it is of critical importance to simultaneously fit for the richness--mass relation as part of the cosmological analysis. Likewise, one may ask to what extent are our cosmological constraints sensitive to the detailed assumptions we have made about the richness--mass relation.  To address this question, we have repeated our analysis with a range of richness--mass relation models as summarized in Figure~\ref{fig:mod_dep}.  The models considered
are: 
\begin{itemize}
\item A model that allows for the intrinsic scatter $\sigma_{intr}$ to vary with mass via
\begin{equation}
\sigma_{intr}(M)=\sigma_{intr,0}(M/(M_1-M_{min}))^\beta \, .
\end{equation}
\item A model that neglects the perturbations on the observed richness due to correlated structures. To this end we set $P(\lambda\ob|\lambda\true,z\true)$ equal to the probability distributions recovered from injecting synthetic clusters at random positions in the survey mask. This calibration provides a very strict lower limit on the scatter of $\lambda\ob$ due to projection effects: clusters do contain correlated large scale structure.  The difference in the posteriors of the cosmological parameters between our fiducial run and the random-point-injection model places a strict upper limit on the systematic associated with our modeling of projection effects. 
\item A model in which the richness--mass relation is a simple power law
 --- $\langle\lambda\true|M \rangle=\lambda_0 (M/10^{14.344})^\alpha$ --- and $P(\lambda\true|M)$ is a log-normal distribution in which the total scatter it the sum of a Poisson-like term and an intrinsic scatter term --- 
$\sigma_{\ln\lambda\true}^2= \sigma_{intr}^2+ (\langle \lambda\true|M \rangle - 1)/\langle \lambda\true|M \rangle^2$.
\item The richness--mass relation model of \citet{Murata2017}.  This model assumes $P(\lambda\ob|M)$ is log-normal, the mean richness--mass relation is a power-law, and the intrinsic scatter is mass-dependent, and given by  $\sigma_{intr}(M)=\sigma_{intr,0}+\beta \ln (M/2.2\times 10^{14})$. According to \citet{Murata2017} all integrals are truncated at $M_{min}=10^{12}\ \hMsun$.  Reassuringly, when we mirror the analysis of \citet{Murata2017} and fix our cosmological parameters to \citet{PlanckXIII2016} we reproduce their results despite significant methodological differences in how the weak lensing data is incorporated into the likelihood.
\end{itemize}


\begin{figure*}
\begin{center}
    \includegraphics[width= \textwidth]{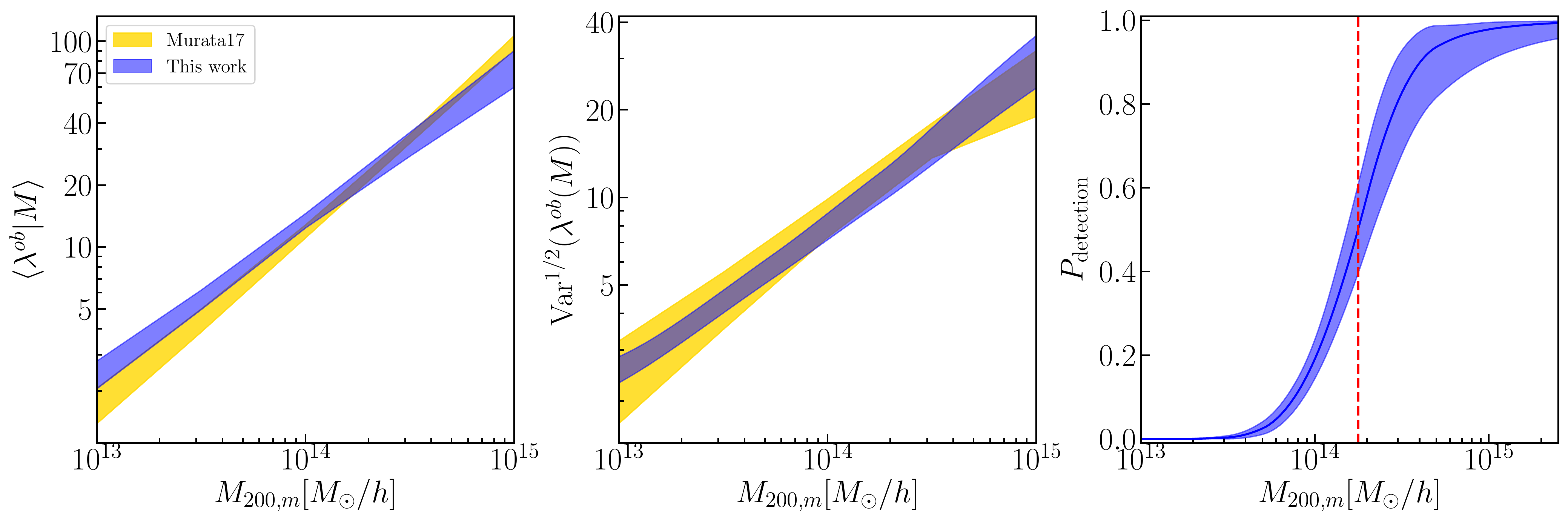}
\end{center}
\caption{Observable--mass relation and mass-selection function of the redMaPPer catalog assuming our reference richness--mass relation model (Eq. \ref{eqn:RMR}) at the mean sample redshift $z=0.22$. {\it Left} panel: Expectation value for the observed richness as a function of mass. {\it Central} panel: Scatter of $\lambda\ob$ -- $\Var^{1/2}(\lambda\ob|M)$ -- as a function of mass. 
{\it Right} panel: detection probability as a function of cluster mass. The \textit{dashed} vertical line correspond to the mass at which the detection probability is $50\%$  ($\log M_{50\%}= 14.24 [M_{\odot}/h])$. 
The \textit{blue} area corresponds to the $68\%$ confidence interval derived for the different quantities in this work. For comparison, the results obtained in \citet{Murata2017} are shown in \textit{yellow} in the two left panels.
}
\label{fig:lob_rels}
\end{figure*}


As can been seen in Figure~\ref{fig:mod_dep} our cosmological posteriors are all consistent with one another.  It is clear that the more restrictive parameterizations (e.g. log-normal + power-law) result in somewhat tighter constraints.  Notably, our standard result --- which we believe is the most appropriate model --- results in the most conservative posteriors.  In particular, we see that opening up the freedom of a mass--dependent intrinsic scatter does not negatively impact the posterior on $S_8$. Interestingly, we noticed that our random-point injection model, which grossly underestimates the impact of projection effects, had a small impact on the posterior of the intrinsic scatter, modifying instead the best fit value for the slope of the richness--mass relation, a degeneracy that should be easily broken via multi-wavelength analyses of the \redmapper\ clusters.   Finally, we note that the Murata-like parameterization has the largest impact on our posteriors, with the systematic shift in $S_8$ being comparable to the width of the posterior.  We will address the origin of this shift in the next section.

\subsection{The Observable--Mass Relation for redMaPPer Clusters}
\label{sec:obsmass}

The left panel of Figure \ref{fig:lob_rels} shows the observed richness--mass relation $\avg{\lambda\ob|M}$
for our fiducial model at the mean sample redshift $z=0.22$. The error bars reflect the posterior of the mean relation at each mass.  These are computed as follows: for each point in our chains we evaluate $\avg{\lambda\ob|M}$ along a grid of masses.  The mean and variance of $\avg{\lambda\ob|M}$ across the chain are then recorded.  In the left panel of Figure~\ref{fig:lob_rels} we use these quantities to plot the 68\% confidence interval for the posterior of the mean of the richness--mass relation. 
The central panel of Figure \ref{fig:lob_rels} is computed in a similar way, only now we show the posterior for the scatter $\Var^{1/2}(\lambda\ob|M)$.\footnote{A reader might find useful to have simple power-law fits to the data shown in Figure~\ref{fig:lob_rels}. We provide such fits below:
\begin{eqnarray}
\avg{\lambda\ob|M} & = & 30.0\left( \frac{M}{3 \times 10^{14} [{\rm M}_\odot h^{-1}]}\right)^{0.75} \nonumber \\
\Var^{1/2}(\lambda\ob|M) & = & 14.7 \left( \frac{M}{3 \times 10^{14} [{\rm M}_\odot h^{-1}]}\right)^{0.54} . \nonumber
\end{eqnarray}
The fits correspond to the best fit-relations.  No errors are provided since these are meant to be ``quick-look'' references.  Detailed quantitative analyses should rely on the full posterior of our model, which will be made available at http://risa.stanford.edu/redmapper/ when the paper is published.
}

For comparison we include in the two panels the richness--mass relation and scatter derived in \citet{Murata2017}, who analyzed this sample of clusters using the same weak lensing shear catalog we employed.  There are significant methodological differences between the two analyses.  Specifically, \citet{Murata2017}:
\begin{itemize}
\item Uses emulators instead of an analytic model for the weak lensing profile of clusters, effectively holding the concentration of the galaxy clusters fixed.  They also place no priors on the miscentering parameters. 
\item Adopts a log-normal model for $P(\lambda\ob|M)$.
\item Adopts a power-law relation for both the mean and variance of $\lambda\ob$ at fixed mass.
\end{itemize}
The \textit{yellow} bands in Figure \ref{fig:lob_rels} are obtained by estimating the 68\% confidence region of their posterior from 10,000 Gaussian random draws of the richness--mass relation parameters quoted in \citet{Murata2017}. Despite methodological differences, the two results are remarkably similar over the mass range probed by the survey ($M \gtrsim 10^{13.5} [{\rm M}_\odot /h]$, see below).
It is especially impressive how well the scatter found in \citet{Murata2017} agrees with our finding, confirming their argument that the mass-trend in the scatter of the richness--mass relation they recovered is due to contamination from projection effects.

\begin{figure}
\begin{center}
    \includegraphics[width=0.45 \textwidth]{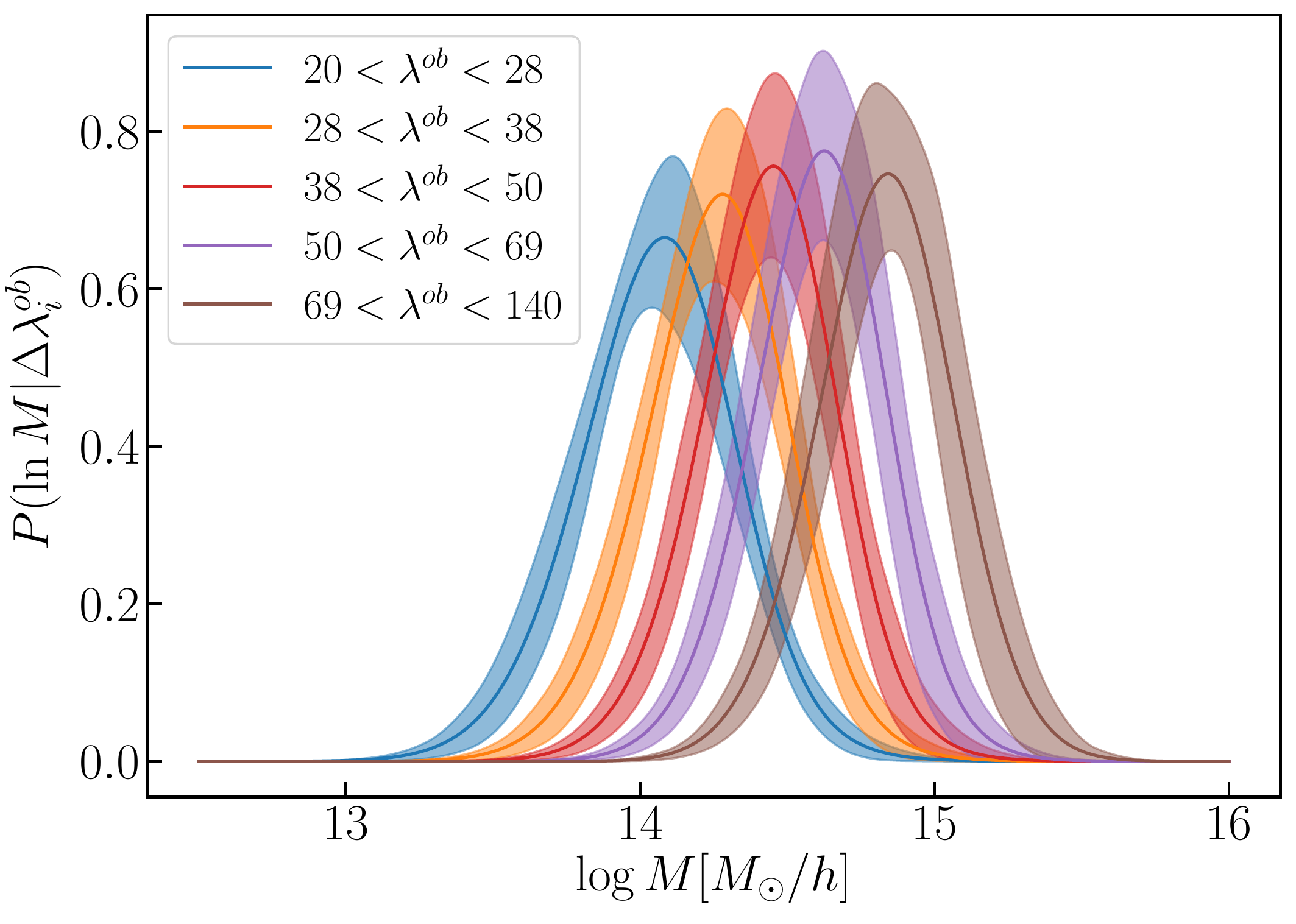}
\end{center}
\caption{Distribution of halo mass for clusters in each of the five richness bins employed in this work, as labeled.  The width of the bands correspond to the 68\% confidence interval of the distribution as sampled from our posterior.}
\label{fig:mdist}
\end{figure}

There is, however, one notable difference between our results and those of \citet{Murata2017}.  Figure 7 of \citet{Murata2017} shows that the mass distribution for clusters in the richness bin $\lambda\ob \in [20,30]$ extends to masses as low as $10^{12}\ \hmsun$, the mass cut imposed in that analysis. Figure~\ref{fig:mdist} shows the posterior of the mass distribution for clusters in our analysis, as labeled. Unlike \citet{Murata2017}, we do not see any evidence for a population of low-mass ($M\leq 10^{13}\ \hmsun$) clusters.  We believe the large number of low-mass halos in the \citet{Murata2017} analysis is driven by the combination of a log-normal model whose scatter increases with decreasing mass, and a model that ignores the central/satellite dichotomy at low masses.  Our model avoids this problem by: 1) enforcing the appropriate Poisson limit in the limit of low $\lambda\true$; and 2) developing a simulation-based model for projection effects that adequately characterizes non-Gaussian tails in the distribution $P(\lambda\ob|\lambda\true)$.

As an independent check of our conclusions, we estimate the halo masses of individual \redmapper\ clusters using the stellar content of the \redmapper\ central galaxy.  Specifically, we fit the photometric SDSS data using a stellar population synthesis model to derive the stellar mass of each of the assigned \redmapper\ central galaxies (Moustakas et al., in preparation).  We then use the UniverseMachine algorithm \citep{behroozietal18} to determine the relation between the stellar mass of the central galaxy of a halo and the mass of the parent halo.  Using the relation between halo mass and the stellar mass of the central galaxy, we can readily assign a mass estimate to each \redmapper\ cluster.  We find that the stellar mass estimates of 95\% of \redmapper\ central galaxies correspond to a halo mass of $2\times 10^{13}\ \hmsun$ or higher.  While the stellar-mass to halo-mass relation of central galaxies is relatively uncertain, this result disfavors the existence of a significant population of low-mass ($M\leq 10^{13}\ \hmsun$) \redmapper\ clusters, in agreement with our results.

We suspect that the tail of low-mass halos recovered in the \citet{Murata2017} model is responsible for the $\sim 1\sigma$ shift in $S_8$ seen in Figure~\ref{fig:mod_dep} when adopting a power-law lognormal model for the richness--mass relation (i.e. the Murata-like analysis): the artificial boost in the abundance of low richness clusters is compensated by a decrease in the predicted halo abundance, which is in turn achieved by lowering the cluster normalization condition.

\begin{figure}
\begin{center}
    \includegraphics[width=0.45 \textwidth]{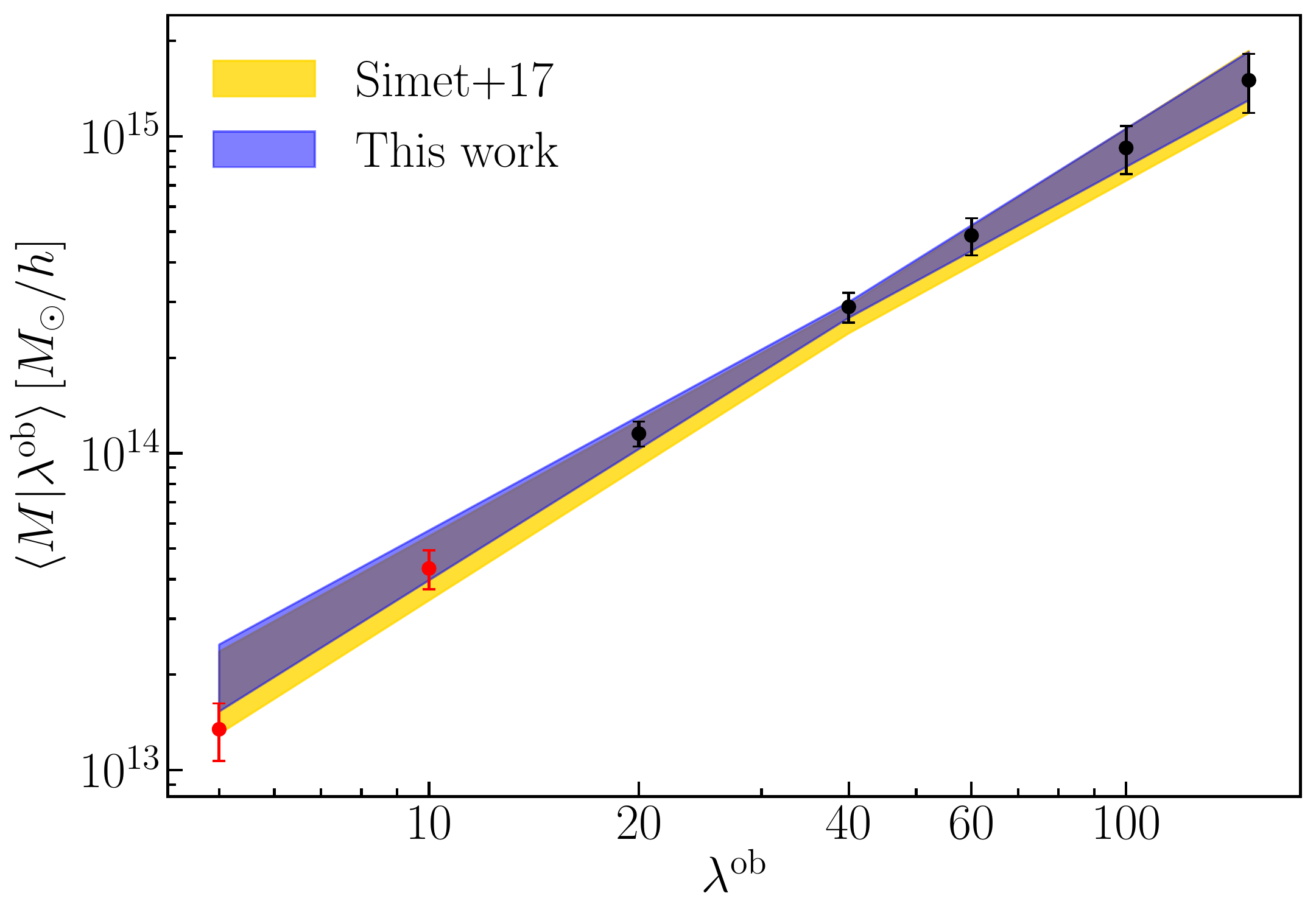}
\end{center}
\caption{Mass--richness relation derived from the redMaPPer  SDSS sample combining cluster abundance and weak lensing data (\textit{blue area}). The data points show the mean mass at a given richness, $\avg{M|\lambda\ob}$, derived from the posterior distributions of our reference model (see text for details). In analogy with the analysis of \citet{Simet2016} only the {\it black} points ($\lambda\ob > 20$) are used to fit the mass--richness relation. For comparison the mass--richness relation derived in \citet{Simet2016} is shown in \textit{yellow} .}
\label{fig:m_rel}
\end{figure}

Finally, the right panel of Figure \ref{fig:lob_rels} shows the mass-selection function of the galaxy clusters selected in our experiment.  That is, it shows the probability $P(M,z)$ that a halo of mass $M$ at redshift $z$ is included in the SDSS \redmapper\ sample.  This probability is given by
\begin{equation}
P(M,z) = \int_{20}^{\infty} d\lambda\ob\ P(\lambda\ob|M,z) .
\end{equation}
The probability $P(M,z)$ is evaluated at a grid of masses for each point in the chain, and the corresponding mean and uncertainty is calculated.  Using linear interpolation over this grid, we find that the mass value for which the detection probability is $1/2$ is $\log M=14.24 [M_{\odot}/h]$.

Using our best-fit cosmological model we can combine the halo mass function with our recovered richness--mass relation to arrive at our best-fit mass--richness relation.  In particular, for each point in the chain we can readily compute $\avg{M|\lambda\ob}$ along a grid of richnesses, and calculate mean and variance of these quantities at each $\lambda\ob$ value as we sample our posterior. We estimated the mean mass at a grid of values, and computed the corresponding covariance matrix, and then fit the data with a power-law to arrive at the corresponding mass--richness relation.   The resulting mass--richness relation is shown in Figure \ref{fig:m_rel} along with the relation derived in \citet{Simet2016}. The posteriors for the mass--richness relation in our analysis is
\begin{equation}
\avg{M|\lambda} = 10^{14.45 \pm 0.03} \left( \frac{\lambda}{40} \right)^{ 1.29 \pm 0.09} \, ,
\end{equation}
where we used only $\lambda\ob>20$ data points (\textit{black} dots in the figure) to fit the power-law relation.  
This is to be compared to the \citet{Simet2016} relation, $\avg{M|\lambda}=10^{14.42\pm0.04} (\lambda/40)^{ 1.3 \pm 0.1}$. Note that the exponent $14.42$ has been obtained correcting the best-fit value $14.37$ derived in \citet{Simet2016} assuming $\Omega_m=0.30$ via Equation~\ref{eqn:MvsOm} using the mean $\Omega_m$ value derived in this analysis. Moreover, we expect a small difference in amplitude due to the updates to our multiplicative shear bias model and the \redmapper\ centering fraction. All together, these corrections should boost the amplitude of the mass--richness relation by $\approx 6\%$, in good agreement with our results.



\begin{figure}
\begin{center}
    \includegraphics[width=0.45 \textwidth]{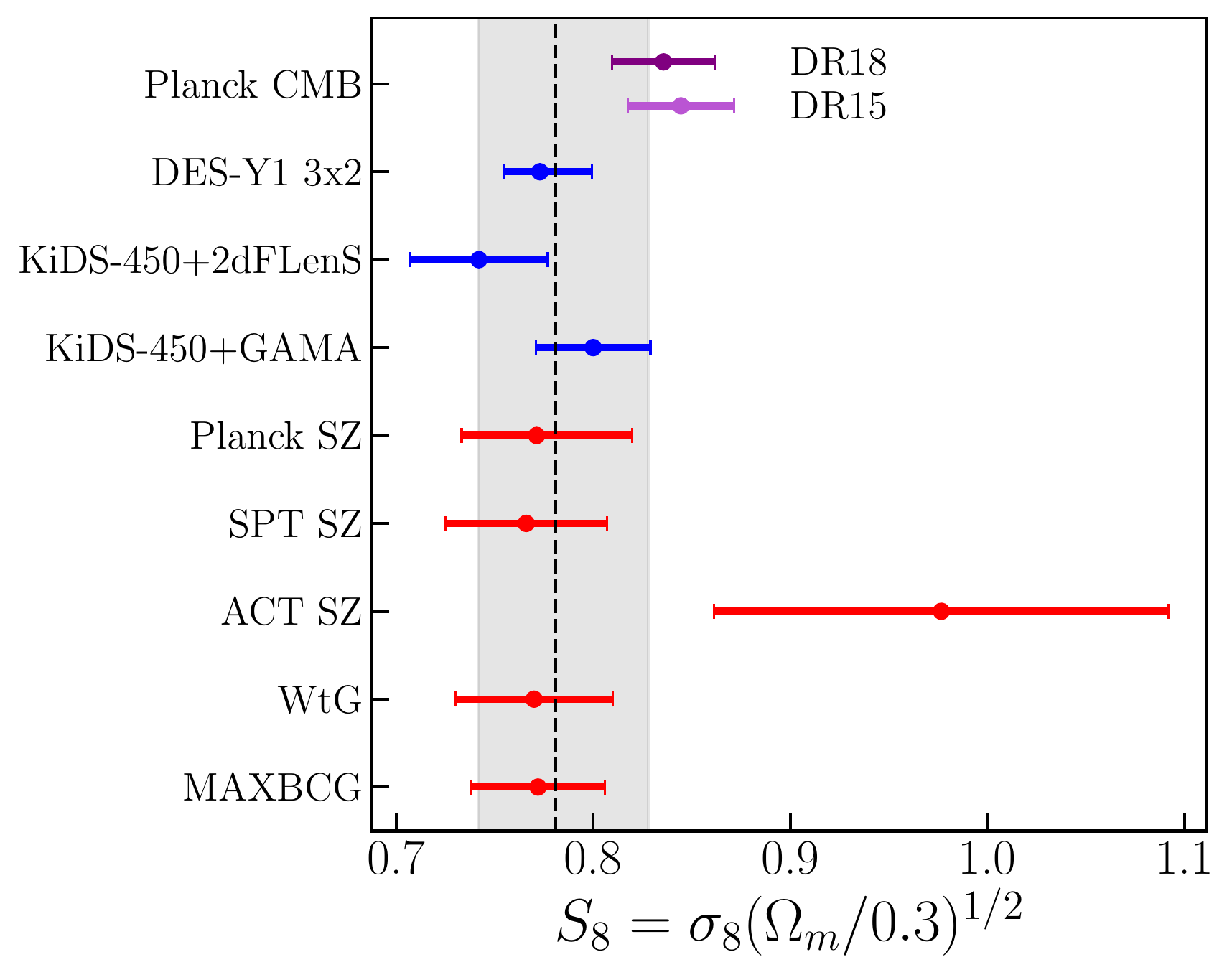}
\end{center}
\caption{Comparison of the $68\%$ confidence level constraint on $S_8$ derived from our baseline model (\textit{shaded gray} area) with other constraints from the literature: \textit{red} error bars for cluster abundance analyses, \textit{blue} error bars for weak lensing and galaxy clustering analyses and \textit{purple} for the CMB constraint. From the bottom to the top: MAXBCG from \citet{Rozo2010}; WtG from \citet{Mantz2015}; ACT SZ from \citet{ACT2013} (BBN+H0+ACTcl(dyn) in the paper); SPT SZ from \citet{SPTSZ2016}; Planck SZ from \citet{PlanckSZ2016} (CCCP+$H_0$+BBN in the paper); KiDs-450+GAMA from \citet{kidsgama2018}; KiDs-450+2dFLens from \citet{kids2df2018}; DES-Y1 3x2 from \citet{des17}; \Planck\ CMB from \citet{PlanckXIII2016} (DR15) and \citet{Planck2018} (DR18). Note that all the constraints but those from DES-Y1 3x2 and Planck CMB have been derived fixing the total neutrino mass either to zero or to $0.06$ eV.}
\label{fig:ext_data}
\end{figure}


\section{Comparison to External Data Sets}
\label{sec:ext}

\begin{table*}
    \centering
    \footnotesize
    \caption{Parameter constraints from the combination of our analysis of the SDSS \redmapper\ cluster abundances with BAO and Planck CMB data sets (see text for details). For reference here are also reported the constraints derived from Planck DR15 CMB and low $l$ polarization data \citep{PlanckXIII2016}, BAO data \citep{BAO6dF,BAOSDSSMain,BAOBoss}, and SDSS data alone. }
    \label{tab:ext_data}
   \begin{tabular}{lcccccccc}
    \hline 
	Data sets			&	$\Omega_m$	& $\sigma_8$ &	$h$ & $S_8$  & $\log M_{min}$ & $\log M_{1}$ & $\alpha$ & $\sigma_{intr}$	\\
    \hline \vspace{-3mm}\\
    Planck15		& $0.328^{+0.039}_{-0.026}$ & $0.81^{+0.03}_{-0.05}$ &$ 0.662^{+0.019}_{-0.028}$ & $0.841 \pm 0.026$&	$-$ &	$-$&	$-$&	$-$	\vspace{0.5mm} \\
    BAO		& $0.373 \pm 0.053$ & $-$ &$ 0.694 \pm 0.033$ & $-$&	$-$ &	$-$&	$-$&	$-$	\vspace{0.5mm} \\
	SDSS		& $0.22^{+0.05}_{-0.04}$ & $0.91^{+0.11}_{-0.10}$ &$ -$ &$0.79^{+0.05}_{-0.04}$	& $11.2\pm 0.2$ & $12.42^{+0.16}_{-0.13}$ & $0.65^{+0.05}_{-0.07}$ & $<0.4$	\vspace{0.5mm} \\
	SDSS+BAO		& $0.316\pm 0.036$ & $0.78\pm 0.06$ &$ 0.662^{+0.019}_{-0.022}$ &$0.792^{+0.039}_{-0.037}$	& $11.38\pm 0.17$ & $12.63 \pm 0.09$ & $0.76\pm 0.05$ & $<0.2$	\vspace{0.5mm} \\
    SDSS+BAO+Planck15		& $0.316^{+0.010}_{-0.008}$ & $0.81\pm 0.02$ &$ 0.671^{+0.006}_{-0.008}$ &$ 0.829^{+0.022}_{-0.020}$	&	$11.42\pm 0.15$ & $12.65 \pm 0.02$ & $0.76 \pm 0.03$ & $<0.2$	\vspace{0.5mm} \\
   \hline \vspace{-3mm}\\   
    \end{tabular}
\end{table*}

Our analysis allows us to place the constraint $S_8=0.79^{+0.05}_{-0.04}$.  A comparison
of our baseline result (\textit{gray shaded} area) to several other constraints from the literature can be seen in Figure \ref{fig:ext_data}. To estimate the level of tension between two analyses A and B we consider the quantity: $T_{A,B}=|\Delta S_8|/\sigma_{\rm tot}$, where $\sigma_{\rm tot}^2=(\sigma_{S_8,A})^2+(\sigma_{S_8,B})^2$. According to our consistency criterion (see item (v) in Section \ref{sec:blind}) all the measurements of $S_8$ from the external data sets we considered are consistent with the one derived in this analysis ($\Delta S_8<3\sigma_{\rm tot}$).  The most significant difference comes with respect to the cluster constraints from the Atacama Cosmology Telescope cluster sample \citep{ACT2013}, though the significance of this difference is still below $2\sigma_{\rm tot}$. As for the \planck\ DR18 CMB $S_8$ constraint, the significance of the difference between the two results is only $1.1\sigma_{\rm tot}$\footnote{The significance of the difference between our $S_8$ posterior and that of the {\it Planck} 2015 analysis is $1.2\sigma_{\rm tot}$.}.

We planned to combine the SDSS \redmapper\ cluster abundances constraints with two distinct external data sets, provided that these data sets were consistent with our results (see item (vii) in section \ref{sec:blind}). Namely we considered:
\begin{enumerate}
\item Baryon Acoustic Oscillation (BAO) data from multiple galaxy
surveys, specifically the Six Degree Field Galaxy Survey \citep[6dF][]{BAO6dF}, 
the SDSS DR 7 Main galaxy sample \citep{BAOSDSSMain}, and data from the Baryon
Oscillation Spectroscopic Survey \citep[BOSS][]{BAOBoss}.
\item CMB data from \Planck\ satellite, including low $l$ polarization data, from the 2015 data release \citep[hereafter Planck DR15; ][]{PlanckXIII2016}.
\end{enumerate}
When combining with BAO data we replace the Gaussian prior on $h$ by a flat prior, while when combining with CMB data we relax all the informative priors on cosmological parameters (i.e. $h$, $\Omega_b h^2$ and $\Omega_\nu h^2 $, see Table \ref{tab:res}) and add the optical depth $\tau$ as a free parameter.

\begin{figure*}
\begin{center}
    \includegraphics[width=0.45 \textwidth]{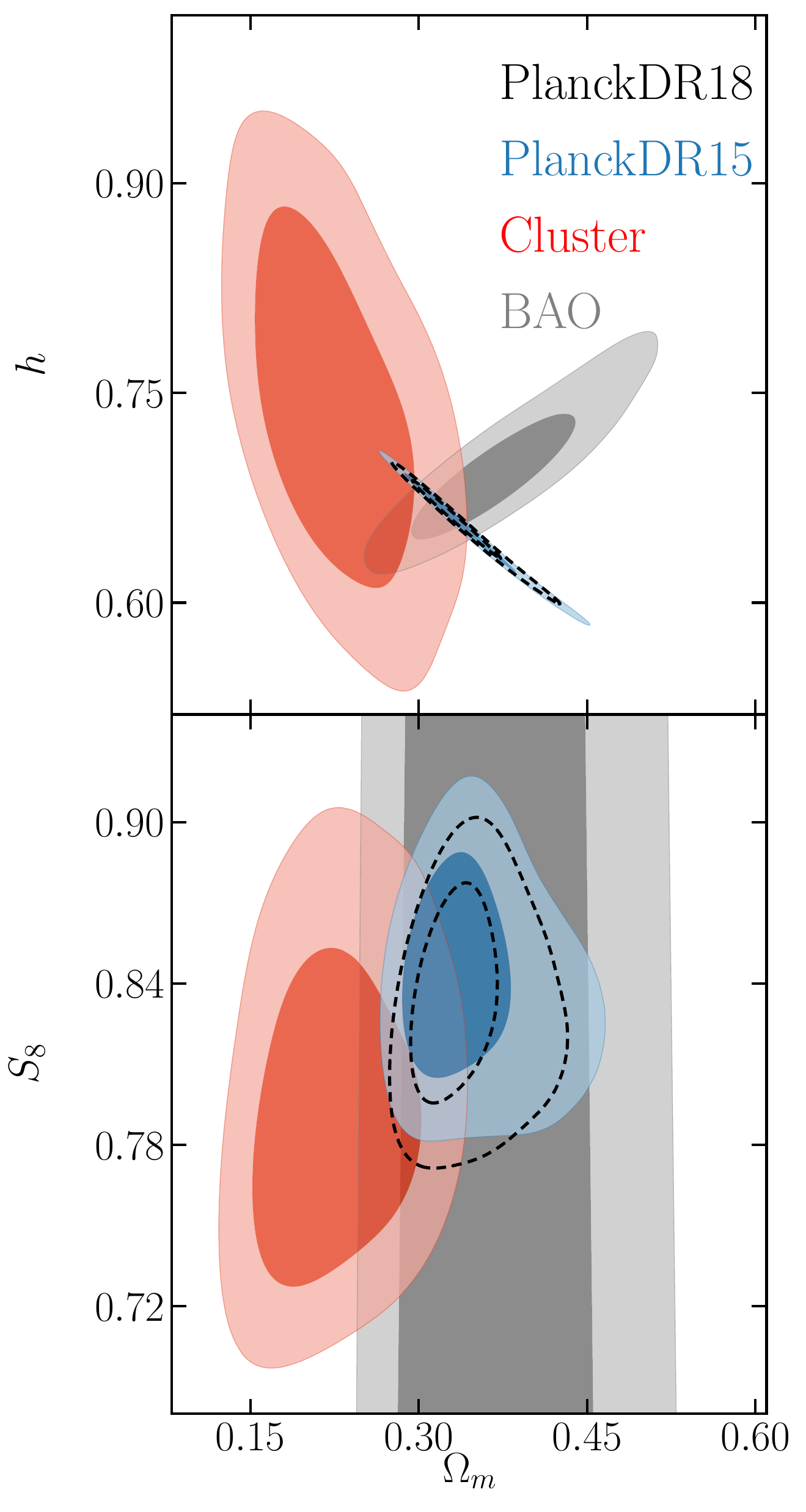}
    \includegraphics[width=0.438 \textwidth]{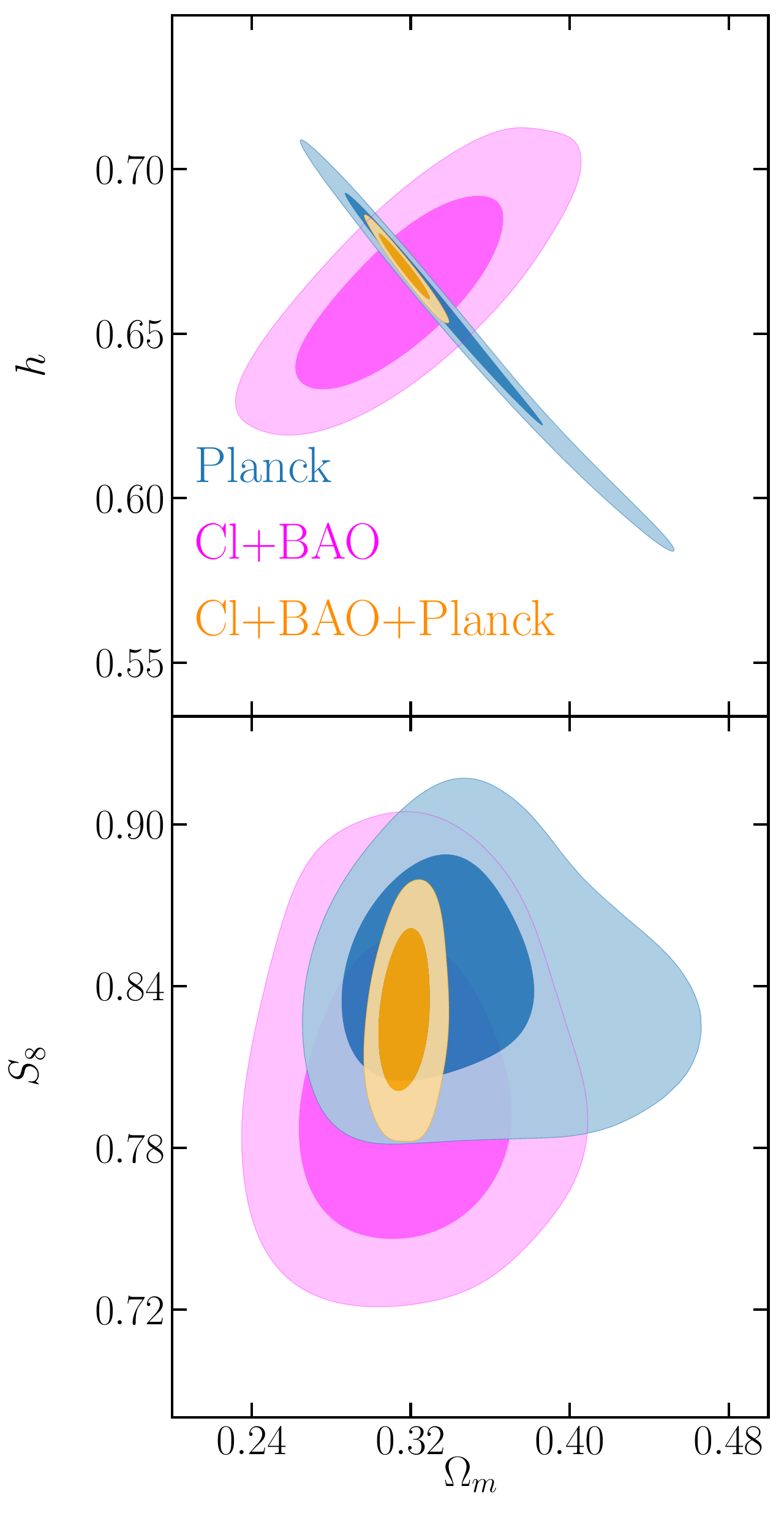}
\end{center}
\caption{$68\%$ and $95\%$ confidence level constraints in the $(S_8,\Omega_m,h)$ plane. \textit{Left} panel: Comparison of the constraints derived from the different data sets considered in this work: \planck\ DR15 (\textit{blue}), BAO (\textit{gray}) and SDSS clusters (\textit{red}). The BAO contours are obtained including the flat prior $\Omega_\nu h^2 \in [0.0006,0.01]$, as in the SDSS cluster analysis. Also shown for comparison the latest \Planck\ results (\citep{Planck2018}; \textit{dashed black lines}). \textit{Right} panel: $68\%$ and $95\%$ confidence contours obtained from the combination of the SDSS cluster sample with BAO data (\textit{magenta}) and from the combination of SDSS clusters, BAO and \planck\ data (\textit{orange}). For comparison, the \textit{blue} contours show the constraints derived from \planck\ CMB data alone. Note the different scales between the \textit{left} and \textit{right} panel.}
\label{fig:SDSS_ext_data}
\end{figure*}

According to the protocol detailed in section \ref{sec:blind}, both data sets passed the consistency criterion required to perform the combined analysis with the SDSS data. Specifically, for the combination of the BAO and SDSS data sets we checked for consistency between $\Omega_m$ posteriors, finding the point $p_{\rm SDSS} - p_{\rm BAO} =0$ to lie within the $96.5\%$ confidence level of the $p_{\rm SDSS} - p_{\rm BAO}$ distribution.  
As for the consistency between Planck DR15 CMB and SDSS data sets we considered the parameters sub-space ($\Omega_m$,$\sigma_8$), for which we found the point $\bm{p}_{\rm SDSS} - \bm{p}_{\rm CMB} =0$ to fall within the $85\%$ confidence level distribution of $\bm{p}_{\rm SDSS} - \bm{p}_{\rm CMB}$\footnote{At the time of performing this analysis the latest \planck\ DR18 likelihood is not publicly available. However, the latest \planck\ results \citep{Planck2018} are consistent with the previous data release \citep{PlanckXIII2016} (see Figure \ref{fig:SDSS_ext_data} for a comparison). According to our consistency criterion, SDSS is consistent also with Planck DR18; specifically we find the $\bm{p}_{\rm SDSS} - \bm{p}_{\rm Planck18} =0$ point to lie within the $82\%$ confidence distribution of $\bm{p}_{\rm SDSS} - \bm{p}_{\rm Planck18}$}.  

Figure~\ref{fig:SDSS_ext_data} shows the 2D marginalized contours for each of the above experiments in the $S_8$--$\Omega_m$--$h$ parameter sub-space (\textit{left} panels), as well as the posterior for a joint clusters+BAO, and clusters+BAO+Planck analysis (\textit{right} panels). The corresponding 1D marginalized posterior are listed in Table \ref{tab:ext_data}.

The combination of galaxy clusters data and BAO measurements results in a precise measurement of the Hubble parameter, $h=0.66\pm 0.02$.
This value is in excellent agreement and competitive with those derived from \Planck\ DR15 CMB data alone $h=0.66^{+0.02}_{-0.03}$.  By contrast, the posterior of $h$ is in $2.7 \sigma$ tension with the one derived by the SH0ES collaboration using type-Ia supernovae data, $h=0.732 \pm 0.017$ \citep[][]{riessetal16}, and in
$2.1\sigma$ tension with the recent strong-lensing based measurement $h=0.725^{+0.021}_{-0.023}$ presented in \citet{birreretal18}.

The further inclusion of \Planck\ DR15 data significantly improve the constraints on all the cosmological parameters considered. Specifically, the errors on $\Omega_m$, $\sigma_8$ and $h$ are reduced compared to the SDSS+BAO analysis by a factor of $4$, $3$ and $2$, respectively.  Nevertheless, the low-redshift Universe contributes a significant amount of new information: the errors on $\Omega_m$, $\sigma_8$ and $h$ for the joint analysis are reduced relative to the {\it Planck} DR15-only constraints by a factor of $3.6$, $2$, and $3$, respectively.

It is also interesting to investigate the impact that the \planck\ cosmological information has on the parameters governing the richness--mass relation of the \redmapper\ clusters.  The error on $\alpha$ is reduced from $\sigma_\alpha=0.06$ to $\sigma_\alpha=0.03$, while the error on $\log M_1$ goes from $\sigma_{\log M_1}=0.09$ to $\sigma_{\log M_1}=0.02$. The factor of four improvement in the $\log M_1$ posterior after adding \planck\ data suggests that the error budget for mass calibration in cluster abundance studies needs to be reduced from the present $\approx 8\%$ to $\approx 2\%$ for \planck\ to add no information to the cluster abundance constraint on $S_8$.
This value can be compared to the $5\%$ mass calibration achieved by the DES collaboration in \citet{McClintock2018}.

Remarkably, neither the BAO nor the \planck\ data sets improve the posterior on the intrinsic scatter. 
This may seem surprising given our earlier discussion on the degeneracy between $S_8$ and $\sigma_{intr}$: if tightening the scatter prior improves the $S_8$ posterior, why does tightening $S_8$ not improve the scatter posterior? The resolution is evident from Figure~\ref{fig:err_b}: the \planck\ data tightens $S_8$ around the value $S_8=0.83$.  This $S_8$ value cuts across the mild $S_8$--$\sigma_{intr}$ degeneracy in such a way that the full range of $\sigma_{intr}$ values is sampled. Had the \planck\ data favored either a higher or lower $S_8$, the posterior on $\sigma_{intr}$ would have been reduced. 

Finally, we find the addition of cluster data has only a modest impact on the posterior on $\sum m_\nu$ from the combination of \planck\ and BAO data. To explore whether future cluster abundance analyses are likely to result in significant improvements we ran chains adopting unrealistically tight 1\% priors on the amplitude and slope of the richness--mass relation, as well as a $\sigma_{intr}\leq 0.3$ prior on the scatter.  Even in this over-optimistic scenario, clusters had only a minor impact on the posterior for $\sum m_\nu$.  This is not entirely unexpected given the small redshift range probed by our cluster catalogue and the fact that the abundance function is only directly sensitive to: 1) $\Omega_{\rm cdm}+\Omega_{\rm b}$, and 2) the amplitude of the dark matter and baryons power spectrum \citep[e.g.][]{Costanzi2013}.  While some sensitivity to $\sum m_\nu$ at a given redshift remains via the volume term in the abundance prediction, the sensitivity to neutrino mass at fixed $\Omega_{\rm cdm}+\Omega_{\rm b}$ and primordial power spectrum amplitude, $A_s$, is relatively mild.


\section{Summary and Conclusion}
\label{sec:summary}

We have performed a joint analysis of the abundance and weak lensing mass measurements of the \redmapper\ clusters identified in the SDSS DR 8 \citep{Aihara2011} to simultaneously constrain cosmology and the richness--mass relation parameters. The cosmological sample consists of $6964$ clusters having richness $\lambda\ob \geq 20$ in the redshift range $0.1<z<0.3$. For the weak lensing mass estimates we employed the results of the stacked weak lensing analysis performed by \citet{Simet2016}, which achieved a 7.7\% precision including both statistical and systematic uncertainties (see Table~\ref{tab:err_budget}).  Our analysis is the first cluster abundance study to be performed whiled blinded to the recovered cosmological parameters. All the modeling choices and validation tests were made before unblinding the cosmological results.  We also verified that our cosmological posteriors are robust to assumptions made about the form and parametrization of the richness--mass relation and systematics associated with the calibration of projection effects (see Figure \ref{fig:mod_dep}).

Assuming a flat  $\Lambda$CDM model with massive neutrinos, and including modest $H_0$ and BBN priors \citep{cookeetal16}, we found $S_{8}=0.79^{+0.05}_{-0.04}$. Our result is in agreement with those obtained by other cluster abundance studies, as well as with constraints derived from the DES Y1 $3\times 2$ analysis \citep{des17} and \planck\ DR18 CMB data \citep{Planck2018}.  The error budget on $S_8$ is not dominated by a single set of observables; while mass calibration uncertainties are typically the dominant source of error in cluster abundance studies, the uncertainty in the scatter of the richness--mass relation  degrades the constraining power of our sample. Since the detailed shape of the abundance function is sensitive to the scatter, the error budget of the abundance data impacts our $S_8$ posterior at a level comparable to that from our mass-calibration uncertainty. Future analyses that accurately measure the scatter in the richness--mass relation --- e.g. from multi-wavelength observations of \redmapper\ clusters  --- will have a significant impact on the cosmological conclusions that can be drawn from optical cluster samples.\footnote{Given an observational mass proxy $X$ that is known a priori to have substantially lower scatter at fixed mass than richness (e.g. gas mass, or SZ decrement), one can use the joint distribution $P(\lambda,X)$ of the detected clusters to infer the scatter in richness at fixed mass.}

Having ascertained the statistical consistency of our data set with \planck\ DR15 CMB data and BAO priors, we combined the SDSS cluster abundance analysis with these two external data sets. From the joint analysis of SDSS and BAO data we obtain $\Omega_m=0.32\pm 0.04$, $\sigma_8=0.78\pm 0.06$ and $h=0.66\pm 0.02$. These constraints are consistent with and of comparable size to the CMB constraints from \planck\ data.  The further inclusion of the \planck\ DR15 data improves the precision of the parameters $S_8$, $h$ and $\alpha$ by a factor of $2$ and $\log M_1$ by a factor of $4.5$. Adding clusters data to \planck+BAO has a negligible impact on the posterior of the sum of the neutrino masses, $\sum m_\nu$. This conclusion holds for local cluster surveys even if cluster mass calibration uncertainties decrease to the per-cent level.  Table~\ref{tab:ext_data} summarizes the posteriors of the SDSS clusters after combining with the BAO and \planck\ DR15 external data sets.

We have also compared our posteriors on the richness--mass relation to other analyses.  In particular, we find that our posterior for the richness--mass relation is in excellent agreement with the results of \citet{Murata2017}, except for the lowest richness bin.  That work claims $\approx 10\%$ of the clusters in our lowest richness bin have masses $M\leq 10^{13}\ \hmsun$, whereas we find all our clusters have mass $M\gtrsim 2\times 10^{13}\ \hmsun$.  We argue that this difference is driven by a theoretical systematic associated with the model adopted in \citet{Murata2017}, rather than systematic uncertainties inherent to the data.  The stellar mass of the central galaxies in our clusters is consistent with this interpretation.  The otherwise excellent agreement between our work and that of \citet{Murata2017} is notable given the significant methodological differences between the two works.

In short, our results are best summarized by saying that the SDSS cluster abundance data is consistent with the best fit flat $\Lambda$CDM cosmology from \planck.  Our results are also consistent with but have somewhat larger errors than current state-of-the-art analysis combining the auto- and cross-correlations of galaxies and shear \citep[e.g.][; see Figure~\ref{fig:ext_data}]{des17,kids2df2018,kidsgama2018}.  Future analyses that reduce the mass-calibration uncertainties, combined with measurements of the scatter of the richness--mass relations, will make cluster abundances studies competitive with these combined-clustering results. The first demonstration of this coming power will be the upcoming analysis of the DES Y1 data set (DES collaboration, in preparation).


\section*{Acknowledgements}
\label{sec:acknowledgements}

We thank R. Mandelbaum for help and comments with early aspects of this work.  This paper has gone through internal review by the DES collaboration. We acknowledge support from the U.S. Department of Energy under contract number DE-AC02-76SF00515. ER was supported by the DOE grant DE-SC0015975, by the Sloan Foundation, grant FG-2016-6443, and the Cottrell Scholar program of the Research Corporation for Science Advancement.  Support for DG was provided by NASA through Einstein Postdoctoral Fellowship grant number PF5-160138 awarded by the Chandra X-ray Center, which is operated by the Smithsonian Astrophysical Observatory for NASA under contract NAS8-03060.  We thank the {\sc Aemulus} collaboration for access to their simulation data products in advance of publication.  This research used simulations that were performed resources of the National Energy Research Scientific Computing Center (NERSC), a U.S. Department of Energy Office of Science User Facility operated under Contract No. DE-AC02-05CH11231.

Funding for the DES Projects has been provided by the U.S. Department of Energy, the U.S. National Science Foundation, the Ministry of Science and Education of Spain, the Science and Technology Facilities Council of the United Kingdom, the Higher Education Funding Council for England, the National Center for Supercomputing Applications at the University of Illinois at Urbana-Champaign, the Kavli Institute of Cosmological Physics at the University of Chicago, 
the Center for Cosmology and Astro-Particle Physics at the Ohio State University,
the Mitchell Institute for Fundamental Physics and Astronomy at Texas A\&M University, Financiadora de Estudos e Projetos, 
Funda{\c c}{\~a}o Carlos Chagas Filho de Amparo {\`a} Pesquisa do Estado do Rio de Janeiro, Conselho Nacional de Desenvolvimento Cient{\'i}fico e Tecnol{\'o}gico and 
the Minist{\'e}rio da Ci{\^e}ncia, Tecnologia e Inova{\c c}{\~a}o, the Deutsche Forschungsgemeinschaft and the Collaborating Institutions in the Dark Energy Survey. 

The Collaborating Institutions are Argonne National Laboratory, the University of California at Santa Cruz, the University of Cambridge, Centro de Investigaciones Energ{\'e}ticas, 
Medioambientales y Tecnol{\'o}gicas-Madrid, the University of Chicago, University College London, the DES-Brazil Consortium, the University of Edinburgh, 
the Eidgen{\"o}ssische Technische Hochschule (ETH) Z{\"u}rich, 
Fermi National Accelerator Laboratory, the University of Illinois at Urbana-Champaign, the Institut de Ci{\`e}ncies de l'Espai (IEEC/CSIC), 
the Institut de F{\'i}sica d'Altes Energies, Lawrence Berkeley National Laboratory, the Ludwig-Maximilians Universit{\"a}t M{\"u}nchen and the associated Excellence Cluster Universe, 
the University of Michigan, the National Optical Astronomy Observatory, the University of Nottingham, The Ohio State University, the University of Pennsylvania, the University of Portsmouth, 
SLAC National Accelerator Laboratory, Stanford University, the University of Sussex, Texas A\&M University, and the OzDES Membership Consortium.

Based in part on observations at Cerro Tololo Inter-American Observatory, National Optical Astronomy Observatory, which is operated by the Association of 
Universities for Research in Astronomy (AURA) under a cooperative agreement with the National Science Foundation.

The DES data management system is supported by the National Science Foundation under Grant Numbers AST-1138766 and AST-1536171.
The DES participants from Spanish institutions are partially supported by MINECO under grants AYA2015-71825, ESP2015-66861, FPA2015-68048, SEV-2016-0588, SEV-2016-0597, and MDM-2015-0509, 
some of which include ERDF funds from the European Union. IFAE is partially funded by the CERCA program of the Generalitat de Catalunya.
Research leading to these results has received funding from the European Research
Council under the European Union's Seventh Framework Program (FP7/2007-2013) including ERC grant agreements 240672, 291329, and 306478.
We  acknowledge support from the Australian Research Council Centre of Excellence for All-sky Astrophysics (CAASTRO), through project number CE110001020.

This manuscript has been authored by Fermi Research Alliance, LLC under Contract No. DE-AC02-07CH11359 with the U.S. Department of Energy, Office of Science, Office of High Energy Physics. The United States Government retains and the publisher, by accepting the article for publication, acknowledges that the United States Government retains a non-exclusive, paid-up, irrevocable, world-wide license to publish or reproduce the published form of this manuscript, or allow others to do so, for United States Government purposes.



\bibliographystyle{mnras}
\bibliography{database.bib} 

\begin{thebibliography}{}
\makeatletter
\relax
\def\mn@urlcharsother{\let\do\@makeother \do\$\do\&\do\#\do\^\do\_\do\%\do\~}
\def\mn@doi{\begingroup\mn@urlcharsother \@ifnextchar [ {\mn@doi@}
  {\mn@doi@[]}}
\def\mn@doi@[#1]#2{\def\@tempa{#1}\ifx\@tempa\@empty \href
  {http://dx.doi.org/#2} {doi:#2}\else \href {http://dx.doi.org/#2} {#1}\fi
  \endgroup}
\def\mn@eprint#1#2{\mn@eprint@#1:#2::\@nil}
\def\mn@eprint@arXiv#1{\href {http://arxiv.org/abs/#1} {{\tt arXiv:#1}}}
\def\mn@eprint@dblp#1{\href {http://dblp.uni-trier.de/rec/bibtex/#1.xml}
  {dblp:#1}}
\def\mn@eprint@#1:#2:#3:#4\@nil{\def\@tempa {#1}\def\@tempb {#2}\def\@tempc
  {#3}\ifx \@tempc \@empty \let \@tempc \@tempb \let \@tempb \@tempa \fi \ifx
  \@tempb \@empty \def\@tempb {arXiv}\fi \@ifundefined
  {mn@eprint@\@tempb}{\@tempb:\@tempc}{\expandafter \expandafter \csname
  mn@eprint@\@tempb\endcsname \expandafter{\@tempc}}}

\bibitem[\protect\citeauthoryear{{Aihara} et~al.,}{{Aihara}
  et~al.}{2011}]{Aihara2011}
{Aihara} H.,  et~al., 2011, \mn@doi [\apjs] {10.1088/0067-0049/193/2/29}, \href
  {http://adsabs.harvard.edu/abs/2011ApJS..193...29A} {193, 29}

\bibitem[\protect\citeauthoryear{{Alam} et~al.}{{Alam} et~al.}{2017}]{BAOBoss}
{Alam} S.,  et~al., 2017, \mn@doi [\mnras] {10.1093/mnras/stx721}, \href
  {http://adsabs.harvard.edu/abs/2017MNRAS.470.2617A} {470, 2617}

\bibitem[\protect\citeauthoryear{{Allen}, {Evrard}  \& {Mantz}}{{Allen}
  et~al.}{2011}]{Allen2011}
{Allen} S.~W.,  {Evrard} A.~E.,   {Mantz} A.~B.,  2011, \mn@doi [\araa]
  {10.1146/annurev-astro-081710-102514}, \href
  {http://adsabs.harvard.edu/abs/2011ARA\%26A..49..409A} {49, 409}

\bibitem[\protect\citeauthoryear{{Anderson} et~al.,}{{Anderson}
  et~al.}{2014}]{anderson14}
{Anderson} L.,  et~al., 2014, \mn@doi [\mnras] {10.1093/mnras/stu523}, \href
  {http://adsabs.harvard.edu/abs/2014MNRAS.441...24A} {441, 24}

\bibitem[\protect\citeauthoryear{{Behroozi}, {Wechsler}  \& {Wu}}{{Behroozi}
  et~al.}{2013}]{Behroozi2013}
{Behroozi} P.~S.,  {Wechsler} R.~H.,   {Wu} H.-Y.,  2013, \mn@doi [\apj]
  {10.1088/0004-637X/762/2/109}, \href
  {http://adsabs.harvard.edu/abs/2013ApJ...762..109B} {762, 109}

\bibitem[\protect\citeauthoryear{{Behroozi}, {Wechsler}, {Hearin}  \&
  {Conroy}}{{Behroozi} et~al.}{2018}]{behroozietal18}
{Behroozi} P.,  {Wechsler} R.,  {Hearin} A.,   {Conroy} C.,  2018, preprint,
  \href {http://adsabs.harvard.edu/abs/2018arXiv180607893B} {} (\mn@eprint
  {arXiv} {1806.07893})

\bibitem[\protect\citeauthoryear{{Berlind} \& {Weinberg}}{{Berlind} \&
  {Weinberg}}{2002}]{berlindweinberg02}
{Berlind} A.~A.,  {Weinberg} D.~H.,  2002, \mn@doi [\apj] {10.1086/341469},
  \href {http://adsabs.harvard.edu/abs/2002ApJ...575..587B} {575, 587}

\bibitem[\protect\citeauthoryear{{Beutler} et~al.}{{Beutler}
  et~al.}{2011}]{BAO6dF}
{Beutler} F.,  et~al., 2011, \mn@doi [\mnras]
  {10.1111/j.1365-2966.2011.19250.x}, \href
  {http://adsabs.harvard.edu/abs/2011MNRAS.416.3017B} {416, 3017}

\bibitem[\protect\citeauthoryear{{Birrer} et~al.,}{{Birrer}
  et~al.}{2018}]{birreretal18}
{Birrer} S.,  et~al., 2018, preprint, \href
  {http://adsabs.harvard.edu/abs/2018arXiv180901274B} {} (\mn@eprint {arXiv}
  {1809.01274})

\bibitem[\protect\citeauthoryear{{Bocquet}, {Saro}, {Dolag}  \&
  {Mohr}}{{Bocquet} et~al.}{2016}]{Bocquet2016}
{Bocquet} S.,  {Saro} A.,  {Dolag} K.,   {Mohr} J.~J.,  2016, \mn@doi [\mnras]
  {10.1093/mnras/stv2657}, \href
  {http://adsabs.harvard.edu/abs/2016MNRAS.456.2361B} {456, 2361}

\bibitem[\protect\citeauthoryear{{Borgani} et~al.,}{{Borgani}
  et~al.}{2001}]{Borgani2001}
{Borgani} S.,  et~al., 2001, \mn@doi [\apj] {10.1086/323214}, \href
  {http://adsabs.harvard.edu/abs/2001ApJ...561...13B} {561, 13}

\bibitem[\protect\citeauthoryear{{Boylan-Kolchin}, {Springel}, {White}  \&
  {Jenkins}}{{Boylan-Kolchin} et~al.}{2010}]{boylankolchinetal10}
{Boylan-Kolchin} M.,  {Springel} V.,  {White} S.~D.~M.,   {Jenkins} A.,  2010,
  \mn@doi [\mnras] {10.1111/j.1365-2966.2010.16774.x}, \href
  {http://adsabs.harvard.edu/abs/2010MNRAS.406..896B} {406, 896}

\bibitem[\protect\citeauthoryear{{Brandbyge}, {Hannestad}, {Haugb{\o}lle}  \&
  {Wong}}{{Brandbyge} et~al.}{2010}]{Brand2010}
{Brandbyge} J.,  {Hannestad} S.,  {Haugb{\o}lle} T.,   {Wong} Y.~Y.~Y.,  2010,
  \mn@doi [\jcap] {10.1088/1475-7516/2010/09/014}, \href
  {http://adsabs.harvard.edu/abs/2010JCAP...09..014B} {9, 014}

\bibitem[\protect\citeauthoryear{{Bullock}, {Wechsler}  \&
  {Somerville}}{{Bullock} et~al.}{2002}]{bullock02}
{Bullock} J.~S.,  {Wechsler} R.~H.,   {Somerville} R.~S.,  2002, \mn@doi
  [\mnras] {10.1046/j.1365-8711.2002.04959.x}, \href
  {http://adsabs.harvard.edu/abs/2002MNRAS.329..246B} {329, 246}

\bibitem[\protect\citeauthoryear{{Burenin} \& {Vikhlinin}}{{Burenin} \&
  {Vikhlinin}}{2012}]{Burenin2012}
{Burenin} R.~A.,  {Vikhlinin} A.~A.,  2012, \mn@doi [Astronomy Letters]
  {10.1134/S1063773712060011}, \href
  {http://adsabs.harvard.edu/abs/2012AstL...38..347B} {38, 347}

\bibitem[\protect\citeauthoryear{{Busch} \& {White}}{{Busch} \&
  {White}}{2017}]{buschwhite17}
{Busch} P.,  {White} S.~D.~M.,  2017, \mn@doi [\mnras] {10.1093/mnras/stx1584},
  \href {http://adsabs.harvard.edu/abs/2017MNRAS.470.4767B} {470, 4767}

\bibitem[\protect\citeauthoryear{{Castorina}, {Sefusatti}, {Sheth},
  {Villaescusa-Navarro}  \& {Viel}}{{Castorina} et~al.}{2014}]{Castorina2014}
{Castorina} E.,  {Sefusatti} E.,  {Sheth} R.~K.,  {Villaescusa-Navarro} F.,
  {Viel} M.,  2014, \mn@doi [\jcap] {10.1088/1475-7516/2014/02/049}, \href
  {http://adsabs.harvard.edu/abs/2014JCAP...02..049C} {2, 049}

\bibitem[\protect\citeauthoryear{{Cataneo} et~al.,}{{Cataneo}
  et~al.}{2015}]{Cataneo2015}
{Cataneo} M.,  et~al., 2015, \mn@doi [\prd] {10.1103/PhysRevD.92.044009}, \href
  {http://adsabs.harvard.edu/abs/2015PhRvD..92d4009C} {92, 044009}

\bibitem[\protect\citeauthoryear{{Charnock}, {Battye}  \& {Moss}}{{Charnock}
  et~al.}{2017}]{charnocketal17}
{Charnock} T.,  {Battye} R.~A.,   {Moss} A.,  2017, \mn@doi [\prd]
  {10.1103/PhysRevD.95.123535}, \href
  {http://adsabs.harvard.edu/abs/2017PhRvD..95l3535C} {95, 123535}

\bibitem[\protect\citeauthoryear{{Cooke}, {Pettini}, {Nollett}  \&
  {Jorgenson}}{{Cooke} et~al.}{2016}]{cookeetal16}
{Cooke} R.~J.,  {Pettini} M.,  {Nollett} K.~M.,   {Jorgenson} R.,  2016,
  \mn@doi [\apj] {10.3847/0004-637X/830/2/148}, \href
  {http://adsabs.harvard.edu/abs/2016ApJ...830..148C} {830, 148}

\bibitem[\protect\citeauthoryear{{Costanzi}, {Villaescusa-Navarro}, {Viel},
  {Xia}, {Borgani}, {Castorina}  \& {Sefusatti}}{{Costanzi}
  et~al.}{2013}]{Costanzi2013}
{Costanzi} M.,  {Villaescusa-Navarro} F.,  {Viel} M.,  {Xia} J.-Q.,  {Borgani}
  S.,  {Castorina} E.,   {Sefusatti} E.,  2013, \mn@doi [\jcap]
  {10.1088/1475-7516/2013/12/012}, \href
  {http://adsabs.harvard.edu/abs/2013JCAP...12..012C} {12, 012}

\bibitem[\protect\citeauthoryear{{Costanzi} et~al.,}{{Costanzi}
  et~al.}{2018}]{costanzi18a}
{Costanzi} M.,  et~al., 2018, preprint, \href
  {http://adsabs.harvard.edu/abs/2018arXiv180707072C} {} (\mn@eprint {arXiv}
  {1807.07072})

\bibitem[\protect\citeauthoryear{{Crocce}, {Fosalba}, {Castander}  \&
  {Gazta{\~n}aga}}{{Crocce} et~al.}{2010}]{Crocce2010}
{Crocce} M.,  {Fosalba} P.,  {Castander} F.~J.,   {Gazta{\~n}aga} E.,  2010,
  \mn@doi [\mnras] {10.1111/j.1365-2966.2009.16194.x}, \href
  {http://adsabs.harvard.edu/abs/2010MNRAS.403.1353C} {403, 1353}

\bibitem[\protect\citeauthoryear{{Cui}, {Borgani}  \& {Murante}}{{Cui}
  et~al.}{2014}]{Cui2014}
{Cui} W.,  {Borgani} S.,   {Murante} G.,  2014, \mn@doi [\mnras]
  {10.1093/mnras/stu673}, \href
  {http://adsabs.harvard.edu/abs/2014MNRAS.441.1769C} {441, 1769}

\bibitem[\protect\citeauthoryear{{DES Collaboration} et~al.,}{{DES
  Collaboration} et~al.}{2017b}]{DESH02017}
{DES Collaboration} et~al., 2017b, preprint, \href
  {http://adsabs.harvard.edu/abs/2017arXiv171100403D} {} (\mn@eprint {arXiv}
  {1711.00403})

\bibitem[\protect\citeauthoryear{{DES Collaboration} et~al.,}{{DES
  Collaboration} et~al.}{2017a}]{des17}
{DES Collaboration} et~al., 2017a, preprint, \href
  {http://adsabs.harvard.edu/abs/2017arXiv170801530D} {} (\mn@eprint {arXiv}
  {1708.01530})

\bibitem[\protect\citeauthoryear{{Dawson} et~al.,}{{Dawson}
  et~al.}{2013}]{Dawson2013}
{Dawson} K.~S.,  et~al., 2013, \mn@doi [\aj] {10.1088/0004-6256/145/1/10},
  \href {http://adsabs.harvard.edu/abs/2013AJ....145...10D} {145, 10}

\bibitem[\protect\citeauthoryear{{DeRose} et~al.,}{{DeRose}
  et~al.}{2018}]{derose18}
{DeRose} J.,  et~al., 2018, preprint, \href
  {http://adsabs.harvard.edu/abs/2018arXiv180405865D} {} (\mn@eprint {arXiv}
  {1804.05865})

\bibitem[\protect\citeauthoryear{{Despali}, {Giocoli}, {Angulo}, {Tormen},
  {Sheth}, {Baso}  \& {Moscardini}}{{Despali} et~al.}{2016}]{Despali2016}
{Despali} G.,  {Giocoli} C.,  {Angulo} R.~E.,  {Tormen} G.,  {Sheth} R.~K.,
  {Baso} G.,   {Moscardini} L.,  2016, \mn@doi [\mnras]
  {10.1093/mnras/stv2842}, \href
  {http://adsabs.harvard.edu/abs/2016MNRAS.456.2486D} {456, 2486}

\bibitem[\protect\citeauthoryear{{Diemer} \& {Kravtsov}}{{Diemer} \&
  {Kravtsov}}{2014}]{diemerkravtsov14}
{Diemer} B.,  {Kravtsov} A.~V.,  2014, \mn@doi [\apj]
  {10.1088/0004-637X/789/1/1}, \href
  {http://adsabs.harvard.edu/abs/2014ApJ...789....1D} {789, 1}

\bibitem[\protect\citeauthoryear{{Diemer} \& {Kravtsov}}{{Diemer} \&
  {Kravtsov}}{2015}]{Diemer2015}
{Diemer} B.,  {Kravtsov} A.~V.,  2015, \mn@doi [\apj]
  {10.1088/0004-637X/799/1/108}, \href
  {http://adsabs.harvard.edu/abs/2015ApJ...799..108D} {799, 108}

\bibitem[\protect\citeauthoryear{{Dietrich} et~al.,}{{Dietrich}
  et~al.}{2014}]{dietrichetal14}
{Dietrich} J.~P.,  et~al., 2014, \mn@doi [\mnras] {10.1093/mnras/stu1282},
  \href {http://adsabs.harvard.edu/abs/2014MNRAS.443.1713D} {443, 1713}

\bibitem[\protect\citeauthoryear{{Eke}, {Cole}, {Frenk}  \& {Patrick
  Henry}}{{Eke} et~al.}{1998}]{Eke1998}
{Eke} V.~R.,  {Cole} S.,  {Frenk} C.~S.,   {Patrick Henry} J.,  1998, \mn@doi
  [\mnras] {10.1046/j.1365-8711.1998.01713.x}, \href
  {http://adsabs.harvard.edu/abs/1998MNRAS.298.1145E} {298, 1145}

\bibitem[\protect\citeauthoryear{{Evrard}, {Arnault}, {Huterer}  \&
  {Farahi}}{{Evrard} et~al.}{2014}]{evrardetal14}
{Evrard} A.~E.,  {Arnault} P.,  {Huterer} D.,   {Farahi} A.,  2014, \mn@doi
  [\mnras] {10.1093/mnras/stu784}, \href
  {http://adsabs.harvard.edu/abs/2014MNRAS.441.3562E} {441, 3562}

\bibitem[\protect\citeauthoryear{{Farahi}, {Evrard}, {Rozo}, {Rykoff}  \&
  {Wechsler}}{{Farahi} et~al.}{2016}]{farahietal16}
{Farahi} A.,  {Evrard} A.~E.,  {Rozo} E.,  {Rykoff} E.~S.,   {Wechsler} R.~H.,
  2016, \mn@doi [\mnras] {10.1093/mnras/stw1143}, \href
  {http://adsabs.harvard.edu/abs/2016MNRAS.460.3900F} {460, 3900}

\bibitem[\protect\citeauthoryear{{Feldmann} et~al.,}{{Feldmann}
  et~al.}{2006}]{feldmannetal2006}
{Feldmann} R.,  et~al., 2006, \mn@doi [\mnras]
  {10.1111/j.1365-2966.2006.10930.x}, \href
  {http://adsabs.harvard.edu/abs/2006MNRAS.372..565F} {372, 565}

\bibitem[\protect\citeauthoryear{{Gelman} \& {Rubin}}{{Gelman} \&
  {Rubin}}{1992}]{Gelman1992}
{Gelman} A.,  {Rubin} D.~B.,  1992, Statistical Science, 7, 457

\bibitem[\protect\citeauthoryear{{Hasselfield} et~al.,}{{Hasselfield}
  et~al.}{2013}]{ACT2013}
{Hasselfield} M.,  et~al., 2013, \mn@doi [\jcap]
  {10.1088/1475-7516/2013/07/008}, \href
  {http://adsabs.harvard.edu/abs/2013JCAP...07..008H} {7, 008}

\bibitem[\protect\citeauthoryear{{Henry}}{{Henry}}{2000}]{Henry2000}
{Henry} J.~P.,  2000, \mn@doi [\apj] {10.1086/308783}, \href
  {http://adsabs.harvard.edu/abs/2000ApJ...534..565H} {534, 565}

\bibitem[\protect\citeauthoryear{{Henry}}{{Henry}}{2004}]{Henry2004}
{Henry} J.~P.,  2004, \mn@doi [\apj] {10.1086/421336}, \href
  {http://adsabs.harvard.edu/abs/2004ApJ...609..603H} {609, 603}

\bibitem[\protect\citeauthoryear{{Henson}, {Barnes}, {Kay}, {McCarthy}  \&
  {Schaye}}{{Henson} et~al.}{2017}]{hensonetal17}
{Henson} M.~A.,  {Barnes} D.~J.,  {Kay} S.~T.,  {McCarthy} I.~G.,   {Schaye}
  J.,  2017, \mn@doi [\mnras] {10.1093/mnras/stw2899}, \href
  {http://adsabs.harvard.edu/abs/2017MNRAS.465.3361H} {465, 3361}

\bibitem[\protect\citeauthoryear{{Hikage}, {Mandelbaum}, {Leauthaud}, {Rozo}
  \& {Rykoff}}{{Hikage} et~al.}{2017}]{hikageetal17}
{Hikage} C.,  {Mandelbaum} R.,  {Leauthaud} A.,  {Rozo} E.,   {Rykoff} E.~S.,
  2017, preprint, \href {http://adsabs.harvard.edu/abs/2017arXiv170208614H} {}
  (\mn@eprint {arXiv} {1702.08614})

\bibitem[\protect\citeauthoryear{{Hinshaw} et~al.,}{{Hinshaw}
  et~al.}{2013}]{wmap9}
{Hinshaw} G.,  et~al., 2013, \mn@doi [\apjs] {10.1088/0067-0049/208/2/19},
  \href {http://adsabs.harvard.edu/abs/2013ApJS..208...19H} {208, 19}

\bibitem[\protect\citeauthoryear{{Hirata} \& {Seljak}}{{Hirata} \&
  {Seljak}}{2003}]{hirataseljak2003}
{Hirata} C.,  {Seljak} U.,  2003, \mn@doi [\mnras]
  {10.1046/j.1365-8711.2003.06683.x}, \href
  {http://adsabs.harvard.edu/abs/2003MNRAS.343..459H} {343, 459}

\bibitem[\protect\citeauthoryear{{Hoekstra}, {Herbonnet}, {Muzzin}, {Babul},
  {Mahdavi}, {Viola}  \& {Cacciato}}{{Hoekstra} et~al.}{2015}]{hoekstraetal15}
{Hoekstra} H.,  {Herbonnet} R.,  {Muzzin} A.,  {Babul} A.,  {Mahdavi} A.,
  {Viola} M.,   {Cacciato} M.,  2015, \mn@doi [\mnras] {10.1093/mnras/stv275},
  \href {http://adsabs.harvard.edu/abs/2015MNRAS.449..685H} {449, 685}

\bibitem[\protect\citeauthoryear{{Hoffmann}, {Bel}  \&
  {Gazta{\~n}aga}}{{Hoffmann} et~al.}{2015}]{Hoffmann2015}
{Hoffmann} K.,  {Bel} J.,   {Gazta{\~n}aga} E.,  2015, \mn@doi [\mnras]
  {10.1093/mnras/stv702}, \href
  {http://adsabs.harvard.edu/abs/2015MNRAS.450.1674H} {450, 1674}

\bibitem[\protect\citeauthoryear{{Hu} \& {Cohn}}{{Hu} \& {Cohn}}{2006}]{Hu2006}
{Hu} W.,  {Cohn} J.~D.,  2006, \mn@doi [\prd] {10.1103/PhysRevD.73.067301},
  \href {http://adsabs.harvard.edu/abs/2006PhRvD..73f7301H} {73, 067301}

\bibitem[\protect\citeauthoryear{{Hu} \& {Kravtsov}}{{Hu} \&
  {Kravtsov}}{2003}]{hukravtsov03}
{Hu} W.,  {Kravtsov} A.~V.,  2003, \mn@doi [\apj] {10.1086/345846}, \href
  {http://adsabs.harvard.edu/abs/2003ApJ...584..702H} {584, 702}

\bibitem[\protect\citeauthoryear{{Jiang} \& {van den Bosch}}{{Jiang} \& {van
  den Bosch}}{2016}]{jiangvandenbosh04}
{Jiang} F.,  {van den Bosch} F.~C.,  2016, \mn@doi [\mnras]
  {10.1093/mnras/stw439}, \href
  {http://adsabs.harvard.edu/abs/2016MNRAS.458.2848J} {458, 2848}

\bibitem[\protect\citeauthoryear{{Joudaki} et~al.,}{{Joudaki}
  et~al.}{2018}]{kids2df2018}
{Joudaki} S.,  et~al., 2018, \mn@doi [\mnras] {10.1093/mnras/stx2820}, \href
  {http://adsabs.harvard.edu/abs/2018MNRAS.474.4894J} {474, 4894}

\bibitem[\protect\citeauthoryear{{Knebe} et~al.,}{{Knebe}
  et~al.}{2013}]{Knebe2013}
{Knebe} A.,  et~al., 2013, \mn@doi [\mnras] {10.1093/mnras/stt1403}, \href
  {http://adsabs.harvard.edu/abs/2013MNRAS.435.1618K} {435, 1618}

\bibitem[\protect\citeauthoryear{{Koester} et~al.,}{{Koester}
  et~al.}{2007}]{maxbcg}
{Koester} B.~P.,  et~al., 2007, \mn@doi [\apj] {10.1086/509599}, \href
  {http://adsabs.harvard.edu/abs/2007ApJ...660..239K} {660, 239}

\bibitem[\protect\citeauthoryear{{Kravtsov} \& {Borgani}}{{Kravtsov} \&
  {Borgani}}{2012}]{Kravtsov2012}
{Kravtsov} A.~V.,  {Borgani} S.,  2012, \mn@doi [\araa]
  {10.1146/annurev-astro-081811-125502}, \href
  {http://adsabs.harvard.edu/abs/2012ARA%26A..50..353K} {50, 353}

\bibitem[\protect\citeauthoryear{{Kravtsov}, {Berlind}, {Wechsler}, {Klypin},
  {Gottl{\"o}ber}, {Allgood}  \& {Primack}}{{Kravtsov}
  et~al.}{2004}]{Kravtsov2004}
{Kravtsov} A.~V.,  {Berlind} A.~A.,  {Wechsler} R.~H.,  {Klypin} A.~A.,
  {Gottl{\"o}ber} S.,  {Allgood} B.,   {Primack} J.~R.,  2004, \mn@doi [\apj]
  {10.1086/420959}, \href {http://adsabs.harvard.edu/abs/2004ApJ...609...35K}
  {609, 35}

\bibitem[\protect\citeauthoryear{{Liu}, {Bird}, {Zorrilla Matilla}, {Hill},
  {Haiman}, {Madhavacheril}, {Petri}  \& {Spergel}}{{Liu}
  et~al.}{2017}]{Liu2017}
{Liu} J.,  {Bird} S.,  {Zorrilla Matilla} J.~M.,  {Hill} J.~C.,  {Haiman} Z.,
  {Madhavacheril} M.~S.,  {Petri} A.,   {Spergel} D.~N.,  2017, preprint, \href
  {http://adsabs.harvard.edu/abs/2017arXiv171110524L} {} (\mn@eprint {arXiv}
  {1711.10524})

\bibitem[\protect\citeauthoryear{{Mana}, {Giannantonio}, {Weller}, {Hoyle},
  {H{\"u}tsi}  \& {Sartoris}}{{Mana} et~al.}{2013}]{Mana2013}
{Mana} A.,  {Giannantonio} T.,  {Weller} J.,  {Hoyle} B.,  {H{\"u}tsi} G.,
  {Sartoris} B.,  2013, \mn@doi [\mnras] {10.1093/mnras/stt1062}, \href
  {http://adsabs.harvard.edu/abs/2013MNRAS.434..684M} {434, 684}

\bibitem[\protect\citeauthoryear{{Mandelbaum}, {Hirata}, {Leauthaud}, {Massey}
  \& {Rhodes}}{{Mandelbaum} et~al.}{2012}]{mandelbaumetal12}
{Mandelbaum} R.,  {Hirata} C.~M.,  {Leauthaud} A.,  {Massey} R.~J.,   {Rhodes}
  J.,  2012, \mn@doi [\mnras] {10.1111/j.1365-2966.2011.20138.x}, \href
  {http://adsabs.harvard.edu/abs/2012MNRAS.420.1518M} {420, 1518}

\bibitem[\protect\citeauthoryear{{Mandelbaum}, {Slosar}, {Baldauf}, {Seljak},
  {Hirata}, {Nakajima}, {Reyes}  \& {Smith}}{{Mandelbaum}
  et~al.}{2013}]{mandelbaumetal13}
{Mandelbaum} R.,  {Slosar} A.,  {Baldauf} T.,  {Seljak} U.,  {Hirata} C.~M.,
  {Nakajima} R.,  {Reyes} R.,   {Smith} R.~E.,  2013, \mn@doi [\mnras]
  {10.1093/mnras/stt572}, \href
  {http://adsabs.harvard.edu/abs/2013MNRAS.432.1544M} {432, 1544}

\bibitem[\protect\citeauthoryear{{Mandelbaum} et~al.,}{{Mandelbaum}
  et~al.}{2017}]{mandelbaumetal2017}
{Mandelbaum} R.,  et~al., 2017, preprint, \href
  {http://adsabs.harvard.edu/abs/2017arXiv171000885M} {} (\mn@eprint {arXiv}
  {1710.00885})

\bibitem[\protect\citeauthoryear{{Mantz}, {Allen}, {Rapetti}  \&
  {Ebeling}}{{Mantz} et~al.}{2010}]{Mantz2010}
{Mantz} A.,  {Allen} S.~W.,  {Rapetti} D.,   {Ebeling} H.,  2010, \mn@doi
  [\mnras] {10.1111/j.1365-2966.2010.16992.x}, \href
  {http://adsabs.harvard.edu/abs/2010MNRAS.406.1759M} {406, 1759}

\bibitem[\protect\citeauthoryear{{Mantz} et~al.,}{{Mantz}
  et~al.}{2015}]{Mantz2015}
{Mantz} A.~B.,  et~al., 2015, \mn@doi [\mnras] {10.1093/mnras/stu2096}, \href
  {http://adsabs.harvard.edu/abs/2015MNRAS.446.2205M} {446, 2205}

\bibitem[\protect\citeauthoryear{{Mao}, {Williamson}  \& {Wechsler}}{{Mao}
  et~al.}{2015}]{maoetal15}
{Mao} Y.-Y.,  {Williamson} M.,   {Wechsler} R.~H.,  2015, \mn@doi [\apj]
  {10.1088/0004-637X/810/1/21}, \href
  {http://adsabs.harvard.edu/abs/2015ApJ...810...21M} {810, 21}

\bibitem[\protect\citeauthoryear{{McClintock} et~al.,}{{McClintock}
  et~al.}{2018b}]{desy1wl}
{McClintock} T.,  et~al., 2018b, preprint, \href
  {http://adsabs.harvard.edu/abs/2018arXiv180500039M} {} (\mn@eprint {arXiv}
  {1805.00039})

\bibitem[\protect\citeauthoryear{{McClintock} et~al.,}{{McClintock}
  et~al.}{2018a}]{McClintock2018}
{McClintock} T.,  et~al., 2018a, preprint, \href
  {http://adsabs.harvard.edu/abs/2018arXiv180405866M} {} (\mn@eprint {arXiv}
  {1804.05866})

\bibitem[\protect\citeauthoryear{{Medezinski} et~al.,}{{Medezinski}
  et~al.}{2018}]{medezinksietal18}
{Medezinski} E.,  et~al., 2018, \mn@doi [\pasj] {10.1093/pasj/psy009}, \href
  {http://adsabs.harvard.edu/abs/2018PASJ...70...30M} {70, 30}

\bibitem[\protect\citeauthoryear{{Melchior} et~al.,}{{Melchior}
  et~al.}{2017}]{Melchior2017}
{Melchior} P.,  et~al., 2017, \mn@doi [\mnras] {10.1093/mnras/stx1053}, \href
  {http://adsabs.harvard.edu/abs/2017MNRAS.469.4899M} {469, 4899}

\bibitem[\protect\citeauthoryear{{Miyatake} et~al.,}{{Miyatake}
  et~al.}{2018}]{miyatakeetal18}
{Miyatake} H.,  et~al., 2018, preprint, \href
  {http://adsabs.harvard.edu/abs/2018arXiv180405873M} {} (\mn@eprint {arXiv}
  {1804.05873})

\bibitem[\protect\citeauthoryear{{Murata}, {Nishimichi}, {Takada}, {Miyatake},
  {Shirasaki}, {More}, {Takahashi}  \& {Osato}}{{Murata}
  et~al.}{2017}]{Murata2017}
{Murata} R.,  {Nishimichi} T.,  {Takada} M.,  {Miyatake} H.,  {Shirasaki} M.,
  {More} S.,  {Takahashi} R.,   {Osato} K.,  2017, preprint, \href
  {http://adsabs.harvard.edu/abs/2017arXiv170701907M} {} (\mn@eprint {arXiv}
  {1707.01907})

\bibitem[\protect\citeauthoryear{{Nakajima}, {Mandelbaum}, {Seljak}, {Cohn},
  {Reyes}  \& {Cool}}{{Nakajima} et~al.}{2012}]{nakajimaetal2012}
{Nakajima} R.,  {Mandelbaum} R.,  {Seljak} U.,  {Cohn} J.~D.,  {Reyes} R.,
  {Cool} R.,  2012, \mn@doi [\mnras] {10.1111/j.1365-2966.2011.20249.x}, \href
  {http://adsabs.harvard.edu/abs/2012MNRAS.420.3240N} {420, 3240}

\bibitem[\protect\citeauthoryear{{Navarro}, {Frenk}  \& {White}}{{Navarro}
  et~al.}{1997}]{NFW}
{Navarro} J.~F.,  {Frenk} C.~S.,   {White} S.~D.~M.,  1997, ApJ, 490, 493.

\bibitem[\protect\citeauthoryear{{Pierpaoli}, {Scott}  \& {White}}{{Pierpaoli}
  et~al.}{2001}]{Pierpaoli2001}
{Pierpaoli} E.,  {Scott} D.,   {White} M.,  2001, \mn@doi [\mnras]
  {10.1046/j.1365-8711.2001.04306.x}, \href
  {http://adsabs.harvard.edu/abs/2001MNRAS.325...77P} {325, 77}

\bibitem[\protect\citeauthoryear{{Planck Collaboration} et~al.,}{{Planck
  Collaboration} et~al.}{2016a}]{PlanckXIII2016}
{Planck Collaboration} et~al., 2016a, \mn@doi [\aap]
  {10.1051/0004-6361/201525830}, \href
  {http://adsabs.harvard.edu/abs/2016A%26A...594A..13P} {594, A13}

\bibitem[\protect\citeauthoryear{{Planck Collaboration} et~al.,}{{Planck
  Collaboration} et~al.}{2016b}]{PlanckSZ2016}
{Planck Collaboration} et~al., 2016b, \mn@doi [\aap]
  {10.1051/0004-6361/201525833}, \href
  {http://adsabs.harvard.edu/abs/2016A%26A...594A..24P} {594, A24}

\bibitem[\protect\citeauthoryear{{Planck Collaboration} et~al.,}{{Planck
  Collaboration} et~al.}{2018}]{Planck2018}
{Planck Collaboration} et~al., 2018, preprint, \href
  {http://adsabs.harvard.edu/abs/2018arXiv180706209P} {} (\mn@eprint {arXiv}
  {1807.06209})

\bibitem[\protect\citeauthoryear{{Reddick}, {Wechsler}, {Tinker}  \&
  {Behroozi}}{{Reddick} et~al.}{2013}]{reddick13}
{Reddick} R.~M.,  {Wechsler} R.~H.,  {Tinker} J.~L.,   {Behroozi} P.~S.,  2013,
  \mn@doi [\apj] {10.1088/0004-637X/771/1/30}, \href
  {http://adsabs.harvard.edu/abs/2013ApJ...771...30R} {771, 30}

\bibitem[\protect\citeauthoryear{{Reiprich} \& {B{\"o}hringer}}{{Reiprich} \&
  {B{\"o}hringer}}{2002}]{Reiprich2002}
{Reiprich} T.~H.,  {B{\"o}hringer} H.,  2002, \mn@doi [\apj] {10.1086/338753},
  \href {http://adsabs.harvard.edu/abs/2002ApJ...567..716R} {567, 716}

\bibitem[\protect\citeauthoryear{{Reyes}, {Mandelbaum}, {Gunn}, {Nakajima},
  {Seljak}  \& {Hirata}}{{Reyes} et~al.}{2012a}]{Reyes2012}
{Reyes} R.,  {Mandelbaum} R.,  {Gunn} J.~E.,  {Nakajima} R.,  {Seljak} U.,
  {Hirata} C.~M.,  2012a, \mn@doi [\mnras] {10.1111/j.1365-2966.2012.21472.x},
  \href {http://adsabs.harvard.edu/abs/2012MNRAS.425.2610R} {425, 2610}

\bibitem[\protect\citeauthoryear{{Reyes}, {Mandelbaum}, {Gunn}, {Nakajima},
  {Seljak}  \& {Hirata}}{{Reyes} et~al.}{2012b}]{reyesetal12}
{Reyes} R.,  {Mandelbaum} R.,  {Gunn} J.~E.,  {Nakajima} R.,  {Seljak} U.,
  {Hirata} C.~M.,  2012b, \mn@doi [\mnras] {10.1111/j.1365-2966.2012.21472.x},
  \href {http://adsabs.harvard.edu/abs/2012MNRAS.425.2610R} {425, 2610}

\bibitem[\protect\citeauthoryear{{Riess} et~al.}{{Riess}
  et~al.}{2016}]{riessetal16}
{Riess} A.~G.,  et~al., 2016, \mn@doi [\apj] {10.3847/0004-637X/826/1/56},
  \href {http://adsabs.harvard.edu/abs/2016ApJ...826...56R} {826, 56}

\bibitem[\protect\citeauthoryear{{Ross}, {Samushia}, {Howlett}, {Percival},
  {Burden}  \& {Manera}}{{Ross} et~al.}{2015}]{BAOSDSSMain}
{Ross} A.~J.,  {Samushia} L.,  {Howlett} C.,  {Percival} W.~J.,  {Burden} A.,
  {Manera} M.,  2015, \mn@doi [\mnras] {10.1093/mnras/stv154}, \href
  {http://adsabs.harvard.edu/abs/2015MNRAS.449..835R} {449, 835}

\bibitem[\protect\citeauthoryear{{Rozo} et~al.,}{{Rozo}
  et~al.}{2010}]{Rozo2010}
{Rozo} E.,  et~al., 2010, \mn@doi [\apj] {10.1088/0004-637X/708/1/645}, \href
  {http://adsabs.harvard.edu/abs/2010ApJ...708..645R} {708, 645}

\bibitem[\protect\citeauthoryear{{Rozo}, {Rykoff}, {Becker}, {Reddick}  \&
  {Wechsler}}{{Rozo} et~al.}{2015}]{Rozo2015b}
{Rozo} E.,  {Rykoff} E.~S.,  {Becker} M.,  {Reddick} R.~M.,   {Wechsler} R.~H.,
   2015, \mn@doi [\mnras] {10.1093/mnras/stv1560}, \href
  {http://adsabs.harvard.edu/abs/2015MNRAS.453...38R} {453, 38}

\bibitem[\protect\citeauthoryear{{Rykoff} et~al.,}{{Rykoff}
  et~al.}{2012}]{rykoffetal12}
{Rykoff} E.~S.,  et~al., 2012, \mn@doi [\apj] {10.1088/0004-637X/746/2/178},
  \href {http://adsabs.harvard.edu/abs/2012ApJ...746..178R} {746, 178}

\bibitem[\protect\citeauthoryear{{Rykoff} et~al.,}{{Rykoff}
  et~al.}{2014}]{Rykoff2014}
{Rykoff} E.~S.,  et~al., 2014, \mn@doi [\apj] {10.1088/0004-637X/785/2/104},
  \href {http://adsabs.harvard.edu/abs/2014ApJ...785..104R} {785, 104}

\bibitem[\protect\citeauthoryear{{Schaller} et~al.,}{{Schaller}
  et~al.}{2015}]{schalleretal15a}
{Schaller} M.,  et~al., 2015, \mn@doi [\mnras] {10.1093/mnras/stv1067}, \href
  {http://adsabs.harvard.edu/abs/2015MNRAS.451.1247S} {451, 1247}

\bibitem[\protect\citeauthoryear{{Sheth} \& {Tormen}}{{Sheth} \&
  {Tormen}}{1999}]{sheth1999}
{Sheth} R.~K.,  {Tormen} G.,  1999, \mn@doi [\mnras]
  {10.1046/j.1365-8711.1999.02692.x}, \href
  {http://adsabs.harvard.edu/abs/1999MNRAS.308..119S} {308, 119}

\bibitem[\protect\citeauthoryear{{Simet}, {McClintock}, {Mandelbaum}, {Rozo},
  {Rykoff}, {Sheldon}  \& {Wechsler}}{{Simet} et~al.}{2017}]{Simet2016}
{Simet} M.,  {McClintock} T.,  {Mandelbaum} R.,  {Rozo} E.,  {Rykoff} E.,
  {Sheldon} E.,   {Wechsler} R.~H.,  2017, \mn@doi [\mnras]
  {10.1093/mnras/stw3250}, \href
  {http://adsabs.harvard.edu/abs/2017MNRAS.466.3103S} {466, 3103}

\bibitem[\protect\citeauthoryear{{Sohn}, {Geller}, {Rines}, {Hwang}, {Utsumi}
  \& {Diaferio}}{{Sohn} et~al.}{2017}]{sohnetal17}
{Sohn} J.,  {Geller} M.~J.,  {Rines} K.~J.,  {Hwang} H.~S.,  {Utsumi} Y.,
  {Diaferio} A.,  2017, preprint, \href
  {http://adsabs.harvard.edu/abs/2017arXiv171200872S} {} (\mn@eprint {arXiv}
  {1712.00872})

\bibitem[\protect\citeauthoryear{{Springel}}{{Springel}}{2005}]{Springel2005}
{Springel} V.,  2005, \mn@doi [\mnras] {10.1111/j.1365-2966.2005.09655.x},
  \href {http://adsabs.harvard.edu/abs/2005MNRAS.364.1105S} {364, 1105}

\bibitem[\protect\citeauthoryear{Springel et~al.,}{Springel
  et~al.}{2017}]{Springel2017}
Springel V.,  et~al., 2017, arXiv, astro-ph.GA

\bibitem[\protect\citeauthoryear{{Suzuki} et~al.,}{{Suzuki}
  et~al.}{2012}]{suzuki12}
{Suzuki} N.,  et~al., 2012, \mn@doi [\apj] {10.1088/0004-637X/746/1/85}, \href
  {http://adsabs.harvard.edu/abs/2012ApJ...746...85S} {746, 85}

\bibitem[\protect\citeauthoryear{{Takada} \& {Spergel}}{{Takada} \&
  {Spergel}}{2014}]{Takada2014}
{Takada} M.,  {Spergel} D.~N.,  2014, \mn@doi [\mnras] {10.1093/mnras/stu759},
  \href {http://adsabs.harvard.edu/abs/2014MNRAS.441.2456T} {441, 2456}

\bibitem[\protect\citeauthoryear{{Tinker}, {Kravtsov}, {Klypin}, {Abazajian},
  {Warren}, {Yepes}, {Gottl{\"o}ber}  \& {Holz}}{{Tinker}
  et~al.}{2008}]{Tinker2008}
{Tinker} J.,  {Kravtsov} A.~V.,  {Klypin} A.,  {Abazajian} K.,  {Warren} M.,
  {Yepes} G.,  {Gottl{\"o}ber} S.,   {Holz} D.~E.,  2008, \mn@doi [\apj]
  {10.1086/591439}, \href {http://adsabs.harvard.edu/abs/2008ApJ...688..709T}
  {688, 709}

\bibitem[\protect\citeauthoryear{{Tinker}, {Robertson}, {Kravtsov}, {Klypin},
  {Warren}, {Yepes}  \& {Gottl{\"o}ber}}{{Tinker} et~al.}{2010}]{tinker2010}
{Tinker} J.~L.,  {Robertson} B.~E.,  {Kravtsov} A.~V.,  {Klypin} A.,  {Warren}
  M.~S.,  {Yepes} G.,   {Gottl{\"o}ber} S.,  2010, \mn@doi [\apj]
  {10.1088/0004-637X/724/2/878}, \href
  {http://adsabs.harvard.edu/abs/2010ApJ...724..878T} {724, 878}

\bibitem[\protect\citeauthoryear{{Velliscig}, {van Daalen}, {Schaye},
  {McCarthy}, {Cacciato}, {Le Brun}  \& {Dalla Vecchia}}{{Velliscig}
  et~al.}{2014}]{Velliscig2014}
{Velliscig} M.,  {van Daalen} M.~P.,  {Schaye} J.,  {McCarthy} I.~G.,
  {Cacciato} M.,  {Le Brun} A.~M.~C.,   {Dalla Vecchia} C.,  2014, \mn@doi
  [\mnras] {10.1093/mnras/stu1044}, \href
  {http://adsabs.harvard.edu/abs/2014MNRAS.442.2641V} {442, 2641}

\bibitem[\protect\citeauthoryear{{Vikhlinin} et~al.,}{{Vikhlinin}
  et~al.}{2009}]{Vikh2009}
{Vikhlinin} A.,  et~al., 2009, \mn@doi [\apj] {10.1088/0004-637X/692/2/1033},
  \href {http://adsabs.harvard.edu/abs/2009ApJ...692.1033V} {692, 1033}

\bibitem[\protect\citeauthoryear{{Villaescusa-Navarro}, {Marulli}, {Viel},
  {Branchini}, {Castorina}, {Sefusatti}  \& {Saito}}{{Villaescusa-Navarro}
  et~al.}{2014}]{Villa2014}
{Villaescusa-Navarro} F.,  {Marulli} F.,  {Viel} M.,  {Branchini} E.,
  {Castorina} E.,  {Sefusatti} E.,   {Saito} S.,  2014, \mn@doi [\jcap]
  {10.1088/1475-7516/2014/03/011}, \href
  {http://adsabs.harvard.edu/abs/2014JCAP...03..011V} {3, 011}

\bibitem[\protect\citeauthoryear{{Wechsler} \& {Tinker}}{{Wechsler} \&
  {Tinker}}{2018}]{wechslertinker18}
{Wechsler} R.~H.,  {Tinker} J.~L.,  2018, preprint, \href
  {http://adsabs.harvard.edu/abs/2018arXiv180403097W} {} (\mn@eprint {arXiv}
  {1804.03097})

\bibitem[\protect\citeauthoryear{{Zehavi} et~al.,}{{Zehavi}
  et~al.}{2011}]{zehavietal11}
{Zehavi} I.,  et~al., 2011, \mn@doi [\apj] {10.1088/0004-637X/736/1/59}, \href
  {http://adsabs.harvard.edu/abs/2011ApJ...736...59Z} {736, 59}

\bibitem[\protect\citeauthoryear{{Zheng} et~al.,}{{Zheng}
  et~al.}{2005}]{zhengetal05}
{Zheng} Z.,  et~al., 2005, \mn@doi [\apj] {10.1086/466510}, \href
  {http://adsabs.harvard.edu/abs/2005ApJ...633..791Z} {633, 791}

\bibitem[\protect\citeauthoryear{{Zu}, {Mandelbaum}, {Simet}, {Rozo}  \&
  {Rykoff}}{{Zu} et~al.}{2017}]{zuetal17}
{Zu} Y.,  {Mandelbaum} R.,  {Simet} M.,  {Rozo} E.,   {Rykoff} E.~S.,  2017,
  \mn@doi [\mnras] {10.1093/mnras/stx1264}, \href
  {http://adsabs.harvard.edu/abs/2017MNRAS.470..551Z} {470, 551}

\bibitem[\protect\citeauthoryear{{de Haan} et~al.,}{{de Haan}
  et~al.}{2016}]{SPTSZ2016}
{de Haan} T.,  et~al., 2016, \mn@doi [\apj] {10.3847/0004-637X/832/1/95}, \href
  {http://adsabs.harvard.edu/abs/2016ApJ...832...95D} {832, 95}

\bibitem[\protect\citeauthoryear{{van Uitert} et~al.,}{{van Uitert}
  et~al.}{2018}]{kidsgama2018}
{van Uitert} E.,  et~al., 2018, \mn@doi [\mnras] {10.1093/mnras/sty551}, \href
  {http://adsabs.harvard.edu/abs/2018MNRAS.476.4662V} {476, 4662}

\bibitem[\protect\citeauthoryear{{von der Linden} et~al.,}{{von der Linden}
  et~al.}{2014a}]{wtgI}
{von der Linden} A.,  et~al., 2014a, \mn@doi [\mnras] {10.1093/mnras/stt1945},
  \href {http://adsabs.harvard.edu/abs/2014MNRAS.439....2V} {439, 2}

\bibitem[\protect\citeauthoryear{{von der Linden} et~al.,}{{von der Linden}
  et~al.}{2014b}]{vdlinden2014}
{von der Linden} A.,  et~al., 2014b, \mn@doi [\mnras] {10.1093/mnras/stu1423},
  \href {http://adsabs.harvard.edu/abs/2014MNRAS.443.1973V} {443, 1973}

\makeatother
\end{thebibliography}



\appendix
\section{Cluster number counts covariance matrix}
\label{app:NC_cov}
The covariance matrix for the cluster number counts is computed adding the different contributions listed in Section~\ref{sec:methods:like}:
\begin{equation}
\label{eqn:cov}
 {\bm C}= {\bm C}^{\rm Poisson}+{\bm C}^{\rm Samp Var}+{\bm C}^{\rm Misc} \, .
\end{equation}
The first two terms, which account for the statistical uncertainty, are computed analytically along the MCMC for the relevant cosmological and model parameters.
The Poisson contribution to the covariance matrix is simply given by the expectation value for the number counts in the specific bin (cf. Equation~\ref{eqn:NC}):
\begin{equation}
 {\bm C}^{\rm Poisson} = \delta_{\rm ii} \langle {\bm N} \rangle_i
\end{equation}
The sample variance terms read \citep{hukravtsov03}
\begin{equation}
{\bm C}_{\rm ij}^{\rm Samp Var} = \langle b N(\Delta \lambda\ob_i, \Delta z\ob_i) \rangle \langle b N(\Delta \lambda\ob_j, \Delta z\ob_j) \rangle \sigma^2(V_i,V_j)
\end{equation}
where the first two terms are defined as:
\begin{multline}
\langle b N(\Delta \lambda\ob_i, \Delta z\ob_i) \rangle =  \int_0^{\infty}  \de z\true\ 
  {\Omega_{\rm mask}} \frac{\de V}{\de z\true \de \Omega}  \int_{\Delta z\ob_i} \de z\ob\ P(z\ob|z\true)_{\Delta\lambda\ob_i} \\
  \int \de M b(M,z)\, n(M,z) \int_{\Delta \lambda\ob_{\rm i}} \de \lambda\ob P(\lambda\ob | M,z)
\end{multline}
and the last one corresponds to the rms variance of the linear density field:
\begin{equation}
\label{eqn:rms}
\sigma^2(V_i,V_j) = \int \frac{d{\bm k}}{(2\pi)^3}\sqrt{ P_L(k,z_i)P_L(k,z_j)} W_i({\bm k}) W_j({\bm k}) \, .
\end{equation}
Here, $b(M,z)$ it the linear halo bias for which we use the \citet{tinker2010} formula, $V_i$ is the comoving volume corresponding to the redshift bin $z_{ob}$, and $W_i({\bm k})$ the Fourier transform of the window function.  Approximating the survey mask with a top-hat window symmetric around the azimuthal axis, and setting the angular area $\Omega_{\rm mask} = 2\pi ( 1-\cos(\theta_s))$ equal to the total survey area, $W_i({\bm k})$ reads:
\begin{multline}
 W_i({\bm k}) = \left( \left. \frac{dV}{d\Omega} \right|_{\substack{\Delta z_i}} \right)^{-1} \int_{\Delta z_i} dz  \frac{dV}{dz d\Omega} 
 4 \pi \sum_{l=0}^{\infty} \sum_{m=-l}^{l} (i)^l j_l (k \chi(z)) Y_{l,m}(\hat{k}) K_{l}
\end{multline}
where $j_l(x)$ are the spherical Bessel functions, $\chi(z)$ is the comoving distance to redshift $z$, $Y_{l,m}(\hat{k})$ are the spherical harmonics, and $K_{l}$ the coefficients of the expansion in spherical harmonics of the angular part of the window function:
\begin{eqnarray}
 {\rm for}\quad  l=0 \quad && K_l=\frac{1}{2\sqrt{\pi}} \\ 
 {\rm for}\quad  l\neq 0  \quad && K_l=\sqrt{\frac{\pi}{2l+1}}\frac{P_{l-1}(\cos(\theta_s))-P_{l+1}(\cos(\theta_s))}{\Omega_{\rm mask}} \nonumber  ,
\end{eqnarray}
where $P_l(x)$ are the Legendre polynomials.

Finally, the term due to the miscentering correction, ${\bm C}^{\rm Misc}$, is estimated numerically from $1000$ realizations of the number counts data (corrected for the miscentering error) obtained sampling the offset distribution parameters from their priors as described in Zhang et al., in preparation.

\section{Skew-normal approximation}

\begin{figure}
\begin{center}
    \includegraphics[width=0.45 \textwidth]{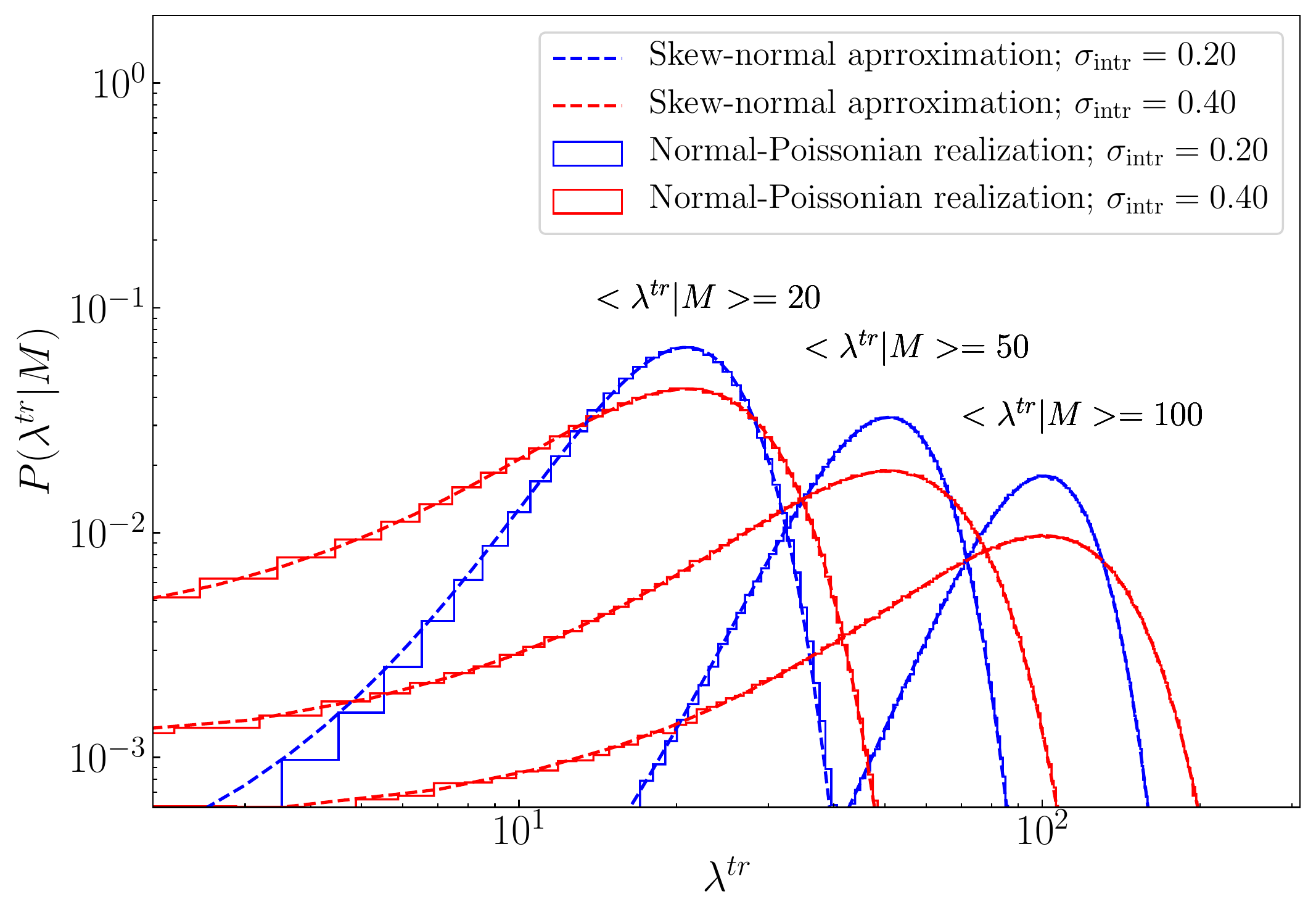}
\end{center}
\caption{Comparison of the convolution of a normal and Poissonian distribution ({\it histograms}) with a skew-normal distribution ({\it solid lines}) for different values of $\langle \lambda^{\rm sat}|M \rangle$ and $\sigma_{\rm intr}$ (see labels).}
\label{fig:skewnorm}
\end{figure}

The richness--mass relation $P(\lambda\true|M)$ is a convolution of a Poissonian and a Gaussian distribution.  In this work, we approximate the resulting convolution with a skew-normal distribution:
\begin{equation}
 P(\lambda\true|M)=\frac{1}{\sqrt{2\pi\sigma^2}} e^{-\frac{(\lambda\true - \langle \lambda^{\rm sat}|M \rangle)^2}{2\sigma^2} } {\rm erfc}\left[ -\alpha \frac{\lambda\true - \langle \lambda^{\rm sat}|M \rangle}{\sqrt{2 \sigma^2}} \right] \, .
\end{equation}
The values of the model parameters $\alpha$ and $\sigma$ vary as a function of the expectation value $\langle \lambda^{\rm sat}|M \rangle$ and intrinsic scatter $\sigma_{\rm intr}$.  We derive these values by fitting the skew-normal distribution to realizations of a normal-Poissonian convolution obtained by varying $\langle \lambda^{\rm sat}|M \rangle$ and $\sigma_{\rm intr}$ along the relevant range of values for this analysis.  Each $P(\lambda\true)$ realization is generated from $10^6$ realizations of the true richness obtained as $\lambda\true=1+\Delta^{\rm Poisson} + \Delta^{\rm Gauss} $, where $\Delta^{\rm Poisson}$ is a random number drawn from a Poisson distribution having mean $\langle \lambda^{\rm sat} \rangle$ and $\Delta^{\rm Gauss}$ a random number drawn from a Gaussian distribution having null mean and scatter equal to $\sigma_{\rm intr}\langle \lambda^{\rm sat} \rangle$.
Figure \ref{fig:skewnorm} compares the histograms obtained from these realizations to the resulting best fit skew-normal distribution.  We calibrate the parameters of the skew-normal distribution along a dense grid in $\langle \lambda^{\rm sat} \rangle$ and $\sigma_{\rm intr}$ and linearly interpolate along this 2D grid to define the skew-normal parameters at every point in parameter space.

\label{app:skewnorm}

\section*{Affiliations}

$^{1}$ Universit\"ats-Sternwarte, Fakult\"at f\"ur Physik, Ludwig-Maximilians Universit\"at M\"unchen, Scheinerstr. 1, 81679 M\"unchen, Germany\\
$^{2}$ Department of Physics, University of Arizona, Tucson, AZ 85721, USA\\
$^{3}$ Department of Physics and Astronomy, University of California Riverside, 900 University Ave, Riverside, CA 92521, USA\\
$^{4}$ Fermi National Accelerator Laboratory, P. O. Box 500, Batavia, IL 60510, USA\\
$^{5}$ Department of Astronomy, University of Michigan, Ann Arbor, MI 48109, USA\\
$^{6}$ Department of Physics, University of Michigan, Ann Arbor, MI 48109, USA\\
$^{7}$ Kavli Institute for Particle Astrophysics \& Cosmology, P. O. Box 2450, Stanford University, Stanford, CA 94305, USA\\
$^{8}$ SLAC National Accelerator Laboratory, Menlo Park, CA 94025, USA\\
$^{9}$ Santa Cruz Institute for Particle Physics, Santa Cruz, CA 95064, USA\\
$^{10}$ Department of Physics, Stanford University, 382 Via Pueblo Mall, Stanford, CA 94305, USA\\
$^{11}$ Department of Physics and Astronomy, Pevensey Building, University of Sussex, Brighton, BN1 9QH, UK\\
$^{12}$ Department of Physics and Astronomy, Stony Brook University, Stony Brook, NY 11794, USA\\
$^{13}$ Department of Physics, Carnegie Mellon University, Pittsburgh, Pennsylvania 15312, USA\\
$^{14}$ Max Planck Institute for Extraterrestrial Physics, Giessenbachstrasse, 85748 Garching, Germany\\
$^{15}$ Excellence Cluster Universe, Boltzmannstr.\ 2, 85748 Garching, Germany\\
$^{16}$ Department of Physics and Astronomy, Pevensey Building,University of Sussex, Brighton, BN1 9QH, UK\\
$^{17}$ Cerro Tololo Inter-American Observatory, National Optical Astronomy Observatory, Casilla 603, La Serena, Chile\\
$^{18}$ Department of Physics \& Astronomy, University College London, Gower Street, London, WC1E 6BT, UK\\
$^{19}$ Department of Physics and Electronics, Rhodes University, PO Box 94, Grahamstown, 6140, South Africa\\
$^{20}$ Institute of Cosmology \& Gravitation, University of Portsmouth, Portsmouth, PO1 3FX, UK\\
$^{21}$ LSST, 933 North Cherry Avenue, Tucson, AZ 85721, USA\\
$^{22}$ Laborat\'orio Interinstitucional de e-Astronomia - LIneA, Rua Gal. Jos\'e Cristino 77, Rio de Janeiro, RJ - 20921-400, Brazil\\
$^{23}$ Observat\'orio Nacional, Rua Gal. Jos\'e Cristino 77, Rio de Janeiro, RJ - 20921-400, Brazil\\
$^{24}$ Department of Astronomy, University of Illinois at Urbana-Champaign, 1002 W. Green Street, Urbana, IL 61801, USA\\
$^{25}$ National Center for Supercomputing Applications, 1205 West Clark St., Urbana, IL 61801, USA\\
$^{26}$ Institut de F\'{\i}sica d'Altes Energies (IFAE), The Barcelona Institute of Science and Technology, Campus UAB, 08193 Bellaterra (Barcelona) Spain\\
$^{27}$ Institut d'Estudis Espacials de Catalunya (IEEC), 08193 Barcelona, Spain\\
$^{28}$ Institute of Space Sciences (ICE, CSIC),  Campus UAB, Carrer de Can Magrans, s/n,  08193 Barcelona, Spain\\
$^{29}$ Centro de Investigaciones Energ\'eticas, Medioambientales y Tecnol\'ogicas (CIEMAT), Madrid, Spain\\
$^{30}$ Faculty of Physics, Ludwig-Maximilians-Universit\"at, Scheinerstr. 1, 81679 Munich, Germany\\
$^{31}$ Department of Astronomy/Steward Observatory, 933 North Cherry Avenue, Tucson, AZ 85721-0065, USA\\
$^{32}$ Jet Propulsion Laboratory, California Institute of Technology, 4800 Oak Grove Dr., Pasadena, CA 91109, USA\\
$^{33}$ Kavli Institute for Cosmological Physics, University of Chicago, Chicago, IL 60637, USA\\
$^{34}$ Instituto de Fisica Teorica UAM/CSIC, Universidad Autonoma de Madrid, 28049 Madrid, Spain\\
$^{35}$ Institute of Astronomy, University of Cambridge, Madingley Road, Cambridge CB3 0HA, UK\\
$^{36}$ Kavli Institute for Cosmology, University of Cambridge, Madingley Road, Cambridge CB3 0HA, UK\\
$^{37}$ Department of Physics, ETH Zurich, Wolfgang-Pauli-Strasse 16, CH-8093 Zurich, Switzerland\\
$^{38}$ Center for Cosmology and Astro-Particle Physics, The Ohio State University, Columbus, OH 43210, USA\\
$^{39}$ Department of Physics, The Ohio State University, Columbus, OH 43210, USA\\
$^{40}$ Harvard-Smithsonian Center for Astrophysics, Cambridge, MA 02138, USA\\
$^{41}$ Australian Astronomical Observatory, North Ryde, NSW 2113, Australia\\
$^{42}$ Departamento de F\'isica Matem\'atica, Instituto de F\'isica, Universidade de S\~ao Paulo, CP 66318, S\~ao Paulo, SP, 05314-970, Brazil\\
$^{43}$ Department of Physics and Astronomy, University of Pennsylvania, Philadelphia, PA 19104, USA\\
$^{44}$ George P. and Cynthia Woods Mitchell Institute for Fundamental Physics and Astronomy, and Department of Physics and Astronomy, Texas A\&M University, College Station, TX 77843,  USA\\
$^{45}$ Department of Astronomy, The Ohio State University, Columbus, OH 43210, USA\\
$^{46}$ Instituci\'o Catalana de Recerca i Estudis Avan\c{c}ats, E-08010 Barcelona, Spain\\
$^{47}$ Brookhaven National Laboratory, Bldg 510, Upton, NY 11973, USA\\
$^{48}$ School of Physics and Astronomy, University of Southampton,  Southampton, SO17 1BJ, UK\\
$^{49}$ Brandeis University, Physics Department, 415 South Street, Waltham MA 02453\\
$^{50}$ Instituto de F\'isica Gleb Wataghin, Universidade Estadual de Campinas, 13083-859, Campinas, SP, Brazil\\
$^{51}$ Computer Science and Mathematics Division, Oak Ridge National Laboratory, Oak Ridge, TN 37831\\


\bsp
\label{lastpage}
\end{document}